\documentclass[aps,prx,twocolumn,superscriptaddress]{revtex4-2}
\usepackage[english]{babel}
\usepackage{amssymb}
\usepackage{amsmath}
\usepackage{mathtools}
\usepackage{txfonts}
\usepackage{mathdots}
\usepackage[normalem]{ulem}
\usepackage{array}
\usepackage{amssymb}
\usepackage{amsfonts}
\usepackage{amsmath}
\usepackage{mathrsfs}
\usepackage{booktabs}
\usepackage{threeparttable}
\usepackage{multirow}
\usepackage[dvips]{graphicx}
\usepackage{epsfig}
\usepackage{graphicx}
\usepackage{subfigure}
\usepackage{threeparttable}
\usepackage{chngpage}
\usepackage{float}
\usepackage[dvipsnames,svgnames,x11names,hyperref]{xcolor}
\usepackage{bm}
\usepackage[toc]{appendix}
\usepackage{tabularx}
\usepackage{adjustbox}
\usepackage{comment}
\usepackage{hyperref}
\usepackage{todonotes}
\usepackage{lineno}
\hypersetup{
	unicode=false,     
	pdftoolbar=false,  
	pdfmenubar=true,   
	pdffitwindow=false, 
	pdfstartview={FitH},
	pdftitle={},    
	pdfauthor={Authors},     
	pdfsubject={},   
	pdfcreator={},   
	pdfproducer={}, 
	pdfkeywords={quantum many-body scars} {superconducting processor} {quantum state tomography}, 
	pdfnewwindow=true,
	colorlinks=true,
	linkcolor=RoyalBlue,
	citecolor=blue, 
	filecolor=magenta,
	urlcolor=blue
}

\begin{document}
\nolinenumbers	
\title{Creating and controlling global Greenberger-Horne-Zeilinger entanglement on quantum processors}

\author{Zehang Bao}\thanks{These authors contributed equally}
\author{Shibo Xu}\thanks{These authors contributed equally}
\affiliation{School of Physics, ZJU-Hangzhou Global Scientific and Technological Innovation Center, and Zhejiang Province Key Laboratory of Quantum Technology and Device, Zhejiang University, Hangzhou, China}
\author{Zixuan Song}
\author{Ke Wang}
\author{Liang Xiang}
\author{Zitian Zhu}
\author{Jiachen Chen}
\author{Feitong Jin}
\author{Xuhao Zhu}
\author{Yu Gao}
\author{Yaozu Wu}
\author{Chuanyu Zhang}
\author{Ning Wang}
\author{Yiren Zou}
\author{Ziqi Tan}
\author{Aosai Zhang}
\author{Zhengyi Cui}
\author{Fanhao Shen}
\author{Jiarun Zhong}
\author{Tingting Li}
\author{Jinfeng Deng}
\author{Xu Zhang}
\author{Hang Dong}
\author{Pengfei Zhang}
\affiliation{School of Physics, ZJU-Hangzhou Global Scientific and Technological Innovation Center, and Zhejiang Province Key Laboratory of Quantum Technology and Device, Zhejiang University, Hangzhou, China}
\author{Yang-Ren Liu}
\affiliation{Kavli Institute for Theoretical Sciences, University of Chinese Academy of Sciences, Beijing 100190, China}
\author{Liangtian Zhao}
\author{Jie Hao}
\affiliation{Institute of Automation, Chinese Academy of Sciences, Beijing 100190, China.}
\author{Hekang Li}
\affiliation{School of Physics, ZJU-Hangzhou Global Scientific and Technological Innovation Center, and Zhejiang Province Key Laboratory of Quantum Technology and Device, Zhejiang University, Hangzhou, China}
\author{Zhen Wang}
\affiliation{School of Physics, ZJU-Hangzhou Global Scientific and Technological Innovation Center, and Zhejiang Province Key Laboratory of Quantum Technology and Device, Zhejiang University, Hangzhou, China}
\affiliation{Hefei National Laboratory, University of Science and Technology of China, Hefei, China}
\author{Chao Song}
\affiliation{School of Physics, ZJU-Hangzhou Global Scientific and Technological Innovation Center, and Zhejiang Province Key Laboratory of Quantum Technology and Device, Zhejiang University, Hangzhou, China}
\author{Qiujiang Guo}
\email{qguo@zju.edu.cn}
\affiliation{School of Physics, ZJU-Hangzhou Global Scientific and Technological Innovation Center, and Zhejiang Province Key Laboratory of Quantum Technology and Device, Zhejiang University, Hangzhou, China}
\affiliation{Hefei National Laboratory, University of Science and Technology of China, Hefei, China}

\author{Biao Huang}
\email{phys.huang.biao@gmail.com}
\affiliation{Kavli Institute for Theoretical Sciences, University of Chinese Academy of Sciences, Beijing 100190, China}

\author{H. Wang}
\email{hhwang@zju.edu.cn}
\affiliation{School of Physics, ZJU-Hangzhou Global Scientific and Technological Innovation Center, and Zhejiang Province Key Laboratory of Quantum Technology and Device, Zhejiang University, Hangzhou, China}
\affiliation{Hefei National Laboratory, University of Science and Technology of China, Hefei, China} 

\begin{abstract}

Greenberger-Horne-Zeilinger (GHZ) states~\cite{Greenberger1990}, also known as two-component Schr\"{o}dinger cats, play vital roles in the foundation of quantum physics and, more attractively, in future quantum technologies such as fault-tolerant quantum computation~\cite{Horodecki2009,Nielsen2011}. Enlargement in size and coherent control of GHZ states are both crucial for harnessing entanglement in advanced computational tasks with practical advantages, which unfortunately pose tremendous challenges as GHZ states are vulnerable to noise~\cite{Pezze2018, Florian2018RMP}. Here we propose a general strategy for creating, preserving, and manipulating large-scale GHZ entanglement, and demonstrate a series of experiments underlined by high-fidelity digital quantum circuits. For initialization, we employ a scalable protocol to create genuinely entangled GHZ states with up to 60 qubits, almost doubling the previous size record~\cite{Moses2023PRX}. For protection, we take a new perspective on discrete time crystals (DTCs)~\cite{Khemani2016,Else2016,Yao2017,Zaletel2023RMP,Khemani2019b,Else2020,Sacha2017, Else2016,Khemani2016,Keyserlingk2016,Yao2017,Else2017,Machado2020}, originally for exploring exotic nonequilibrium quantum matters, and embed a GHZ state into the eigenstates of a tailor-made cat scar DTC~\cite{Huang2023} to extend its lifetime. For manipulation, we switch the DTC eigenstates with in-situ quantum gates to modify the effectiveness of the GHZ protection. Our findings establish a viable path towards coherent operations on large-scale entanglement, and further highlight superconducting processors as a promising platform to explore nonequilibrium quantum matters and emerging applications.
\end{abstract}

\maketitle 

\begin{figure*}[ht]
    \includegraphics[width=17cm]{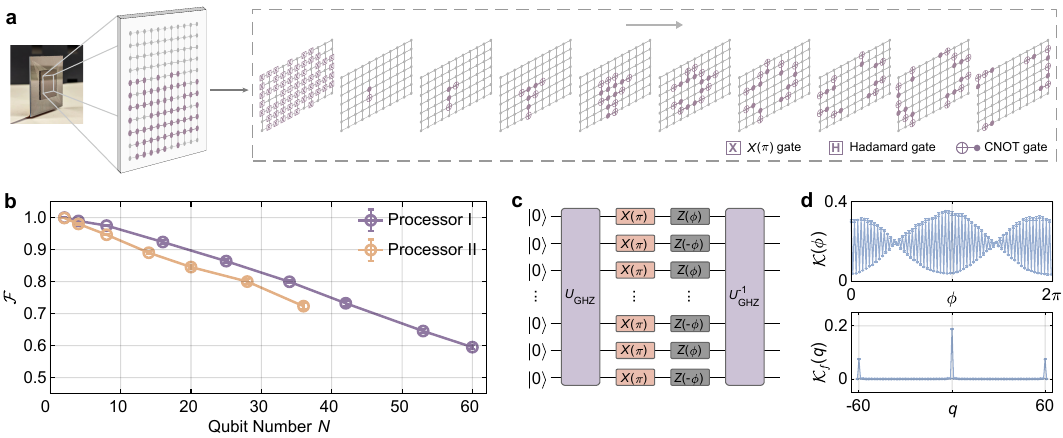}
    \caption{\label{fig:1} {\bf Generation and characterization of GHZ states.} 
        {\bf a}, Illustration of the superconducting quantum processor I and that of a general entangling protocol based on a set of quantum gates, the latter of which is further compiled into experimentally accessible elementary gates to generate the 60-qubit GHZ state. 
        {\bf b}, Measured GHZ state fidelity $\mathcal{F}$ as functions of qubit number $N$ for Processors I and II. The higher $\mathcal{F}$ for Processor I is likely due to its slightly better single-qubit gates.
        {\bf c}, MQC circuit diagram based on $Z(\pm\phi)$ and reversal of $U_\textrm{GHZ}$. $X(\pi)$ is a spin-echo pulse for preserving the qubit coherence, and virtual $Z(\phi)$ [$Z(-\phi)$] is applied to individual qubits in 0 (1) as recorded in basis $|\boldsymbol{s}\rangle$.
        {\bf d}, Measured $\mathcal{K}(\phi)$ for the 60-qubit GHZ state and its Fourier spectrum~$\mathcal{K}_f(q)$. Slow sinusoidal envelope results from sparse sampling~\cite{Wei2020PRB}, which does not affect our analysis. Error bars in all figures throughout the text, if shown, are obtained by repeated measurements. See Supplementary Information for more details.}
\end{figure*}

The ability to generate, preserve, and manipulate highly entangled quantum states is a long-term goal for building practical quantum computers that can outperform classical machines~\cite{Amico2008}. Among various multipartite entangled states, GHZ states constitute a peculiar class showing the strongest nonlocal entanglement for $N$ particles~\cite{Gisin1998}. On the other hand, they are the most fragile entangled states. External perturbations on any single particle can destroy the entanglement and thermalization can arise internally through many-body dynamics if interactions exist~\cite{Abanin2019RMP}. Therefore, creating high-quality GHZ states with larger size and higher fidelity is a standard benchmark for showing the performance of quantum hardware~\cite{Moses2023PRX, Graham2022Nature, Bluvstein2023Nature, Omran2019Science}. Although multipartite entanglement of tens of particles has been created across different physical platforms~\cite{Cao2023Nature, Graham2022Nature, Bluvstein2023Nature, Pogo2021PRX, Moses2023PRX, Song2019Science, Omran2019Science, Wang2018PRL}, the generation of maximally entangled GHZ states, achieving state fidelity of $\mathcal{F}>0.5$ which can verify $N$-particle entanglement, has so far been limited to $N\approx30$~\cite{Moses2023PRX, Mooney2021JPC, Song2019Science, Omran2019Science, Wang2018PRL}. Heading towards the more challenging realm of preserving and manipulating such fragile states, a fully-fledged experiment is still pending~\cite{Florian2018RMP}.

Preserving GHZ states using a discrete time crystal (DTC) is an uncharted territory.
Previously, DTC has attracted broad scientific interest as an exotic nonequilibrium matter~\cite{Zaletel2023RMP,Khemani2019b,Else2020,Sacha2017}, which extends the fundamental concept of spontaneous symmetry breaking to time translations~\cite{Wilczek2012,Wilczek2013}. Ergodicity breaking mechanisms of many-body localization (MBL)~\cite{Khemani2016,Else2016,Yao2017} and prethermalization~\cite{Else2017,Machado2020} have been employed to induce time-crystalline dynamics of product states across a wide range of physical platforms~\cite{Randall2021,Frey2022,Mi2022,Zhang2017,Choi2017,Kyprianidis2021,Beatrez2023,Stasiuk2023,Kongkhambut2022}. DTCs are also considered as potential candidates to accommodate GHZ states by their robust cat eigenstate pairs~\cite{Choi2018,Zaletel2023RMP}. However, this intriguing application has never been achieved. MBL DTC could generate numerous cat eigenstates, but the presence of disorders may lead to unpredictable instability~\cite{Sels2022,Morningstar2022a,Leonard2023}. Meanwhile, prethermal DTC is disorder-free, but the strong diffusion restrict cats eigenstates to be spatially homogeneous ones~\cite{Luitz2020,Khemani2019b,Ippoliti2021,Kyprianidis2021}.
By contrast, the third venue~\cite{Zaletel2023RMP} of weak ergodicity breaking~\cite{Turner2018,Maskara2021,Bernien2017,Bluvstein2021} by a cat scar DTC, where a few Fock-space localized cat eigenstates (cat scars) are deterministically engineered to define a subspace with time-crystalline ordering that is analytically tractable~\cite{Huang2023}, has come to the fore as a potential solution.

In this Article, we report a series of experiments evidencing the possibility of creating, preserving, and manipulating GHZ entanglement on superconducting quantum processors.
We first generate up to 60-qubit GHZ states with fidelities $\mathcal{F}$ all far above 0.5, unambiguously verifying genuine global entanglement. Creating these unprecedentedly large entanglement is enabled by the high fidelity of around 0.999 and 0.995 for single- and two-qubit gates respectively, and an efficient entangling scheme along radial path scalable in two dimensions (2D). We further digitally implement the cat scar DTC with thousands of quantum gates to protect the created GHZ state and manipulate its dynamics. To quantify the protection of DTC, we develop a quantum sensing protocol and observe a subharmonic temporal response for the macroscopic coherent phase of the GHZ state. Remarkably, the phase oscillation is observed throughout 30 cycles under generic perturbation, indicating a DTC lifetime longer than those under non-interacting Rabi drivings and under free decay. The oscillation amplitudes are unaffected even if we further manipulate both the GHZ state and cat scars during evolution, accomplishing a smooth in-situ switch of protection between different GHZ states.\\

\begin{figure*}[t]
    \includegraphics[width=17cm]{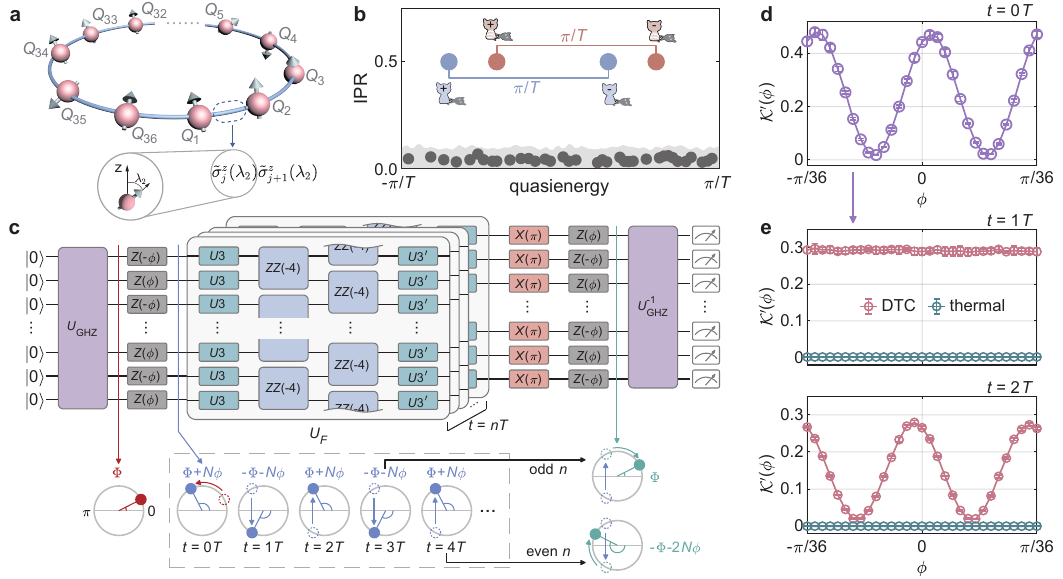} 
    \caption{\label{fig:2}
        {\bf Cat scar DTC and Schr\"{o}dinger cat interferometry.} 
        {\bf a}, Schematic representation of the 36-qubit Ising chain. Neighboring qubits are coupled by a perturbed Ising interaction.
        {\bf b}, Eigenstructure of a cat scar DTC. With strong Ising interaction $|J_j|=J\sim 1/T \gg|\lambda_1|,|\lambda_2|$, two pairs of cat scars ($|\Phi, \boldsymbol{s}\rangle_N$ with $\boldsymbol{s}=0101\dots$ and $0000\dots$, shown as blue and red dots, respectively, with IPR$\to0.5$) remain localized in Fock space under generic perturbations, in contrast to the majority thermal eigenstates (gray dots with IPR$\to0$). Two cat scars within each pair are separated by a quasienergy gap $\pi/T$.
        {\bf c}, Schr\"{o}dinger cat interferometry. The circuit is similar to the MQC protocol in Fig.~\ref{fig:1}{\bf c}, but with an extra layer of reversed phase rotations to detect the phase oscillations of a GHZ state. In the DTC unitary $U_{\rm F}$, $U3$ is the single-qubit rotation with 3 Euler angles and $ZZ(-4)=\exp(-{\rm i}\sigma^z_j\sigma^z_{j+1})$. Lower panel: Evolution of a GHZ state viewed on the $xy$ plane with the poles defined by $|\boldsymbol{s}\rangle$ and $|\bar{\boldsymbol{s}}\rangle$. The initial GHZ state picks up a phase due to $Z(\pm\phi)$ and becomes $|\Phi+N\phi, \boldsymbol{s}\rangle_N$. Under DTC evolutions inside the gray dashed box, phase oscillation occurs as the GHZ state alternates between $|\pm(\Phi+N\phi), \boldsymbol{s}\rangle_N$. Afterwards, the echo and reversed phase rotation double~(or cancel) the coherent phase $N\phi$ for even~(or odd) driving cycles. $U^{-1}_{\rm GHZ}$ disentangles the qubits and $\mathcal{K}^\prime(\phi,t)$ in Eq.~\eqref{eq:kphi} is measured by the ground state probability. 
        {\bf d}-{\bf e}, Exemplary measurements of $\mathcal{K}^\prime(\phi,t)$ at three consecutive instants for an initial GHZ state evolved by DTC~(red circles) or thermal unitaries~(green circles).
    }
\end{figure*}

\noindent \textbf{\MakeUppercase{Generating GHZ state}}

\noindent We first demonstrate the generation of $N$-qubit GHZ states
\begin{equation}
    \label{eq:GHZ_N}
|\Phi, \boldsymbol{s}\rangle_N = \left(|\boldsymbol{s}\rangle + e^{-{\rm i}\Phi}|\bar{\boldsymbol{s}}\rangle\right)/{\sqrt{2}},
\end{equation}
where $|\boldsymbol{s}\rangle$ is an $N$-bit Fock basis, with each bit encoding a qubit in either ground (0) or excited (1) state, and $|\bar{\boldsymbol{s}}\rangle$ is that with all bits of $|\boldsymbol{s}\rangle$ flipped. In this experiment, we choose $|\boldsymbol{s}\rangle$ to be of antiferromagnetic ordering, i.e., $|0101\dots\rangle$. The phase factor $\Phi$ quantifies the coherence between Fock bases $|\boldsymbol{s}\rangle$ and $|\bar{\boldsymbol{s}}\rangle$. 

To create $|\Phi, \boldsymbol{s}\rangle_N$ among qubits in 2D, we design an efficient protocol based on a set of unitaries including $X(\pi)$, Hadamard and CNOT gates~(see Methods). As illustrated in Fig.~\ref{fig:1}{a}, after a layer of single-qubit gates, this protocol starts with a CNOT on two qubits around the center of the qubit layout, and then radially entangles peripherals stepwise by appending layers of CNOTs. In the realization, we compile the set of unitaries in Fig.~\ref{fig:1}{a} into a digital quantum circuit composed of experimentally accessible single-qubit rotational and two-qubit controlled $\pi$-phase gates, whose combined effect is denoted with a unitary $U_\textrm{GHZ}$. Running similar digital quantum circuits we can entangle up to 60 qubits on Processor I and achieve genuine multipartite entanglement with $\mathcal{F}=\rm 0.595\pm0.008$ for $N=60$~(Fig.~\ref{fig:1}{b}). We emphasize that our protocol is universal as it can be adapted to any particular qubit layout topology in 2D. In a parallel effort, we entangle all $6 \times 6$ qubits on Processor II~\cite{Yao2023NP} with $\mathcal{F}=0.723\pm 0.010$ for $N=36$. Numerical simulations suggest that the reported $\mathcal{F}$ values are consistent with our calibrated gate fidelities~(see Methods and Supplementary Information).

We attempt to measure major elements of the GHZ density matrix to obtain $\mathcal{F}$. Two diagonal elements $P_{\boldsymbol{s}}$ and $P_{\bar{\boldsymbol{s}}}$, the probabilities of finding the qubits in Fock bases $|\boldsymbol{s}\rangle$ and $|\bar{\boldsymbol{s}}\rangle$ respectively, can be directly probed. Several methods such as measuring parity oscillation~\cite{Sackett2000Nature, Omran2019Science, Song2019Science} and sliced Wigner function~\cite{Rundle2017PRA} can be used to probe off-diagonal elements, but here we resort to the more scalable multiple quantum coherence~(MQC) protocol~\cite{Baum1985JCP,Martin2017NP,Doronin2003PRA,Wei2020PRB}.
With the MQC circuit shown in Fig.~\ref{fig:1}{c}, the appropriate phase gates $Z(\pm \phi)$~(see Methods) on individual qubits imprint an enhanced phase of $N \phi$, resulting in $|-(\Phi+ N\phi), \boldsymbol{s}\rangle_N$. Subsequent reversal of $U_{\rm GHZ}$, referred to as $U_{\rm GHZ}^{-1}$, disentangles these $N$ qubits and steer them back to ground state $|0000\dots\rangle$ with a probability $\mathcal{K}(\phi)$, which displays fast sinusoidal oscillations at a rate $\propto N$. Figure~\ref{fig:1}{d} exemplifies such measured $\mathcal{K}(\phi)$ signal for $N=60$ qubits and the corresponding Fourier amplitude, in which the Fourier peak $\mathcal{K}_f(q=N)$ characterizes the off-diagonal elements. As such, GHZ state fidelity is given by $\mathcal{F}=(P_{\boldsymbol{s}} + P_{\bar{\boldsymbol{s}}})/2 + \sqrt{\mathcal{K}_f(N)}$~\cite{Wei2020PRB} (Fig.~\ref{fig:1}{b}). We emphasize that the MQC protocol tends to underestimate $\mathcal{F}$ since the detection is not instantaneous but involves a long sequence of gates in $U_{\rm GHZ}^{-1}$~(see Supplementary Information).\\

\noindent \textbf{\MakeUppercase{Cat scar DTC}}

\noindent For a GHZ state defined in Eq.~\eqref{eq:GHZ_N}, where $|\boldsymbol{s}\rangle$ can be more generic than the antiferromagnetic pattern with alternating 0 and 1, we are able to design and realize a cat scar DTC model that naturally accommodates the entanglement. Here and below we focus on Processor II with 36 qubits for proof-of-principle experiments. As illustrated in Fig.~\ref{fig:2}{a}, we construct a perturbed Ising chain ($N=36$) of periodic boundary on Processor II. Under the periodic driving, the Floquet unitary $U_{\rm F}=U_{2}U_{1}$ per cycle is given by
\begin{align} \nonumber
    U_1 &= \left(\prod_{j=1}^N e^{{-{\rm i} \varphi_1 \sigma^z_j/2}} e^{{{\rm i}\lambda_1 \sigma^y_j/2}} e^{{-{\rm i} \varphi_2 \sigma^z_j/2}}\right)e^{-{\rm i}\pi\sum_{j=1}^N \sigma^x_j/2}
    \\ \label{eq:uf}
    U_2 &= e^{-{\rm i}\sum_{j=1}^N J_j\tilde{\sigma}^z_j(\lambda_2)\tilde{\sigma}^z_{j+1}(\lambda_2)},
\end{align}
where $\sigma^{x,y,z}_j$ are Pauli matrices on $Q_j$, $\varphi_1$ and $\varphi_2$ are introduced to break the integrability of the model while avoiding fine-tuned echoes, 
and $\lambda_1$ is the single-qubit perturbing strength. $U_2$ characterizes the perturbed Ising interaction with $\tilde{\sigma}^z_j(\lambda_2)=\cos(\lambda_2) \sigma^z_j + \sin(\lambda_2) \sigma^x_j$.
The strong Ising interaction $|J_j|=J$ comparable with Floquet driving frequency $1/T$ and the qubit-flip pulses $ e^{-{\rm i}\pi\sum_{j=1}^N \sigma^x_j/2}$ are essential ingredients.

In the unperturbed limit $\lambda_1,\lambda_2=0$, 
Ising interaction structures all eigenstates to be degenerate doublets $|\boldsymbol{s}\rangle, |\bar{\boldsymbol{s}}\rangle$, while spin-flip pulses further combine them into cat eigenstates. In particular, there are two pairs of cat eigenstates isolated from all the others by large quasienergy or qubit pattern differences, such that they, as cat scars, remain robust when all perturbations are turned on~\cite{Huang2023}, as illustrated in Fig.~\ref{fig:2}b. 
Here, the inverse participation ratio for a Floquet eigenstate $|\epsilon_m\rangle$, i.e.  $U_{\rm F}|\epsilon_m\rangle=e^{{\rm i}\epsilon_m}|\epsilon_m\rangle$, reads IPR($\epsilon_m$)=$\sum_{\boldsymbol{s}} |\langle\epsilon_m|\boldsymbol{s}\rangle|^4$.
A larger value of IPR indicates stronger Fock space localization, and therefore better quality of a cat eigenstate to store and protect a GHZ state.
It is seen that two pairs of cat scars (IPR$\to0.5$) stand out, based on which we experimentally choose the homogeneous case $J_j = +1$ so that one of the two pairs naturally accommodate the generated GHZ state.

We implement $U_{\rm F}$ in the DTC regime with perturbations $\lambda_1=\lambda_2=0.05$ and strong detuning from echoes $\varphi_1=-\pi/2, \varphi_2=\pi/2-0.6$. This is realized by a digital quantum circuit~(Fig.~\ref{fig:2}{c}). 
To quantify a dynamical GHZ state, we design a quantum sensing protocol 
dubbed Schr\"{o}dinger cat interferometry. As shown in Fig.~\ref{fig:2}c, only an extra layer of reversal phase rotation $Z(\pm\phi)$ is introduced here, such that the scalability of MQC protocol is fully inherited. 
The ground state probability measured at the end of the circuit in Fig.~\ref{fig:2}c corresponds to the physical quantity
\begin{equation}
    \label{eq:kphi}
    \mathcal{K}^\prime(\phi, t) = \left| \langle-(\Phi +N \phi), \boldsymbol{s}| U_{\rm F}^{t/T} |\Phi+N \phi,\boldsymbol{s}\rangle \right|^2.
\end{equation}
GHZ state oscillations $U_{\rm F}^{t/T}|\Phi + N\phi, \boldsymbol{s}\rangle \sim |(-1)^{t/T} (\Phi + N\phi),\boldsymbol{s}\rangle$
are then sharply revealed by the alternation of $\mathcal{K}^\prime(\phi,t)$ between constructive $\sim \cos(2N\phi+\Phi)$ and total destructive $\sim 1$ interference for $\phi$-dependence at consecutive driving periods. As exemplified in Fig.~\ref{fig:2}{d}, for an initial GHZ state~($N=36$), the measured $\mathcal{K}^\prime(\phi,t=0)$ exhibits an evident period-$\pi/36$ oscillation (Fig.~\ref{fig:2}{d}). Under the DTC dynamics, the oscillation vanishes at odd period $t=1T$~(the upper panel in Fig.~\ref{fig:2}{e}) and reappears in the subsequent even period $t=2T$~(the lower panel in Fig.~\ref{fig:2}{e}). In contrast, a thermal system modeled by large perturbations~($\lambda_1=0.3, \lambda_2=0.4$) quickly erases the initial global entanglement, leaving a vanishing $\mathcal{K}^\prime(\phi,t)$ (Fig.~\ref{fig:2}{e})~(see Methods and Supplementary Information).\\

\noindent \textbf{\MakeUppercase{Preserving GHZ state}}

\begin{figure}[t]
        \includegraphics[width=8.5cm]{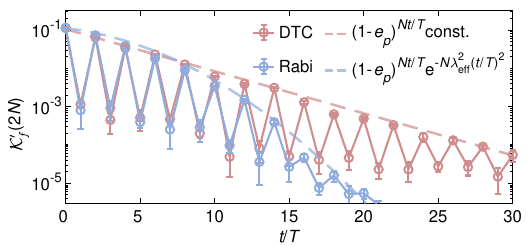}
        \caption{\label{fig:3}
        {\bf GHZ dynamics preserved in cat scar DTC.} 
        Measured $\mathcal{K}^\prime_f(2N,t)$ dynamics in DTC~(red circles) and the benchmark against that under non-interacting Rabi drivings~(blue circles). Dashed lines are analytical results.
        Here the constant in legend is $\sqrt{2}\cdot\sqrt{\textrm{IPR}}\approx 0.92$ with the IPR of cat scar given by analytical perturbation theory, while for the Rabi driving case $\lambda_{\text{eff}}\approx 0.0239$, the coefficient parametrizing the combined effect of all perturbative factors, can be rigorously obtained~(see Supplementary Section 5). The effective cycle error per qubit is estimated based on an apparent match between the analytical results and experimental data, which yields $e_p=0.007$ in DTC and $e_p=0.003$ for the Rabi driving case. 
        }
\end{figure}

To illustrate the long-time dynamics and benchmark the protective effects of DTC, we perform Fourier transformation of $\mathcal{K}^\prime(\phi,t)$ on $\phi$ for $t$ from $0$ to $30T$, 
where the Fourier peak $\mathcal{K}_f^\prime(q=2N,t)$ exhibits a period-$2T$ oscillation of DTC orders~(Fig.~\ref{fig:3}), corresponding to the pattern alternations as shown in Fig.~\ref{fig:2}e. In comparison, we perform a parallel measurement for $\mathcal{K}_f^\prime(2N,t)$ in a non-interacting Rabi model, which amounts to turning off the two-qubit gates for the DTC while keeping all single-qubit Rabi drivings intact, i.e., $U_{\rm F} = U_1$ in Eq.~\eqref{eq:uf}. A qualitative difference emerges in Fig.~\ref{fig:3}. In DTC, $\mathcal{K}_f^\prime(2N,t)$ is chiefly damped by external noise effects, leading to an exponential decay $\sim e^{-t}$. In contrast, the Rabi driving case suffers from an additional term $\sim e^{-t^2}$ due to the fact that $\lambda_1,\lambda_2 \neq 0$, which signals the delocalization of a GHZ state from the original Fock bases. This apparent difference in Fig.~\ref{fig:3} indicates that the cat scar DTC integrates both dynamical decoupling of Rabi drivings~\cite{Suter2016RMP} and strong Ising interactions, achieving an improved protection on GHZ states. Note for the free-decay case without any protection, measured $\mathcal{K}_f^\prime(q=2N,t)$ drops approximately three times more quickly than that in DTC (data not shown).\\

\begin{figure*}[t]
    \includegraphics[width=17cm]{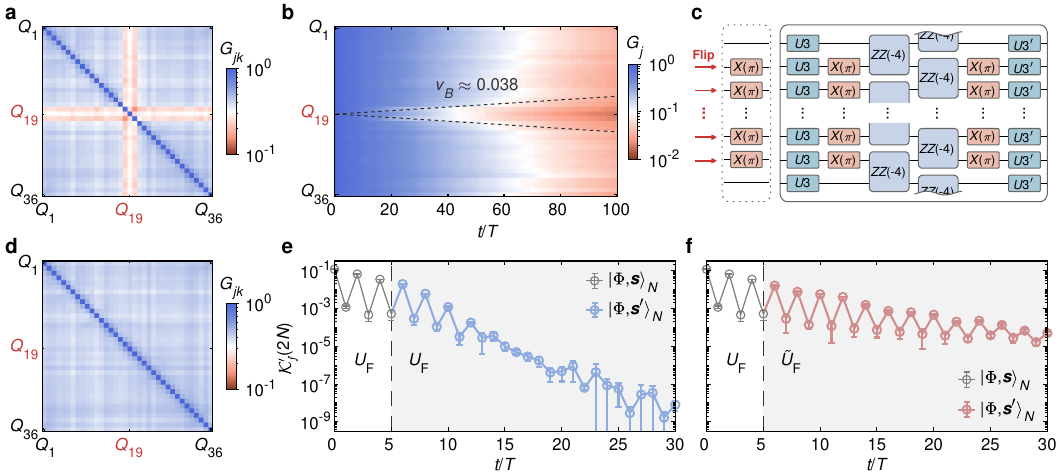}
    \caption{{\bf Manipulating cat scar DTC to protect the switched GHZ states.} 
    {\bf a}, $G_{jk}(t)$ measured, e.g., at $t=24T$ under the evolution of $U_{\rm F}$ for a 36-qubit GHZ state, which is created by flipping $Q_{19}$ of $|\Phi, \boldsymbol{s}\rangle_N$ at $t=0$.
    {\bf b}, Measured $G_{j}(t)$ dynamics for the same initial GHZ state as in {\bf a}, where a light cone emerges around the flipped site $Q_{19}$. Dashed lines are the analytical predictions of the thermalizing light cone with the mean butterfly velocity $v_B\approx0.038$~(see Supplementary Section 5).
    {\bf c}, Exemplary quantum circuit diagram of $\tilde{U}_{\rm F}$, which illustrates a scheme to edit the original $U_\textrm{F}$ for effectively reversing the sign of local Ising interaction with $X(\pi)$ gates.
    Exact circuit layout of $\tilde{U}_{\rm F}$ depends on the spin pattern of the generic GHZ state, which is produced by the left-most layer of $X(\pi)$ flip gates acting on $|\Phi, \boldsymbol{s}\rangle$. {\bf d}, $G_{jk}(t)$ measured at $t=24T$ under the evolution of a compatible $\tilde{U}_{\rm F}$ for the same initial GHZ state as in {\bf a}. In {\bf a}, {\bf b}, and {\bf d}, experimental results are sample-averaged including 10 random $\varphi_1$ to exclude the effects of possible single-qubit echoes, and for each $\varphi_1$ the flipped spin is sampled over six physical qubits to reduce the detrimental effect of the qubit non-uniformity. 
    {\bf e}, Measured $\mathcal{K}_f^\prime(2N)$ dynamics under the evolution of the original $U_{\rm F}$. The GHZ state is switched from the initial $|\Phi, \boldsymbol{s}\rangle_N$ to $|\Phi, \boldsymbol{s}^\prime\rangle_N$ by flipping 18 qubits at $t=5T$.
    {\bf f}, Measured $\mathcal{K}_f^\prime(2N)$ dynamics for conditions similar to those in {\bf e}, except that the DTC unitary is switched from $U_\textrm{F}$ to a compatible $\tilde{U}_{\rm F}$ at $t=5T$. The cat scar DTC timely catches up with spin flips, so that protection is kept effective at longer times compared with that in {\bf e}.}
    \label{fig:4}
\end{figure*}

\noindent \textbf{\MakeUppercase{Manipulating GHZ dynamics}}

\noindent In previous experiments, we have fixed the antiferromagnetic qubit pattern in a GHZ state and focused on the dynamics of coherent phase $\Phi$. Practically, it is desirable to switch the scarred subspace such that it becomes compatible with a generic GHZ state, even better if the switch takes place seamlessly during evolution. To identify the method of editing scarred subspace, we first note that thermalization in a cat scar DTC occur in a structured way. Specifically, under strong and uniform Ising interaction $J_j=1$, a spin can only be flipped by perturbations if it is sandwiched by {\it anti-parallel} neighbors, i.e., $011\leftrightarrow 001$, because such a process conserves the Ising energy. Contrarily, antiferromagnetic patterns (i.e., $0101\dots$) are immune to perturbations, while global anti-ferromagnetic states constitute scarred subspace. Such a constraint is revealed by the site-resolved detection of the connected correlation function
\begin{align}
    G_{jk} (t) =
    |\langle 
    \sigma^z_j(t)\sigma^z_k(t) \rangle 
    -
    \langle \sigma^z_j(t) \rangle \langle{\sigma^z}_k(t)\rangle|,
\end{align}
where ${\sigma^z_j}(t) = (U_{\rm F}^{t/T})^\dagger{\sigma^z_j}U_{\rm F}^{t/T}$. 
It approaches 1 for a perfect GHZ state, while $G_{jk}(t)\rightarrow0$ if the $Q_j$-$Q_k$ pair is disentangled. In Figs.~\ref{fig:4}a and b, we initialize a 36-qubit GHZ state $|\Phi, \boldsymbol{s}\rangle_N$ and flip $Q_{19}$ immediately, so that the 5-qubit chain in $\boldsymbol{s}$, $Q_{17}Q_{18}Q_{19}Q_{20}Q_{21}$, changes from ``01010'' to ``01110''. Then, thermalization is ignited at $Q_{18}$ and $Q_{20}$ according to the kinetic constraint, as we see in Fig.~\ref{fig:4}a a cross-shaped thermal region centering around $Q_{19}$ occurs for $G_{jk}(t=24T)$. Taking an average $G_{j}(t)=(1/35)\sum_{k\neq j}G_{jk}(t)$, we observe a light-cone $|j-19|=v_Bt$ propagating from $Q_{18}$ and $Q_{20}$ in Fig.~\ref{fig:4}b, with the analytical $v_B\approx 0.038$ approximately obtained under the kinetic constraint condition. The fact that thermalization occurs locally strongly indicates that a local dressing can also hinder such a process.

We exemplify the modification of cat scarred subspace in a new $\tilde{U}_{\rm F}$, where we locally reverse the sign of Ising interaction at $J_{18}=J_{19}=-1$ while keeping all other $J_j=+1$ unchanged. Such a sign reversal is experimentally realized by inserting a pair of $X(\pi)$ gates on the flipped $Q_{19}$, which are located around the $ZZ(-4)$ gate of the original $U_{\rm F}$ sequence as illustrated in Fig.~\ref{fig:4}c. Then, the kinetic constraint is modified locally for $Q_{18}$ and $Q_{20}$, so that each of them is only vulnerable to perturbative flips if it is sandwiched by {\em parallel} neighbors. Contrarily, processes like ``011''$\leftrightarrow$``001'' for spin chain $Q_{17}Q_{18}Q_{19}$, now violates the conservation of local Ising energy, i.e. $-J_{17} + J_{18} = -2 \neq  +J_{17} - J_{18}=+2$, with  $J_{17}=-J_{18}=1$, and therefore cannot occur. Thus, the source of thermalization in Figs.~\ref{fig:4}a and b is extinguished, and the new GHZ state, $|\Phi, \boldsymbol{s}\rangle_N$ with $Q_{19}$ flipped, now resides inside the new scarred subspace. Correspondingly, we recover $G_{jk}(t)$ in Fig.~\ref{fig:4}d for the previous cross-shaped thermal region. 

To demonstrate dynamical switching and benchmark the efficiency of editing scarred subspace with $X(\pi)$ gates,
we consider the GHZ state with pattern $\boldsymbol{s}^\prime = 00110011\dots$, which
is obtained by generating $|\Phi, \boldsymbol{s}\rangle_N$ and then flipping qubits with the indices $j=4m+2\,\&\,4m+3,\,\textrm{for}\,m=0,1,\dots,N/4-1$. Such a generic GHZ state thermalizes most rapidly under the original $U_{\rm F}$ with $J_j=+1$, because every single qubit is a source for thermalization. A compatible $\tilde{U}_{\rm F}$ involves $N/2$ pairs of $X(\pi)$ gates, with one pair for each flipped qubit. We start with the 36-qubit GHZ state, $|\Phi, \boldsymbol{s}\rangle_N$, which oscillates in a compatible cat scar DTC ($U_{\rm F}$) for 5 cycles as in Fig.~\ref{fig:3}. Then, we flip appropriate qubits to produce $|\Phi, \boldsymbol{s^\prime}\rangle_N$, and continue the evolution under two conditions: The GHZ state is evolved in the original $U_{\rm F}$, verifying a rapid decay after the switch (Fig.~\ref{fig:4}e); in contrast, the phase oscillation persists resulting from a simultaneous switching from $U_{\rm F}$ to the new $\tilde{U}_{\rm F}$ (Fig.~\ref{fig:4}f), witnessing a similar amplitude as in Fig.~\ref{fig:3}a.\\

\noindent\textbf{\MakeUppercase{conclusion and outlook}}

\noindent Here, a set of concepts and protocols to preserve, control, and detect macroscopic quantum coherence in nonequilibrium many-body dynamics is developed, opening a new avenue for exploring large-scale GHZ states and practical applications of nonequilibrium quantum matters~\cite{Zaletel2023RMP, Choi2017}. We not only create an unprecedented 60-qubit GHZ state with genuine global entanglement, but also push the research front towards preserving its coherence and controlling its dynamics. Meanwhile, for the studies of DTC, our findings offer the long-sought-after direct evidence of spectral-paired cat eigenstates, which establishes a new perspective of using nonequilibrium eigenstructures to steer unconventional quantum dynamics.

In a broader spectrum, our findings bridge central topics in quantum computation with those in the emergent nonequilibrium quantum many-body physics~\cite{Zaletel2023RMP}. A tantalizing direction is to engineer the eigenstate structure of a wider range of exotic nonequilibrium matters as control knobs to steer multipartite entanglement~\cite{Maskara2021}. In addition to DTC, long-range entangled eigenstates also exist in Floquet spin liquids~\cite{Sun2023,Kalinowski2023}, dynamical scars in fracton matters~\cite{Pai2019}, and string-net models~\cite{Liu2022a}, based on which, a further development of our platform to larger size and higher fidelity, provides an ideal testbed to design new frameworks for versatile applications in quantum information, quantum metrology, and error correction.\\

\noindent\textbf{\MakeUppercase{Methods}}

\noindent Our experiments are carried out on two superconducting processors featuring transmon qubits arranged on a square lattice, one with 60 qubits selected~(Processor I in Fig.~\ref{fig:1}{a}) and the other one with 36 qubits (Processor II). Each qubit can be individually excited by microwave pulse for rotation of its state around an arbitrary axis in the $xy$ plane of the Bloch sphere, e.g., $x$-axis by an angle $\theta$, noted as $X(\theta)$; phase gate $Z(\theta)$ is virtually applied by recording the phase $\theta$ in subsequent microwaves. Single-qubit rotation with 3 Euler angles, referred to as $U3(\alpha,\beta,\theta)$ in the main text, is effectively a rotation gate plus a virtual $Z(\theta)$. Any two neighboring qubits have a tunable coupler, so that controlled $\phi$-phase gates can be dynamically implemented, which are used to assemble controlled $\pi$-phase gates for creating GHZ states and the two-qubit ZZ interaction required in cat scar DTC. For both processors, all physical single- and two-qubit gates are calibrated to be of high precision, with average gate fidelity around 0.999 and 0.995 respectively. As such, we are able to observe relevant experimental features even by executing digital quantum circuits with more than 300 layers in depth, which consist of about 7,000 quantum gates (see Supplementary Information for more details).\\

\noindent{\bf Acknowledgments:}  We thank Rubem Mondaini, Dong-Ling Deng for helpful discussion and Siwei Tan for technical support. The device was fabricated at the Micro-Nano Fabrication Center of Zhejiang University. Numerical simulation of quantum circuits was performed using MindSpore Quantum framework. We acknowledge support from the National Natural Science Foundation of China (Grants No. 92065204, 12274368, 12174389, U20A2076, 12174342, 12274367, and 12322414), the National Key Research and Development Program of China (Grant No. 2023YFB4502600), Innovation Program for Quantum Science and Technology (Grant No. 2021ZD0300200), the Zhejiang Provincial Natural Science Foundation of China (Grants No. LR24A040002 and LDQ23A040001), and Zhejiang Pioneer (Jianbing) Project (Grant No. 2023C01036). 
\\
\noindent{\bf Author contributions:}  B.H. and Q.G. conceived the idea; Z.B. and S.X. carried out the experiments and analyzed the experimental data under the supervision of Q.G., C.S., and H.W.; B.H. conducted the theoretical analysis; H.L. and J.C. fabricated the device supervised by H.W.; Q.G., B.H., Z.B., and H.W. co-wrote the manuscript. All authors contributed to the experimental setup, and/or the discussions of the results and the writing of the manuscript. 
\\
\noindent{\bf Competing interests:} The authors declare no competing interests. 
\\
\noindent{\bf Data and materials availability:} The data presented in the figures and that support the other findings of this study will be publically available upon its publication. All the relevant source codes are available from the corresponding authors upon reasonable request.
\\
\\

\bibliographystyle{naturemag}
\bibliography{catdtc.bib}
	
\end{document}


\nolinenumbers	
\title{Supplementary Information for\\
``Creating and controlling global Greenberger-Horne-Zeilinger entanglement on quantum processors"
}
\maketitle

\tableofcontents
\beginsupplement

\section{Experimental setup}
\label{secs:experimental_details}
\begin{figure}[t]
    \includegraphics[width=18cm]{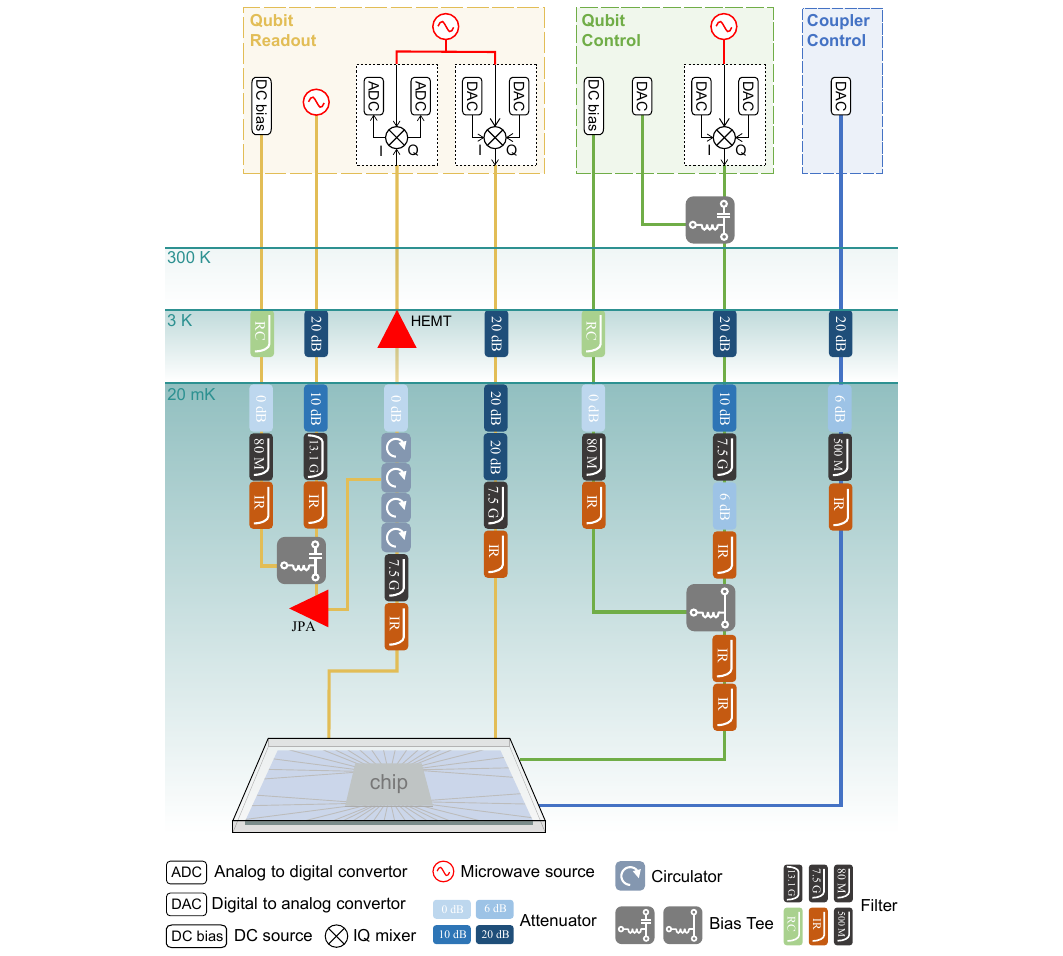}
    \caption{\textbf{Illustration of experimental setup.} Room-temperature electronics for controlling and measuring qubits on the processor include three parts: coupler control~(blue box), qubit control~(green box) and readout~(orange box). Fast Z control pulses for each qubit or coupler are generated with a single digital-to-analog converter (DAC) channel. Microwave pulses for either qubit readout or XY control are synthesized by mixing the local oscillator signal, generated by a microwave source, with two sideband signals generated by two DAC channels. DC bias outputs slow Z control pulses for biasing either the qubit idle frequencies or the operation frequencies of the Josephson parametric amplifiers~(JPAs). All control pulses are transmitted through a series of attenuators and filters at different temperatures, for isolation of noises, before being injected into the flip-chip superconducting quantum processor, which is mounted on the mixing chamber plate of a dilution refrigerator. 
    }
    \label{fig:wiring}
\end{figure}

Our experiments include generating large-scale GHZ states and observing the phase oscillation of the GHZ state in the cat scar discrete time crystal~(DTC). The first part is performed on two different quantum processors, labeled as I~\cite{Xu2023CPB} and  II~\cite{Yao2023NP}, respectively. Processor I~(II) is a two-dimensional~(2D) flip-chip superconducting processor consisting of $121$ ($36$) frequency-tunable transmon qubits and 200~(60) tunable couplers. The 121~(36) qubits are arranged as an $11\times 11$ ($6\times 6$) square lattice and each nearest-neighbor (NN) qubit pair is connected by a tunable coupler in order to tune the NN coupling strength. Each processor is mounted on the mixing chamber plate of a dilution refrigerator~(DR) with a temperature of $\sim$20~mK~(Fig.~\ref{fig:wiring}). Experimental setups for controlling and measuring the two processors are similar, which are illustrated in Fig.~\ref{fig:wiring}.

\section{Device performance}
\begin{figure}[t]
    \includegraphics[width=18cm]{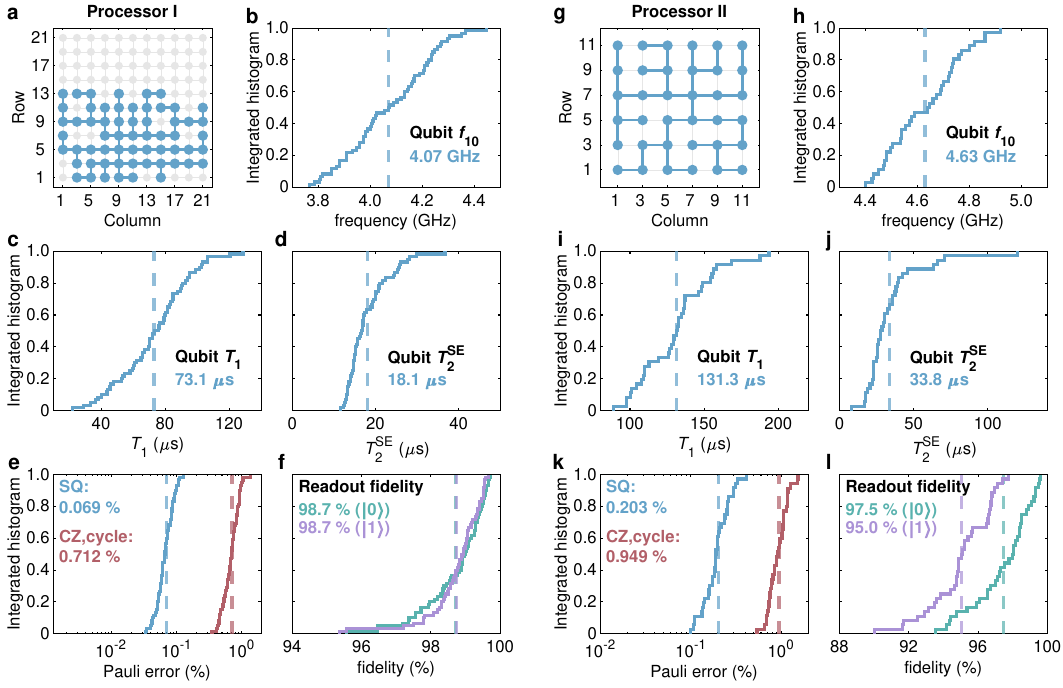}
    \caption{{\bf Device performance for Processor I ({\bf a}-{\bf f}) and Processor II ({\bf g}-{\bf l}).} 
    {\bf a}, {\bf g}, Qubit layout for the two processors. Processor I~(II) consists of an $11\times11$~($6\times6$) qubit array where we use up to 60~(36) qubits to generate GHZ states. Blue dots and lines denote the actively-used qubits and couplers for the generation of GHZ states, respectively. {\bf b}, {\bf h}, Distribution of qubit idle frequency $f_{10}$. {\bf c}, {\bf i}, Distribution of qubit relaxation time $T_1$ measured at the idle frequency. {\bf d}, {\bf j}, Distribution of qubit pure dephasing time $T_2^{\rm SE}$ measured with spin echo sequence at the idle frequency. {\bf e}, {\bf k}, Cycle Pauli error for single-qubit gates~(blue line) and two-qubit CZ gates~(red line) for generating GHZ states. Single-qubit errors are measured with XEB sequences running on 60~(36) qubits simultaneously for Processor I~(II). For single-qubit XEB, a cycle only consists of one single-qubit gate. Two-qubit cycle errors are obtained by performing simultaneous XEB for each two-qubit layer in the GHZ state generation circuit, which is plotted in Fig.~\ref{fig:circuit_ghz_processor_1} (Fig.~\ref{fig:circuit_ghz_processor_2}) for Processor I~(II). During the simultaneous XEB for each two-qubit layer, those qubits that are not involved in CZ gates are also applied with single-qubit XEB sequences. For the two-qubit CZ gate, a cycle consists of one CZ gate and two single-qubit gates. {\bf f}, {\bf l}, Qubit readout fidelity for $|0\rangle$ (green line) and $|1\rangle$ (purple line). Vertical dashed lines in these panels denote the mean values.}
    \label{fig:sq_params}
\end{figure}

We select 60 qubits~(Fig.~\ref{fig:sq_params}a) on Processor I to generate GHZ states due to limited wirings. For Processor II, all the 36 qubits are available and thus we can generate global entanglement across the whole device. Figures~\ref{fig:sq_params}a-f~(g-l) display the basic performance of all the actively-used qubits on Processor I~(II), including qubit idle frequency $f_{10}$, energy relaxation time $T_1$, spin echo dephasing time $T_2^{\rm SE}$,  single-qubit~(SQ) gate cycle error, two-qubit CZ gate cycle error, and readout fidelity. Notably, the average Pauli errors benchmarked with simultaneous cross entropy benchmarking~(XEB)~\cite{Boixo2018NP, Arute2019Nature} are $0.069\%$ ($0.203\%$) for single-qubit gates and $0.574\%$ ($0.543\%$) for two-qubit CZ gates on Processor I~(II). We use single-qubit gates and two-qubit CZ gates to realize the parallel quantum circuits for generating GHZ state in Fig.~1a in the main text.

\section{GHZ states calibration}

Multipartite entanglement plays a pivotal role in quantum information processing, serving as a crucial resource to speed up quantum algorithms~\cite{Buluta2019Science} and realize error-correcting codes~\cite{Knill2005Nature}, etc. Among various entangled states, Greenberger-Horne-Zeilinger (GHZ) states are particularly significant since they possess global entanglement. However, they are also extremely vulnerable to noise, which makes the generation of large-scale GHZ states a formidable challenge. Experiment efforts on generating large-scale GHZ states verifying genuine multipartite entanglement with fidelity $\mathcal{F}>0.5$ have been made in a range of quantum platforms including Rydberg atom arrays~\cite{Omran2019Science} (20 qubits), trapped ions~\cite{Pogo2021PRX, Moses2023PRX}~(up to 32 qubits), photons~\cite{Wang2018PRL}~(18 qubits), and superconducting qubits~\cite{Song2019Science, Mooney2021JPC}~(up to 27 qubits). 

In our experiments, we exploit the  2D architecture of our superconducting quantum chips and design efficient digital quantum circuits to generate $N$-qubit GHZ states. Leveraging on the high-fidelity single- and two-qubit gates on our device~(Fig.~\ref{fig:sq_params}), we achieve a 60-qubit GHZ state with a fidelity $\mathcal{F}=0.595\pm0.008$, unambiguously proving genuine multipartite entanglement with $\mathcal{F}>0.5$.

\begin{figure*}[t]
    \includegraphics[width=18cm]{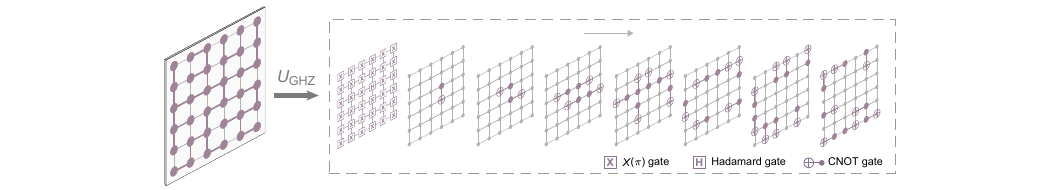}
    \caption{{\bf Illustration of the quantum circuit for generating the 36-qubit GHZ state on Processor II.} 
    Left panel illustrates the radial path to build up entanglement, where purple dots and lines denote the actively-used qubits and couplers, respectively. Right panel bounded by dashed lines shows the full sequence of circuits to generate the anti-ferromagnetic GHZ state stepwise. 
    }
    \label{fig:ghz_cartoon_circuit}
\end{figure*}

\subsection{GHZ state generation circuits}\label{Sec: preparation of GHZ states}
The paradigmatic scheme to generate an $N$-qubit GHZ state $(|0101\dots01\rangle + |1010\dots10\rangle)/{\sqrt{2}}$ in one dimension follows a procedure starting from the Fock state $|0101\dots01\rangle$: The first qubit is initialized to a superposition state $(1/{\sqrt{2}})(|0\rangle+|1\rangle) \otimes |101\dots 01\rangle$ by a Hadamard gate, after which a CNOT gate is applied to entangle it with its neighbor originally in $|1\rangle$ to produce the Bell state $(1/{\sqrt{2}}) (|01\rangle+|10\rangle)\otimes |01\dots 01\rangle$; sequential CNOT gates are then applied to include more neighboring qubits one by one to achieve the global multipartite entanglement. The procedure described above is time-consuming, as it requires $N-1$ layers of CNOT gates to create an $N$-qubit GHZ state. However, since the qubits are arranged in 2D for our chips, each qubit can be entangled with more neighbors simultaneously and the circuit depth can be shortened. In this way, we generate the GHZ states using the protocols illustrated in Fig.~1a of the main text and Fig.~\ref{fig:ghz_cartoon_circuit}, where the first single-qubit layer is designed to determine the spin pattern of the final GHZ state, and the following 9 (7) layers of CNOT gates build up the entanglement along a radial path for 60 (36) qubits. 

\begin{figure*}[t]
    \includegraphics[width=18cm]{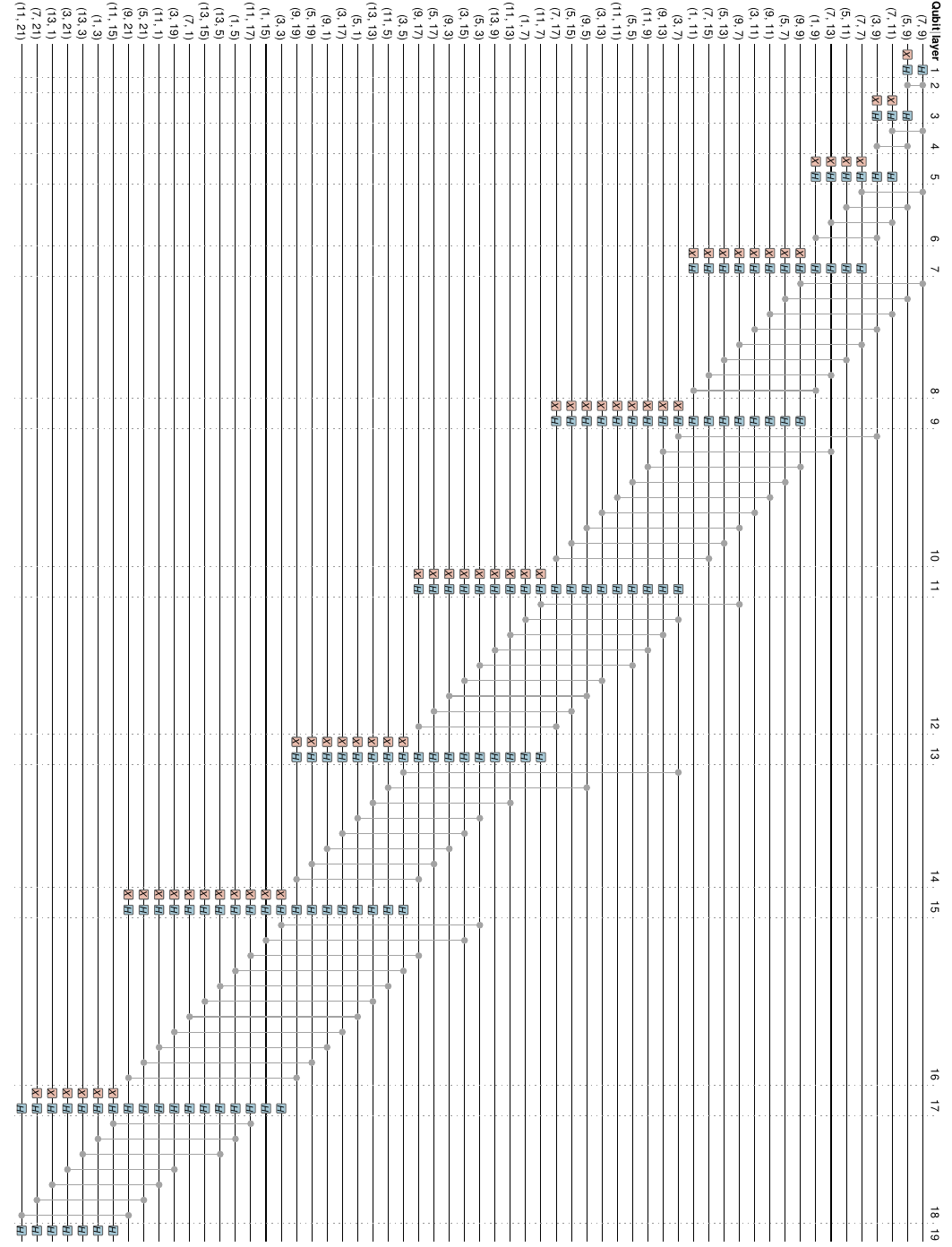}
    \caption{{\bf Experimental circuit for generating the 60-qubit GHZ state on Processor I.} This circuit diagram is obtained by compiling the circuit in Fig.~1a of the main text into single-qubit gates [$X(\pi)$, Hadamard gate] and two-qubit CZ gates. The whole circuit contains 19 layers, including 10 layers of parallel single-qubit gates and 9 layers of two-qubit gates. Each qubit is labeled by its column and row numbers according to the index map defined in Fig.~\ref{fig:sq_params}a. 
    }
    \label{fig:circuit_ghz_processor_1}
\end{figure*}

\begin{figure*}[t]
    \includegraphics[width=18cm]{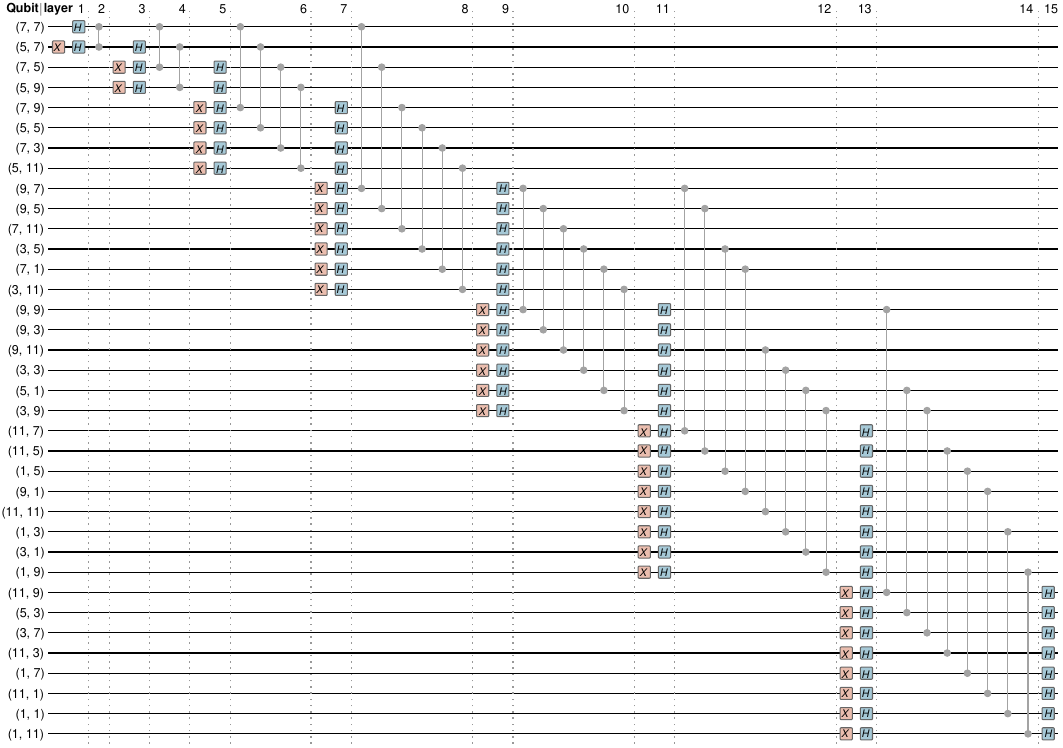}
    \caption{{\bf Experimental circuit for generating the 36-qubit GHZ state on processor II.} Similar to Fig.~\ref{fig:circuit_ghz_processor_1}, this circuit is obtained by compiling the circuit in Fig.~\ref{fig:ghz_cartoon_circuit}, which includes 8 layers of parallel two-qubit CZ gates and 7 layers of simultaneous single-qubit gates. Each qubit is labeled by its column and row numbers according to the index map defined in Fig.~\ref{fig:sq_params}g.
    }
    \label{fig:circuit_ghz_processor_2}
\end{figure*}

In the experimental realization, we further compile these circuits using the experimentally accessible single- and two-qubit CZ gates on our processors. These gates are calibrated with high fidelities~(see Figs.~\ref{fig:sq_params}e and k), which makes it possible to generate sizable GHZ states with considerably high fidelities. The compiled circuit for generating the GHZ state of 60 (36) qubits on Processor I~(II) is shown in Fig.~\ref{fig:circuit_ghz_processor_1} (Fig.~\ref{fig:circuit_ghz_processor_2}), which contains 19 (15) physical circuit layers directly implemented for the state generation experiments. 

\subsection{Characterization of GHZ states}
To obtain the fidelity of a GHZ state, we need to measure two diagonal and two off-diagonal terms of its density matrix (see Eqs.~\eqref{smeq:GHZ} and \eqref{smeq:fidelitya}), from which we can calculate its fidelity. Typically, the fidelity $\mathcal{F}>0.5$ witnesses the genuine $N$-particle entanglement~\cite{Sackett2000Nature}. The two diagonal terms can be obtained by directly measuring the population of the two composing Fock states. The two off-diagonal terms, which describe the quantum coherence between the two components, can be assessed by parity~\cite{Sackett2000Nature} or multiple quantum coherence (MQC) measurements~\cite{Wei2020PRB, Baum1985JCP}, with the latter being more suitable for systems equipped with digital gates, especially when the system size scales up. In this work, we adopt MQC to characterize the GHZ state~(Fig.~1d of the main text). As a sanity check, we also benchmark the GHZ state fidelity by measuring the parity oscillations for the system size $N\leq20$, with the results shown in Fig.~\ref{fig:ghz_scaling_sim}.

\subsubsection{Multiple quantum coherence~(MQC)}
\begin{figure}[ht]
    \includegraphics[width=18cm]{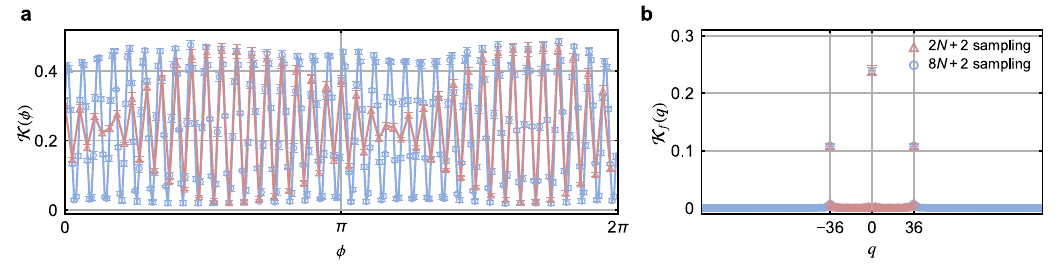}
    \caption{{\bf Effect of sampling size on MQC spectra--sparse sampling vs. dense sampling.} 
    {\bf a}, Experimentally measured $\mathcal{K}({\phi})$ for the 36-qubit GHZ state. Red triangles denote the results by the sparse sampling of $2N+2$ points, and blue circles show those obtained by the dense sampling with $8N + 2$ points. 
    {\bf b}, Fourier spectra of $\mathcal{K}({\phi})$.
	}
    \label{fig:mqc_detailed}
\end{figure}

MQC spectra are originally developed in nuclear magnetic resonance experiments~\cite{Baum1985JCP} to characterize the many-body correlations. Its applications have been extended to probe out-of-time-order correlations~\cite{Martin2017NP}, localization effects~\cite{Alvarez2010PRL}, and many-body entangled states~\cite{Wei2020PRB}. MQC only requires measuring the ground state probability and thus is regarded as a scalable metric for experimentally verifying many-body entanglement~\cite{Wei2020PRB}. 
Nevertheless, the protocol requires the implementation of additional reversed circuits to disentangle a GHZ state, which demands high-fidelity, versatile programmability, and long coherence time of quantum gates that can be satisfied by our system.
In the following, we employ the procedure in Ref.~\cite{Wei2020PRB} as the main scheme to benchmark the macroscopic quantum coherence of the generated GHZ states, which we recapitulate below~(the circuit is shown in Fig.~1c of the main text.)

\begin{enumerate}
    \item Starting from $N$-qubit ground state $|00\dots0\rangle$, we apply a quantum circuit $U_{\text{GHZ}}$~(Fig.~\ref{fig:ghz_cartoon_circuit}) to prepare the GHZ state: $U_{\text{GHZ}}|00\dots0\rangle=|\Phi, \boldsymbol{s}\rangle_N = \frac{1}{\sqrt{2}}(|0101\dots01\rangle+ e^{-{\rm i}\Phi} |1010\dots10\rangle)$, where the spin pattern $\boldsymbol{s}=0101\dots01$.
    \item For each qubit, we apply an $X(\pi)$ gate as the spin echo pulse to suppress the dephasing noise and a subsequent phase rotation $\phi$ (or $-\phi$) at the even (or odd) sites. The GHZ state then gains a collective phase as $|\Phi + N\phi, \boldsymbol{s}\rangle_N=\frac{1}{\sqrt{2}}(|0101\dots01\rangle+e^{-{\rm i} (\Phi +N\phi)}|1010\dots10\rangle)$.
    \item We disentangle the GHZ state $|\Phi+N\phi, \boldsymbol{s}\rangle_N$ with a reversed circuit $U_{\rm GHZ}^{-1}$ and obtain $\frac{1}{\sqrt{2}}[(1+e^{-iN\phi})|0\rangle + (1-e^{-iN\phi})|1\rangle]\otimes |0000\dots00\rangle$.
    \item In the ideal case, the ground state probability $\mathcal{K}({\phi})$ is given by $\frac{1}{2}[1+\cos(N\phi)]$.
\end{enumerate}
$\mathcal{K}(\phi)$ can also be expressed by the following form
\begin{equation}
    \mathcal{K}({\phi}) = |\langle 00\dots0|U_{\text{GHZ}}^\dagger U_{Z}(\phi)U_{X}U_{\text{GHZ}}|00\dots0\rangle|^2 =  \text{Tr}(\rho({\phi})\rho), 
\end{equation}
where $U_Z({\phi}) = \prod_{j=1}^N Z_j [(-1)^{s_j}\phi]$, $U_{X}=\prod_{j=1}^N X_j(\pi)$, $\rho = |\Phi,\boldsymbol{s}\rangle_N{}_N\langle \Phi,\boldsymbol{s}|$, $\rho({\phi}) =U_{Z}(\phi)U_{X}\rho U_{X}^\dagger U^\dagger_{Z}(\phi)$. The density matrix $\rho$ can be written into blocks as $\rho = \sum_{m_{\boldsymbol{s}},m_{\boldsymbol{s'}}}\rho_{m_{\boldsymbol{s}},m_{\boldsymbol{s'}}}|m_{\boldsymbol{s}}\rangle\langle m_{\boldsymbol{s'}}| = \sum_q\rho_q$, where $\rho_q = \sum_{m_{\boldsymbol{s}}}\rho_{m_{\boldsymbol{s}},m_{\boldsymbol{s}}-q}|m_{\boldsymbol{s}}\rangle\langle m_{\boldsymbol{s}}-q|$, $m_{\boldsymbol{s}}=\sum_js_j$, and $|{\boldsymbol{s}}\rangle=|s_j\rangle^{\otimes N}$. Due to $U_Z(\phi)\rho_q U_Z(\phi)^\dagger = e^{iq\phi}\rho_{q}$ and $U_X\rho_q U_X^\dagger = \rho_{-q}$, we have 
\begin{equation}
   \mathcal{K}({\phi}) = \text{Tr}(\rho({\phi})\rho) = \text{Tr}(\sum_{q} e^{iq\phi}\rho_{-q}\sum_p\rho_p) = \sum_{q}e^{iq\phi}\text{Tr}(\rho_{-q}\rho_{q}).
\end{equation}
By performing the discrete Fourier transformation to the experimentally measured $\mathcal{K}({\phi})$, we can obtain $\mathcal{K}_f(q) = \mathcal{N}_s^{-1}|\sum_{\phi} e^{iq\phi}\mathcal{K}({\phi})|$, where the normalization constant $\mathcal{N}_s$ is the sampling number of $\phi$. Thus, the off-diagonal term of the GHZ state $|\rho_{0101\dots,1010\dots}|^2=\text{Tr}(\rho_{N}\rho_{-N}) =\text{Tr}(\rho_{N}\rho_{N}^\dagger)$ is given by $\mathcal{K}_f(N)$. 

\begin{figure*}[t]
    \includegraphics[width=18cm]{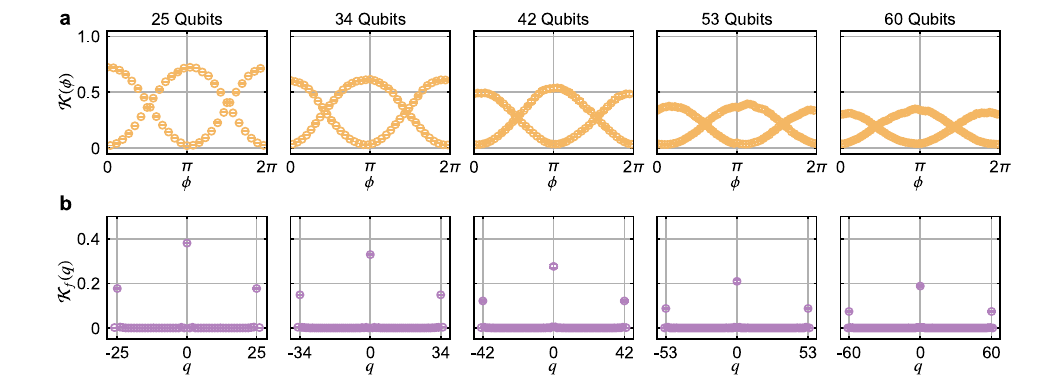}
    \caption{{\bf MQC results for different system sizes on processor I.} 
    {\bf a}, Experimentally measured $\mathcal{K}({\phi})$. 
    {\bf b}, Amplitudes of the Fourier spectrum for $\mathcal{K}({\phi})$. The highest three peaks appear at $q=0$ and $q=\pm N$, where $N$ is the number of qubits. The absolute value of the major off-diagonal elements for the GHZ states can be calculated by the Fourier amplitude $\mathcal{K}_f(q=N)$. }
    \label{mqc_processor_1}
\end{figure*}

\begin{figure*}[t]
    \includegraphics[width=18cm]{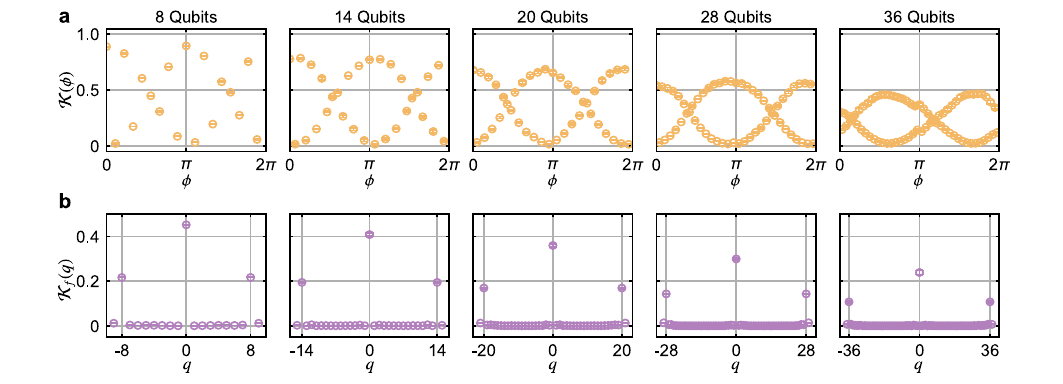}
    \caption{{\bf MQC results for different system sizes on processor II.} 
    {\bf a}, Experimentally measured $\mathcal{K}({\phi})$. 
    {\bf b}, Amplitudes of the Fourier spectrum for $\mathcal{K}({\phi})$. The highest three peaks appear at $q=0$ and $q=\pm N$, where $N$ is the number of qubits. The absolute value of the major off-diagonal elements for the GHZ states can be calculated from the MQC amplitude $\mathcal{K}_f(q=N)$. }
    \label{mqc_processor_2}
\end{figure*}

To accurately measure $\mathcal{K}_f(N)$ without sacrificing the efficiency, we follow the convention of Ref.~\cite{Wei2020PRB} and measure $\mathcal{K}({\phi})$ with a sparse sampling on $\phi$ with $\phi = \frac{\pi j}{N+1}$~($j=0,1,2,...,2N+1$).
We compare the results of sparse sampling~($\mathcal{N}_s=2N+2$)  with a dense one~($\mathcal{N}_s=8N+2$) for the 36-qubit GHZ state, which are shown in Fig.~\ref{fig:mqc_detailed}. In spite of less sample points for $\phi$~(Fig.~\ref{fig:mqc_detailed}b), the measured MQS spectra $\mathcal{K}_f(q)$ for sparse sampling $\mathcal{N}_s=2N+2$ match well with that of $\mathcal{N}_s=8N+2$. They show almost the same peak value at $q=36$, which signifies the existence of the $N$-particle entanglement for the prepared GHZ state. Therefore, we can safely measure the $\mathcal{K}({\phi})$ signals by sampling $2N+2$ points~($\phi = \frac{\pi j}{N+1}$, $j=0,1,2,...,2N+1$) in our experiments. For each $\mathcal{K}({\phi})$, we take $60,000$ measurement shots on Processor I and $72,000$ shots on Processor II. Since MQC only requires measuring the ground state probabilities, readout errors occur mostly in the low-excitation subspace, which enables us to correct the measurement errors in a scalable way~\cite{Wei2020PRB}. We rank the Fock bases according to their measured raw probabilities from high to low, and pick up the first 256 Fock bases for readout corrections. The measured $\mathcal{K}(\phi)$ after readout correction and its Fourier spectra $\mathcal{K}_f(q)$ are shown in Fig.~\ref{mqc_processor_1} for Processor I and Fig.~\ref{mqc_processor_2} for Processor II. In addition, we use the same readout correction scheme to eliminate readout errors in measuring diagonal terms $P_{\boldsymbol{s}}$, $P_{\bar{\boldsymbol{s}}}$.

\begin{figure}[t]
    \includegraphics[width=18cm]{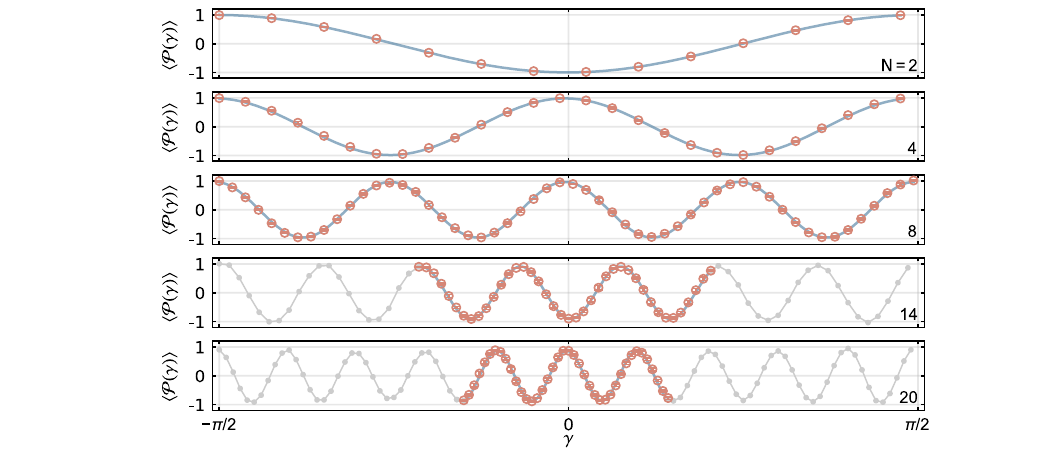}
    \caption{{\bf Parity oscillations.} Measured $\langle\mathcal{P}\rangle$ on Processor II for the bell state and the GHZ states of 4, 8, 14, and 20 qubits. Data with error bars~(orange dots) are repeatedly measured five times and error bars stem from their standard deviations, while gray dots connected by gray lines are experimental data measured only once. Blue lines are fitted with the function $(-1)^{N/2}\times A\cos(N\gamma+C)$.
    }
    \label{fig:parity_shots_scaling}
\end{figure}
\subsubsection{Parity oscillations}
Parity oscillations have been widely used to measure the off-diagonal terms of GHZ states~\cite{Sackett2000Nature, Leibfried2005Nature, Monz2011PRL, Wang2018PRL, Song2019Science, Omran2019Science}. Parity $\mathcal{P}(\gamma)$ of a GHZ state $\frac{1}{\sqrt{2}}\left(|0101\dots\rangle+\exp{(i\Phi)}|1010\dots\rangle\right)$ for an even qubit number is given by
\begin{equation}
    \langle \mathcal{P}(\gamma)\rangle = \langle \otimes_{j=1}^{N}(-1)^{j}\left[ \cos{((-1)^{j}\gamma)\sigma^y_j+\sin{((-1)^{j}\gamma)}}\sigma^x_j\right] \rangle = (-1)^{N/2}\times2|\rho_{0101\dots,1010\dots}|\cos{(N\gamma+\Phi)},
\end{equation}
where $\gamma$ is the rotation angle. $\mathcal{P}(\gamma)$ is an $N$-body operator involving the probabilities of a large number of Fock bases, which requires a significantly large number of measurement shots and limits the size of the system that can be characterized. Therefore, as a cross-validation, we characterize the Bell state and the GHZ states for small system sizes with $N=4, 8, 14, 20$. The measured results are illustrated in Fig.~\ref{fig:parity_shots_scaling}, where we conduct about $2^{20}$, 30000, 6000 measurement shots for system sizes of $N=20$, $N=14$, and $N<14$, respectively. Because the probabilities can be distributed over the whole Fock basis of the Hilbert space for measuring the parity, we cannot truncate the number of the basis to eliminate the measurement errors. Here, instead, we apply the full readout correction matrix to the raw probabilities~\cite{Song2018PRL,Song2019Science} for the parity measurements, which is a tensor product of the measured single-qubit correction matrices.

\subsection{Numerical results}\label{sec:ghz_numerical}
We numerically simulate the generation of the GHZ states using Monte Carlo wavefunction method \cite{Mlmer1993MonteCarlo,Xiang2024arxiv}. In the simulation, error operations are applied following each quantum gate, with probabilities determined by their noise models. The expectation value of an observable is obtained by averaging over the outcomes of an ensemble of noisy quantum circuits. Here, the numerical simulation is performed using MindSpore Quantum~\cite{mq_2021}, which is an open-source quantum computation framework.

\begin{figure*}[th]
    \includegraphics[width=18cm]{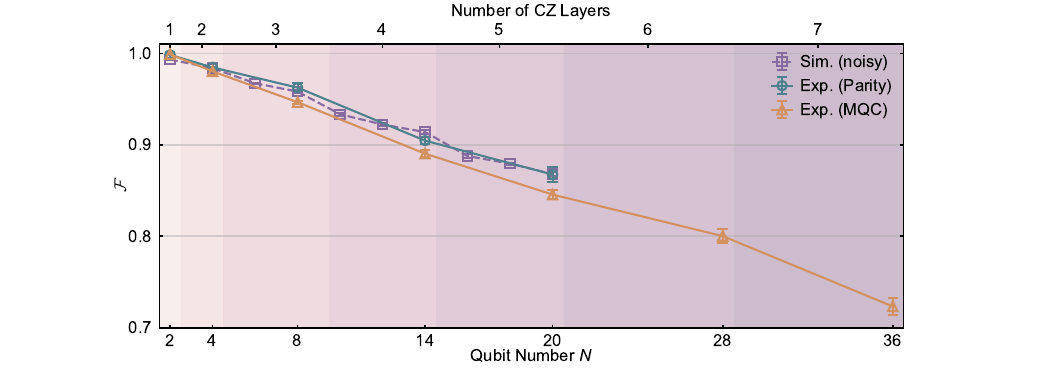}
    \caption{{\bf GHZ state fidelity on Processor II -- experiment vs. simulation.}
    Experimentally measured fidelities of GHZ states are obtained using parity (circles) and MQC (triangles) methods. Numerical results~(dashed lines with squares) are obtained with the Monte Carlo wavefunction method. Colored backgrounds indicate the number of CZ layers for different system sizes. Numerical results are averaged over $30,000$ random samples of noise realizations for good convergence. Experimental~(numerical) error bars correspond to the standard deviations of five repetitions of measurements~(five random seeds to generate random noise samples).
}
\label{fig:ghz_scaling_sim}
\end{figure*}

We simulate the fidelity scaling of the GHZ states on Processor II for $N\leq20$. Note that there are 7 layers of CZ gates in our circuit~(Fig.~\ref{fig:circuit_ghz_processor_2}) for generating 36-qubit GHZ states. For simulating a $N$-qubit GHZ state, we consider the depolarization, energy relaxation, and dephasing as our noise models. The error rates of the depolarization model for different system sizes are listed in Table.~\ref{tab:ghz_sim_error}, which are assessed based on the experimentally measured gate errors subtracting the contribution of the decoherence errors. The error rates of the energy relaxation and dephasing models are calculated based on the average energy relaxation time $T_1$~(Fig.~\ref{fig:sq_params}j), dephasing time $T^{\rm SE}_2$~(Fig.~\ref{fig:sq_params}i), and gate length ($24$~ns single-qubit gate and $60$~ns two-qubit gate). Figure~\ref{fig:ghz_scaling_sim} shows the comparison between the experimental results and the numerical simulations. Numerical simulations agree well with the experimental GHZ fidelities measured with parity method. Note that experimentally measured GHZ fidelities with MQC show a lower fidelity than parity method due to extra gate errors and decoherence errors introduced by the imperfect reversal unitary.
\begin{table}
    [ht]
    \caption{\label{tab:ghz_sim_error} Gate errors for simulating GHZ states.}
    \begin{tabular}{cllllllllll}
        \hline\hline
        Qubit number $N$ &  2  &  4  &   8    &  14  &   20  
        \\ \hline
        number of CZ layers & 1 & 2  &3&4&  5 \\
        $e_p$ of single-qubit gates ($\%$) & 0.049 & 0.031 & 0.045 &  0.064 &  0.056 \\
        $e_p$ of CZ gates ($\%$) & 0.256 & 0.148 &  0.199 & 0.225 & 0.227
        \\ \hline\hline 
    \end{tabular}
\end{table}

\section{Realization of a cat scar DTC}\label{Sec: quantum gates}
\subsection{Digital circuit of DTC}
\begin{figure*}[t]
    \includegraphics[width=18cm]{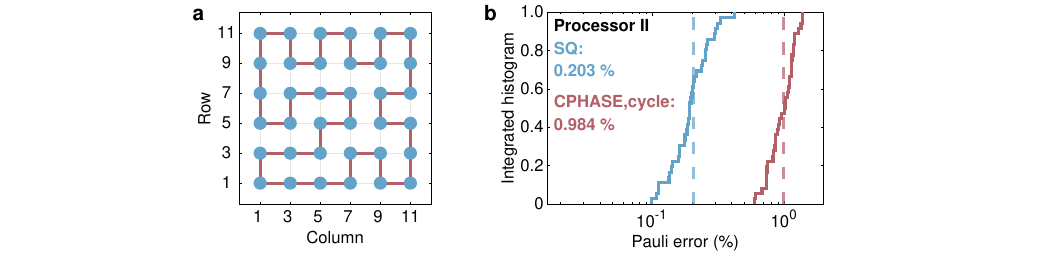}
    \caption{{\bf Realization of an Ising ring.} 
    {\bf a}, Construction of a 36-qubit Ising ring on Processor II. Blue dots denote qubits, which emulate the spins. Red lines represent tunable couplers, which realize Ising interactions between spins. {\bf b}, Cycle Pauli errors of single-qubit gates and CPHASE(-4) gates, which are obtained by performing simultaneous XEB for each layer in the DTC circuits.}
    \label{fig:processor_1_cp}
\end{figure*}

To realize a perturbed Ising model of periodical boundary described by equation (2) of the main text, we construct a 36-qubit ring on Processor II~(Fig.~\ref{fig:processor_1_cp}a). The quantum circuit to realize a cycle of the Floquet unitary $U_{\rm F}$ is shown in Fig.~2c of the main text, which includes two kinds of gates, $U3(\alpha, \beta, \gamma)$ and $ZZ(-4)$. $U3(\alpha, \beta, \gamma)$ is a single-qubit rotation parameterized with three Euler angles, given by 

\begin{equation}\label{eqs:single_gate}
\begin{array}{cc}
U3(\alpha, \beta, \theta) = \left(
\begin{array}{cc}
\cos \left(\frac{\alpha}{2}\right) & -i e^{-i \beta} e^{i \theta} \sin \left(\frac{\alpha}{2}\right) \\
-i e^{i \beta} \sin \left(\frac{\alpha}{2}\right) & e^{i \theta} \cos \left(\frac{\alpha}{2}\right)
\end{array}
\right)
=\left(
\begin{array}{cc}
1 & 0 \\
0 & e^{i \theta}
\end{array}
\right)
\left(
\begin{array}{cc}
\cos \left(\frac{\alpha}{2}\right) & -i e^{-i (\beta-\theta)} \sin \left(\frac{\alpha}{2}\right) \\
-i e^{i (\beta-\theta)}\sin \left(\frac{\alpha}{2}\right) &  \cos \left(\frac{\alpha}{2}\right)
\end{array}
\right)
\end{array}.
\end{equation}
$U3(\alpha, \beta, \gamma)$ gate is realized by a microwave pulse, which rotates qubit state around $(\beta-\theta)$-axis in the $xy$ plane by an angle $\alpha$, and a subsequent virtual phase gate $Z(\theta)$. Two layers of two-qubit $ZZ(\phi)$ gates emulate the Ising interactions on the chain, which are realized by a controlled $\phi$-phase gate~[CPHASE($\phi$)] and two single-qubit phase rotations  
$$
ZZ(\phi) = e^{i\frac{\phi}{4}Z_jZ_k} = Z_j(-\phi/2)Z_k(-\phi/2)\mathrm{CPHASE}(\phi).
$$
Note that CZ gate used in preparing GHZ state is a CPHASE($\phi$) gate with $\phi=\pi$. 

In our experiments, we set interaction strength $J=1$ and cycle period $T=1$, thus, the controlled phase $\phi=-4JT=-4$. Two single-qubit rotations, $U3(\alpha, \beta, \gamma)$  and $U3^{\prime}(\alpha^{\prime}, \beta^{\prime}, \gamma^{\prime})$ encode single-qubit Rabi drives and the generic perturbations in $U_{\rm F}$, whose mapping to the parameters $\varphi_1$,$\varphi_2$, $\lambda_1$, and $\lambda_2$ in $U_{\rm F}$ are given in Section \ref{sec:theory}B. We experimentally benchmark single-qubit gates and two-qubit CPHASE(-4) gates with simultaneous XEB sequences, with the measured Pauli errors per cycle shown in Fig.~\ref{fig:processor_1_cp}b.

\begin{figure}[t]
    \includegraphics[width=18cm]{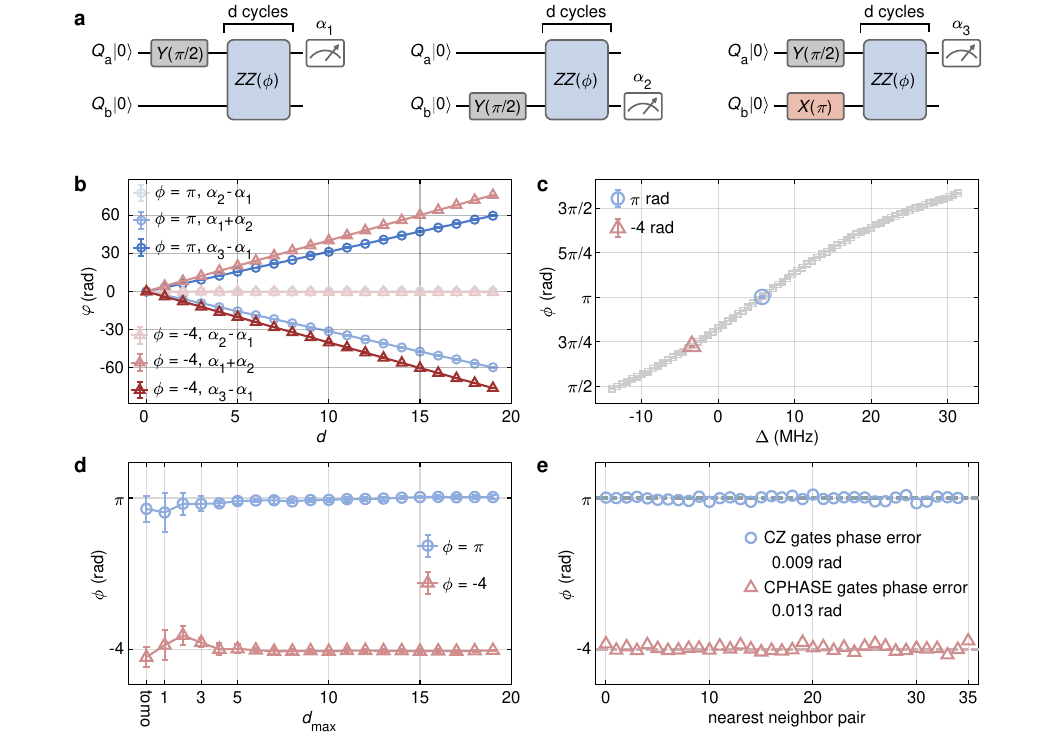}
    \caption{ {\bf Calibration of Floquet circuits.}
    {\bf a}, Quantum circuits for calibrating $\alpha_1$, $\alpha_2$, and $\alpha_3$.
    {\bf b}, Experimental results of the calibrated  $\alpha_1$, $\alpha_2$, and $\alpha_3$ for $\phi=\pi$~(red) and $\phi=-4$~(blue) as functions of cycle $d$. Markers are experimental results and lines are linear fits.
    {\bf c}, Measured controlled phase $\phi$ as a function of detuning $\Delta$.
    {\bf d}, Convergence of $\phi$ with the increase of the maximum cycle $d_{\max}$ for $\phi=\pi$~(blue) and $\phi=-4$~(red). 
    {\bf e}, Measured controlled phase $\phi$ for the gates used in our experiments. Dashed lines indicate the target controlled phase. Marks are experimental results.
	}
    \label{floquet_cali}
\end{figure}
\subsection{CPHASE gate calibration}

We implement two-qubit CPHASE($\phi$) gates by tuning the coupling strength between qubits~\cite{Yan2018PRApp}. In the general case, the tunable couplers allow us to realize a Fermionic Simulation gate~\cite{Foxen2020PRL}, which is given by 
\begin{equation}
\left(\begin{array}{cccc}
1 & 0 & 0 & 0 \\
0 & e^{i\left(\Delta_{+}+\Delta_{-}\right)} \cos \theta & -i e^{i\left(\Delta_{+}-\Delta_{-, \text {off }}\right)} \sin \theta & 0 \\
0 & -i e^{i\left(\Delta_{+}+\Delta_{-, \text {off }}\right)} \sin \theta & e^{i\left(\Delta_{+}-\Delta_{-}\right)} \cos \theta & 0 \\
0 & 0 & 0 & e^{i\left(2 \Delta_{+}+\phi\right)}
\end{array}\right),
\end{equation}
where $\phi$ is the controlled phase on $|11\rangle$, $\theta$ is the swap angle between $|01\rangle$ and $|10\rangle$~[$\theta=0$ for CPHASE($\phi$) gate], and $\Delta_+$,$\Delta_-$ and $\Delta_{-,\text{off}}$ are single-qubit phases. In our experiments, we needs CPHASE($\phi$) gates with two specific angles, $-4$ (ZZ interaction) and $\pi$ (CZ gate). We use Floquet calibration method~\cite{Neill2021Nature, Mi2022} to tune up the parameters $\phi$, $\Delta_+$, and $\Delta_-$ to realize a target CPHASE($\phi$) gate. The controlled phase $\phi$ and single-qubit phases $\Delta_+$, $\Delta_-$ are calibrated using periodic circuits sharing a similar structure as shown in Fig.~\ref{floquet_cali}{a}. We initialize one qubit along the $x$-axis of the Bloch sphere and keep the other one in $|0\rangle$ or $|1\rangle$ state. $d$ cycles of $ZZ(\phi)$ gates are then applied to improve the sensitivity. Finally, we perform tomographic measurements on the first qubit and obtain  $\alpha=\arctan{(\langle Y \rangle/\langle X\rangle)}$. In the case of $\theta\approx0$,we can extract these phases by $\alpha_1+\alpha_2 = 2\Delta_+ d$, $\alpha_2-\alpha_1=2\Delta_- d$, $\alpha_3-\alpha_1 = \phi d$. Whereas the measured phase may be not accurate for each sequence with a specific $d$, we can estimate it more faithfully by fitting the accumulated phase as a function of the cycle $d$~({Fig.~\ref{floquet_cali}b}). As shown in Fig.~\ref{floquet_cali}{d}, as $d_{\max}$~($d=1,2,...,d_{\rm max}$) increases, the fitted $\phi$ converges to a stable value. The controlled phase $\phi$ depends on the detuning $\Delta$ ($f_{10} - f_{21}$) of the two qubits. We carefully calibrate $\phi$ with $d_{\max}=20$ for each $\Delta$. As shown in Fig.~\ref{floquet_cali}c, we can control $\phi$ in a range of $[\pi/2,3\pi/2]$ by tuning $\Delta$. We show the calibrated values of $\phi$ for CZ gate ($\phi=\pi$) and CPHASE(-4) gate in Fig.~\ref{floquet_cali}{e}. The average phase error is $\sim 0.01$ rad.

\section{Theoretical details and additional experimental data}\label{sec:theory}
    
\subsection{General scheme of engineering cat eigenstates and comparison of different DTCs}

Discrete time crystals (DTCs) are featured by a rigid reduced periodicity $nT, n\geqslant 2 \in \mathbb{Z}$ for the oscillation of physical observables, compared with driving period $T$. Such a phenomenon fulfills the concept of spontaneous breaking of discrete time translation symmetry. Here, for our purpose of protecting and controlling GHZ states, we focus on a particular class of DTCs that can host pairwise cat eigenstates, which are catalyzed by Ising interactions and global spin-flip pulses. The functionality of these two driving terms can be most intuitively understood at the unperturbed anchor point
\begin{align}\label{smeq:uf_finetune}
    U_F^{(0)} = U_2^{(0)} U_1^{(0)},
    \qquad
    \text{Ising: }U_2^{(0)} = e^{-i\sum_{j,k}J_{jk} \rzj \rzk },
    \quad 
    \text{spin-flip: }U_1^{(0)} = e^{-i\pi \sum_{j=1}^N (\rxj/2)} = (-i)^N \prod_{j=1}^N \rxj .
\end{align}
Here, $\sigma^{x,y,z}_j$ are Pauli operators acting on a spin-$1/2$ state $|\sbbs_j \rangle$ at site $j=1,2,\dots, N$, where $\sbbs_j = 1,0$ for spin $\uparrow, \downarrow$. In the Floquet operator $U_F^{(0)}$, Ising interaction in $U_2^{(0)}$ structures the eigenstates as Fock states,
\begin{align}
    |\SBBS \rangle \equiv |s_1s_2\dots s_N\rangle, 
    \qquad 
    \qquad\qquad
    U_2^{(0)} |\SBBS\rangle = e^{-iE_{\text{Ising}}(\SBBS)T} |\SBBS\rangle, 
\end{align}
with Ising interaction energy $E_{\text{Ising}} (\SBBS) = \sum_{jk} J_{jk} (2s_j-1) (2s_k-1)$.
Importantly, two opposite Fock states  $|\SBBS\rangle$ and $|\bar{\SBBS}\rangle \equiv \prod_{j=1}^N \rxj |\SBBS\rangle =  |(1-s_1)(1-s_2)\dots (1-s_N)\rangle$ share the same $E_{\text{Ising}}(\SBBS) = E_{\text{Ising}}(\bar{\SBBS})$ and form a degenerate doublet. Then, the global spin-flip pulse $U_1^{(0)} |\boldsymbol{s}\rangle \propto |\bar{\boldsymbol{s}}\rangle$ lifts the degeneracy and generates cat eigenstates,
\begin{align}\label{smeq:cateigs}
    |\scarpm \rangle = \frac{|\SBBS \rangle + e^{i\Phi_\pm} |\bar{\SBBS}\rangle}{\sqrt{2}};
    \qquad
    U_F |\scarpm \rangle =  (-i)^N e^{-i\left( E_{\text{Ising}}(\SBBS) + \Phi_\pm \right)T } |\scarpm \rangle,
    \quad 
    \Phi_+ = 0, \,\, \Phi_- = \pi/T. 
\end{align}
Two cat eigenstates in the same Fock subspace $|\Phi_+,\SBBS\rangle, |\Phi_-,\SBBS \rangle$ are separated by a large quasienergy gap $\Phi_- - \Phi_+ = \pi/T$. 

The relation between cat eigenstate pairs and previous experiments on DTCs can be illustrated below. Consider an initial state $|\SBBS\rangle$, which simultaneously overlaps with both cat eigenstates $|\Phi_\pm,\SBBS\rangle$. This Fock state will alternate between $|\SBBS\rangle$ and $|\bar{\SBBS}\rangle$ during evolution with period $2T$ due to $\pi/T$ spectral gap, corresponding to magnetic order oscillations in experiments. Nevertheless, a subtle issue is whether the observed magnetic order oscillation is attributable to slow-relaxation in diffusive systems or to localized cat eigenstates~\cite{Khemani2019b}. To distinguish the two scenarios, recent DTC experiments have measured two-body correlations in addition to one-body magnetic orders~\cite{Mi2022,Stasiuk2023}. Alternatively, GHZ states generated in this work offer a unique $N$-body observable, the macroscopic quantum coherence, to directly benchmark the underlying cat eigenstates. 

Focusing on cat eigenstate engineering, a crucial question is how robust are these cat eigenstates facing perturbations. To quantify the robustness, for generic Floquet eigenstates $|\epsilon_m\rangle$,  $U_F|\epsilon_m\rangle = e^{i\epsilon_m}|\epsilon_m\rangle $, we exploit the inverse participation ratio (IPR)
\begin{align}
    \text{IPR}_m = \sum_{\SBBS} |\langle \SBBS | \epsilon_m\rangle |^4.
\end{align}
A larger value of IPR corresponds to stronger localization in Fock space and indicates better quality of a cat eigenstate.
For instance, ideal cat eigenstates in Eq.~\eqref{smeq:cateigs} takes the IPR value $1/2$. Contrarily, an eigenstate in the ergodic limit with equal amplitudes on all Fock states $|\epsilon_m\rangle = (1/\sqrt{2^N})\sum_{\boldsymbol{s}}|\boldsymbol{s}\rangle$ gives vanishing $\text{IPR}_m = 1/2^N \rightarrow0$ as the system size $N$ grows.

\begin{table}
    [b]
    \caption{\label{tab:dtc_compare} Comparison of features for different mechanisms to engineer cat eigenstates in DTCs using kicked Ising model in Eq.~\eqref{smeq:uf}. In the examples with $N=16$ qubits, we take a random target GHZ state subspace spanned by $|\SBBS\rangle = |101001010101000001\rangle$ and $ |\bar{\SBBS}\rangle = |0101101010111110\rangle $. The overlap of an eigenstate $|\epsilon_m\rangle$ with GHZ state subspace is computed by $ |\langle \epsilon_m|\SBBS\rangle|^2 + |\langle \epsilon_m|\bar{\SBBS}\rangle|^2 $. }
    \begin{tabular}{cccc}
        \hline\hline
        mechanisms & many-body localization & 
        \parbox{4.5cm}{prethermalization with Landau's symmetry breaking} & {\bf cat scars}
        \\ \hline
        \parbox{1.9cm}{
            Conditions of Ising interaction $e^{-iT \sum_{jk}J_{jk}\sigma^z_j \sigma^z_k}$}
        &
        \parbox{5cm}{
            \begin{flushleft}
                1, Disordered {\em interaction} $J_{j,j+1}\in[J-\frac{W}{2},J+\frac{W}{2}]$, ($JT, WT\sim 1$) needed, 
                as spin $\pi$-pulse $e^{-i\pi\sum_j (\sigma^x_j/2)}$ cancels longitudinal fields every two periods\\
                2, 1D with short-range interaction
            \end{flushleft}
        }
        &
        \parbox{5cm}{
            \begin{flushleft}
                1, High frequency driving $J_{jk}T\ll 1$ for prethermal lifetime $\sim e^{1/J_{jk}T}$, and  weak perturbation $ \lambda < J_{jk} $ for ordering \\
                2, In 1D, long-range interaction is needed to circumvent the instability predicted by Mermin-Wagner theorem
            \end{flushleft}
        }
        &
        \parbox{5cm}{
            \begin{flushleft}
                With strong interaction $JT\sim 1$ and the signs $J_{j,j+1} = (2\sbbs_j-1)(2\sbbs_{j+1}-1) J$, two pairs of scars are engineered as 
                $ |\sbbs_1\sbbs_2\dots\rangle \pm e^{-i\gamma} |\bar{\sbbs}_1 \bar{\sbbs}_2 \dots \rangle $ and 
                $ |\sbbs_1 \bar{\sbbs}_2\dots\rangle \pm e^{+i\gamma'} |\bar{\sbbs}_1 \sbbs_2 \dots \rangle $. IPR of scars can be estimated by perturbation theory
            \end{flushleft}
        }
        \\
        \hline 
        \quad & \quad & \quad & \quad \\
        \parbox[b]{2cm}{Schematic illustration of relevant cat eigenstates in the spectrum \\ \quad }
        & 
        \parbox[b]{5cm}{\includegraphics[width=4cm]{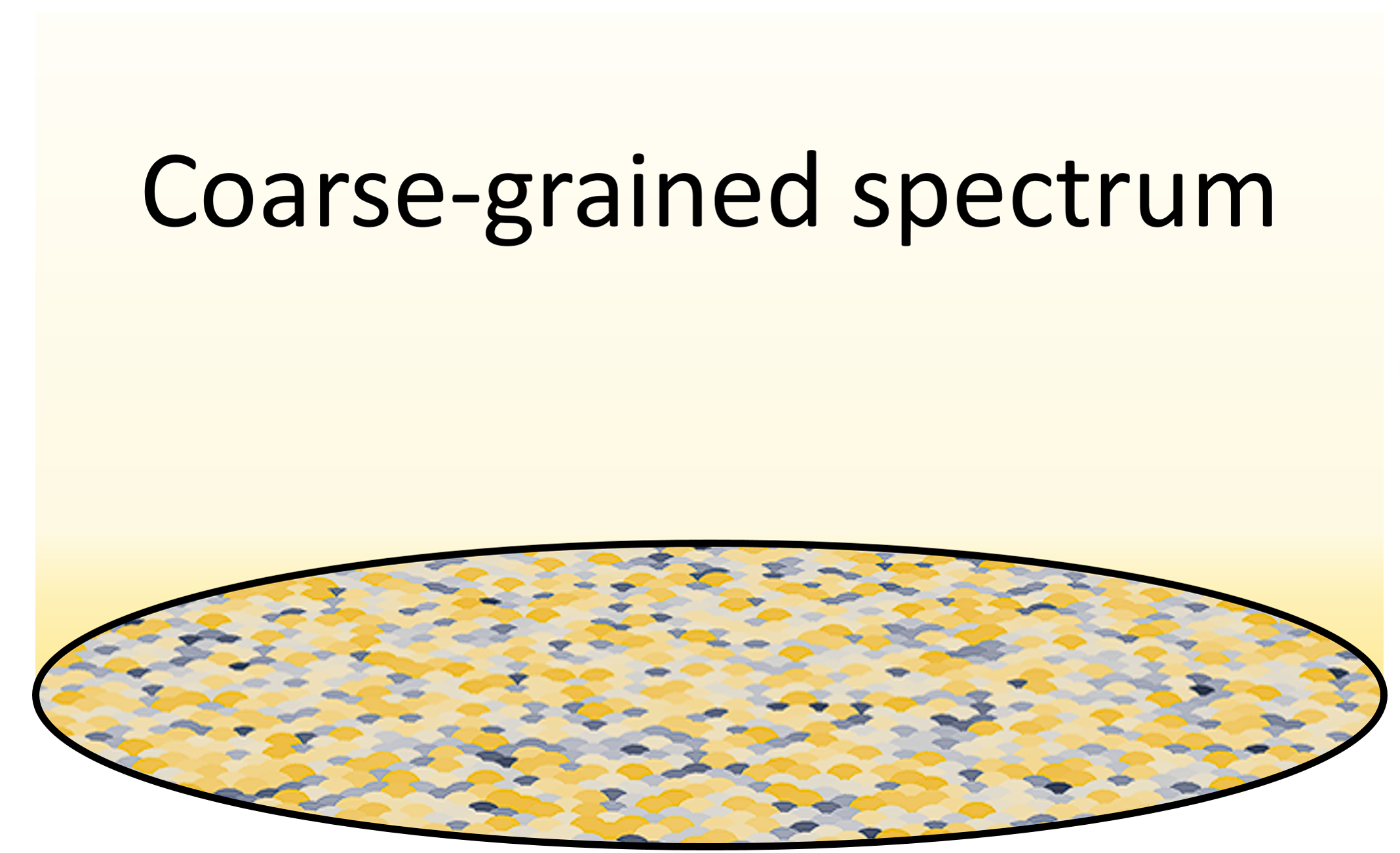}}
        & 
        \parbox[b]{5cm}{\includegraphics[width=4cm]{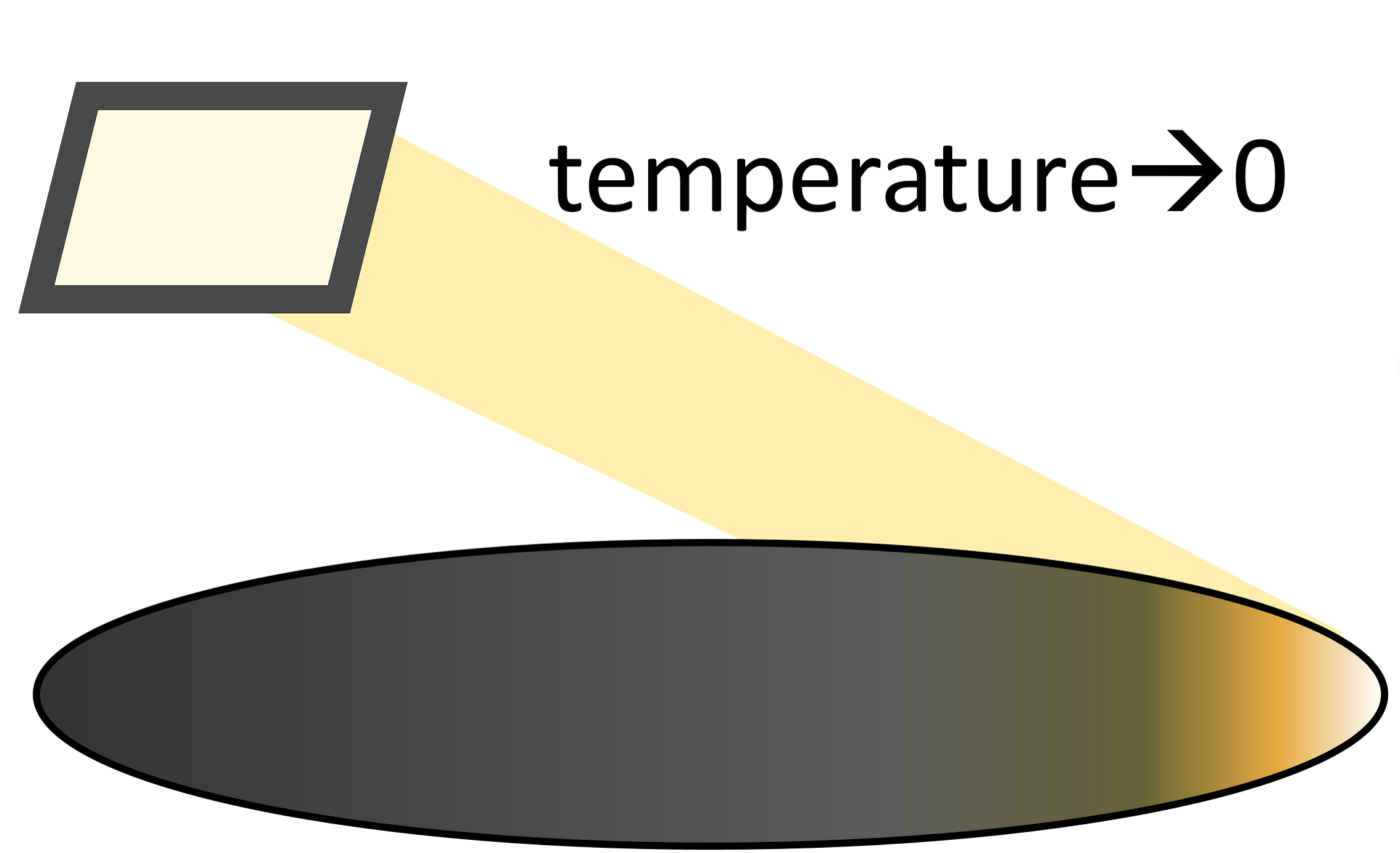}}
        & 
        \parbox[b]{5cm}{\includegraphics[width=4cm]{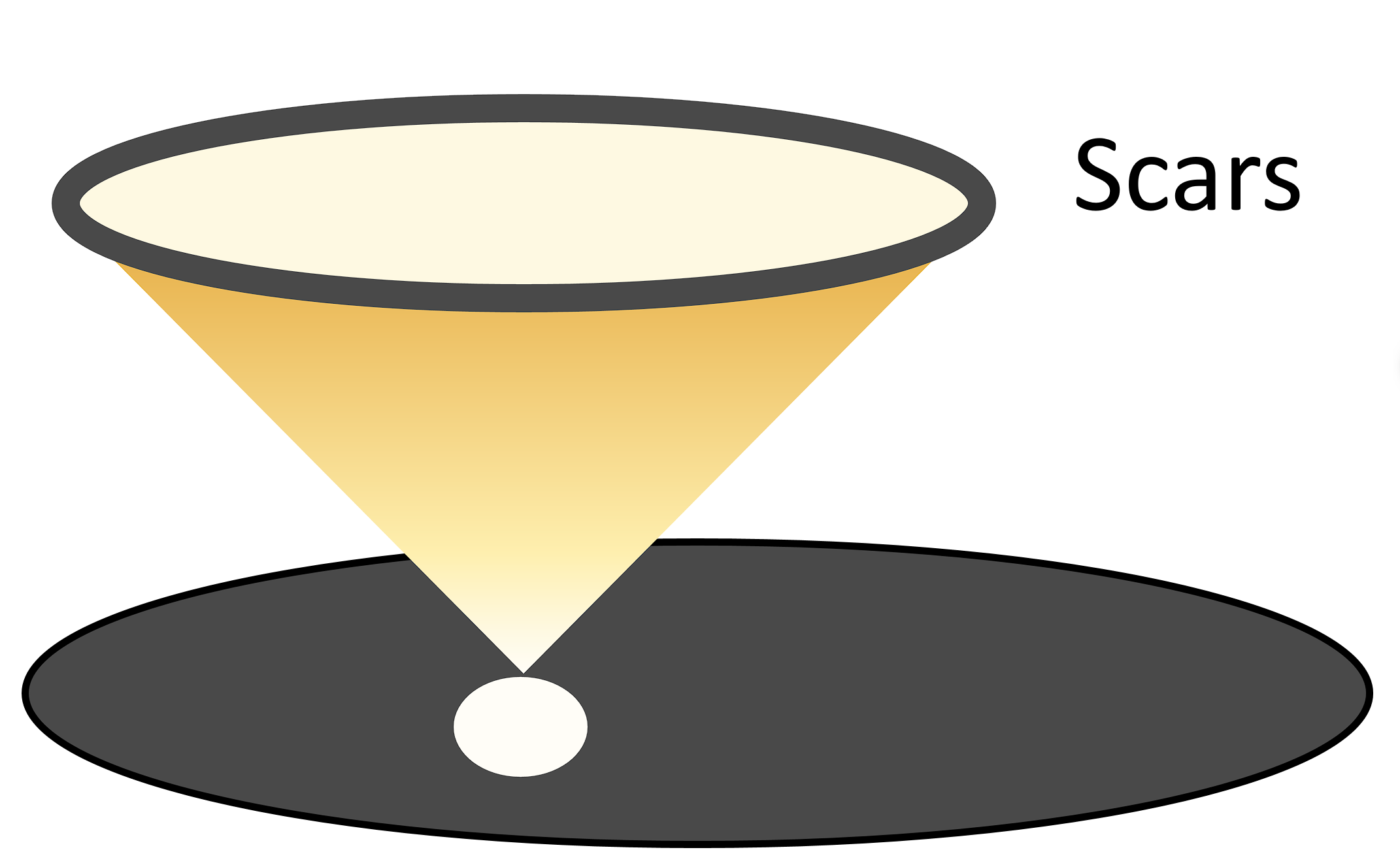}
        }
        \\ \hline
        \parbox{2cm}{Examples\\ \quad \\ Parameters: $ N=16$,\\ $\lambda_{1,2}T = 0.05$,  \\ $ \yuanphi T = -\pi/2$, \\ $\yuanhz T = \pi/2-0.6$
            \\ \quad \\ \quad	\\ \quad \\
            }
        &
        \parbox{5cm}{
            \includegraphics[width=5cm]{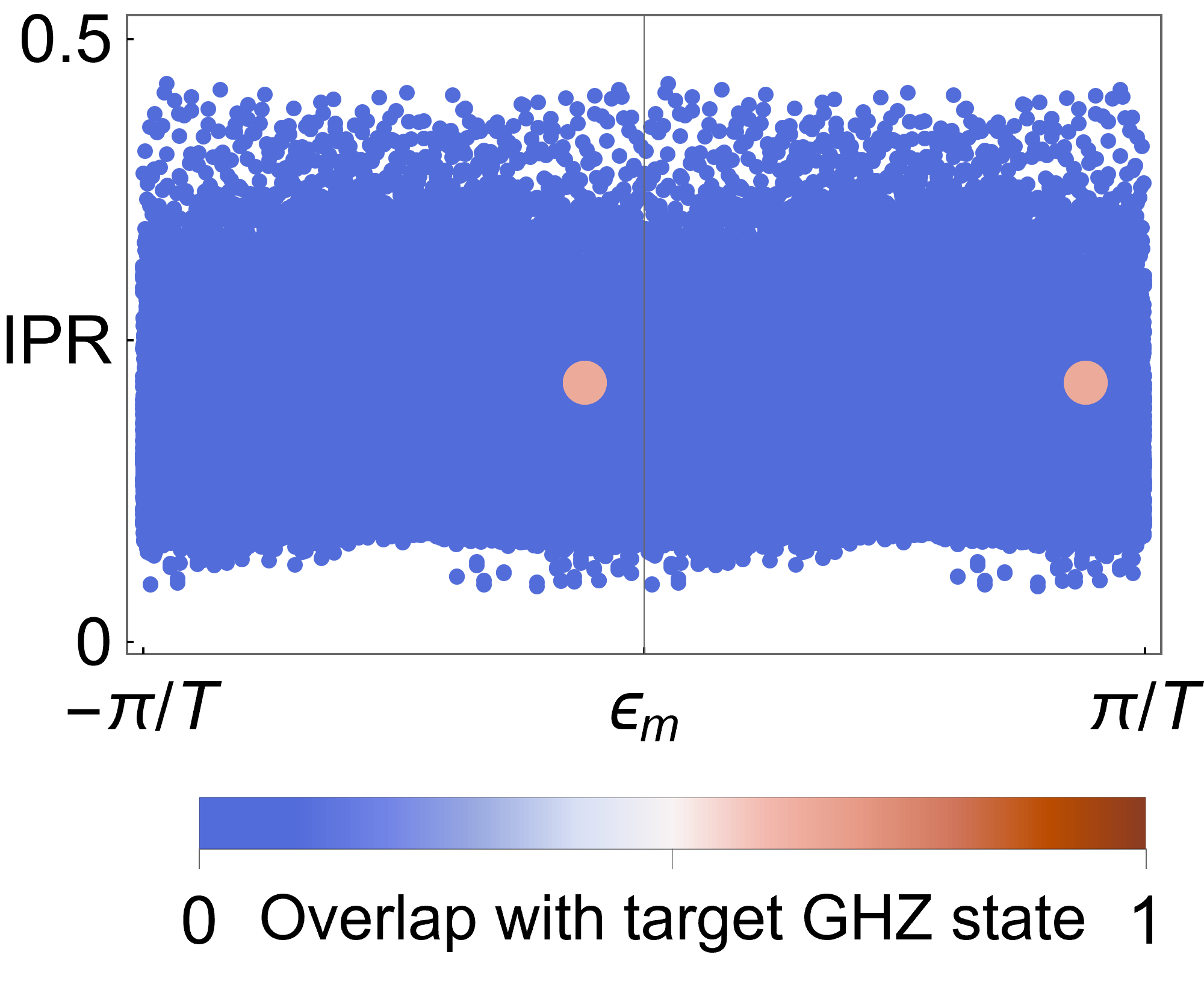}
            }
        &
        \parbox{5cm}{
            \includegraphics[width=5cm]{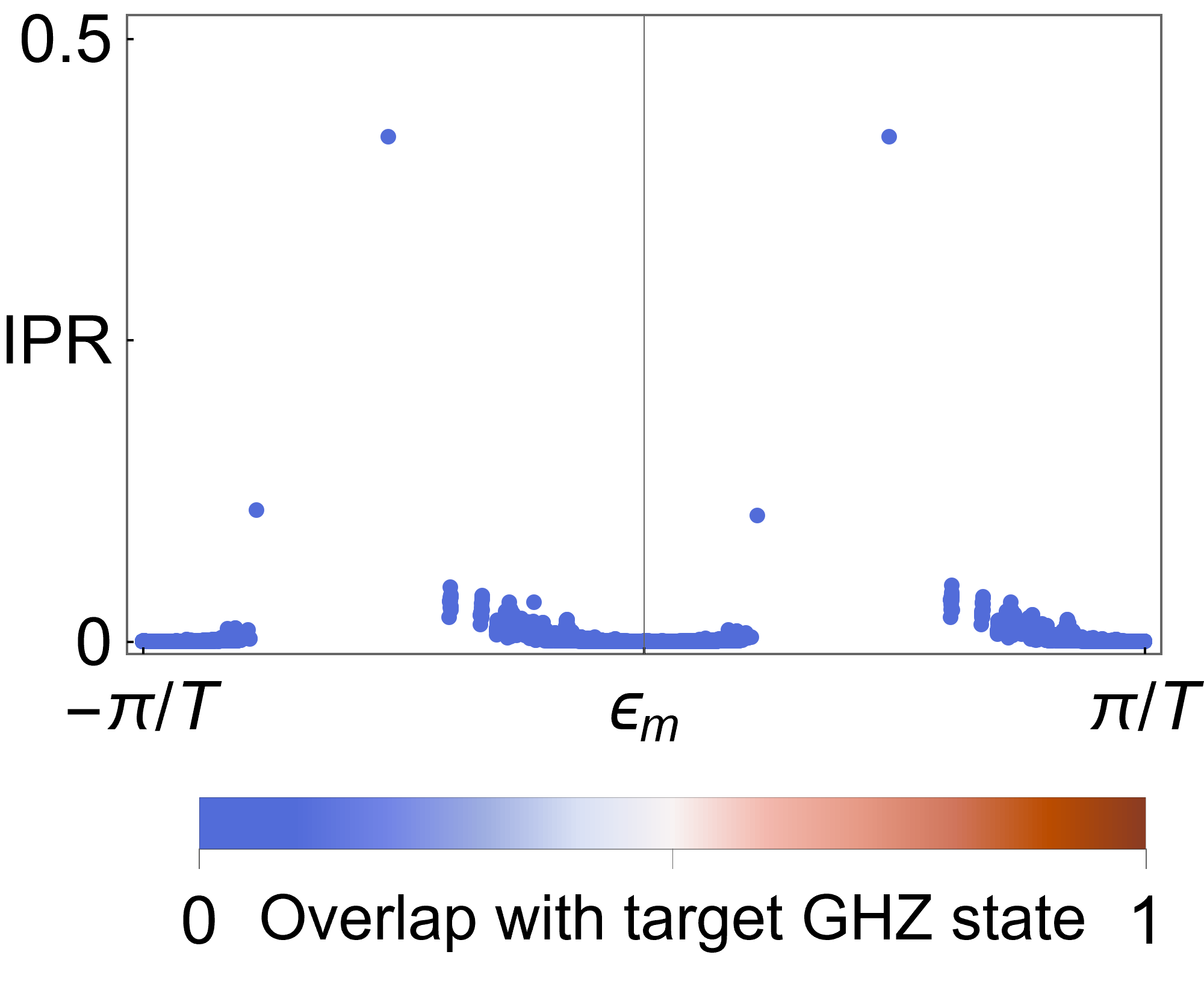}
            }
        &
        \parbox{5cm}{
            \includegraphics[width=5cm]{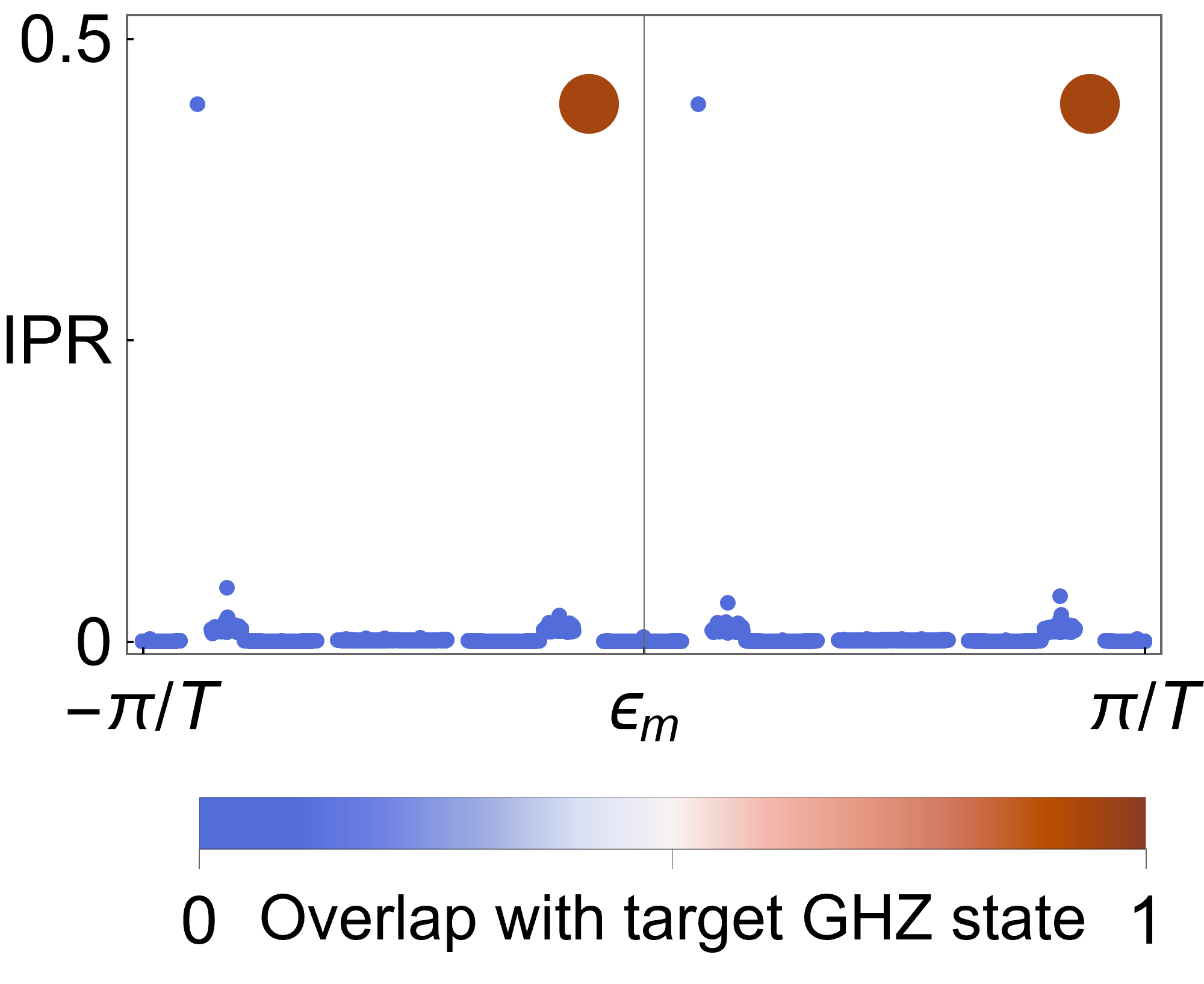}
            }
        \\ 
        Interaction & 
        $JT = WT = \pi/4$
        &
        $J_{jk} = J/|j-k|^{1.5}, JT=0.05 $
        &
        $JT = 1$
        \\ 
        Highest IPR state &
        Depends on disorder realization &
        Ferromagnetic pattern &
        Designed on-demand
        \\ \hline\hline
    \end{tabular}
\end{table}

To date, three mechanisms to stabilize cat eigenstates against perturbations have been proposed, namely, many-body localization, prethermalization with Landau's symmetry breaking, and cat scars. They chiefly differ in the Ising interactions employed, which result in different number and type of robust cat eigenstates as summarized in Table~\ref{tab:dtc_compare}. For unified comparison, we consider a model similar to Eq.~(2) in the main text,
\begin{align} \nonumber
    &
    U_F = U_2(\lambda_2) U_1(\lambda_1),
    \qquad\qquad
    U_1(\lambda_1) = U_{P}(\yuanphi, \lambda_1, \yuanhz) e^{-i\pi \sum_{j=1}^N (\rxj/2) }, 
    \qquad
    U_2(\lambda_2) = e^{-iT\sum_{j=1}^N J_{jk}\tilde{\sigma}^z_j(\lambda_2)\tilde{\sigma}^z_{k}(\lambda_2)},
    \\ \label{smeq:uf}
    & U_{P}(\yuanphi, \lambda_1, \yuanhz) = \prod_{j=1}^N e^{-i \yuanphi \rzj/2} e^{i\lambda_1 \ryj/2} e^{-i \yuanhz \rzj/2},
    \qquad
    \tilde{\sigma}^z_j(\lambda_2)=\cos(\lambda_2) \rzj + \sin(\lambda_2) \rxj.
\end{align}
Here, to test the robustness of cat eigenstates against generic perturbation, we add to Eq.~\eqref{smeq:uf_finetune} both single-qubit and two-qubit perturbations parameterized by $\lambda_1, \lambda_2 \ll1$ respectively. Two longitudinal fields $\yuanphi, \yuanhz$ are adopted to ensure both breaking the integrability of the model and avoiding single-qubit echoes that may interfere with the benchmark of many-body effects. These perturbations are chosen to be the same for all three cases as specified in the first column of Table~\ref{tab:dtc_compare}. On the other hand, different Ising interactions $J_{jk}$ prescribe three distinct regimes listed in the second to fourth columns in Table~\ref{tab:dtc_compare}.
\begin{enumerate}
    \item 
    {\bf Many-body localization (MBL).} \\
    In DTCs, the global qubit flip $ e^{-i\pi\sum_j (\sigma^x_j/2)} = (-i)^N \prod_{j=1}^N \rxj$ can largely cancel the effect of hypothetical random longitudinal fields, i.e., $h^z_j\sigma^z_j$, that are often exploited to engineer MBL, because in every two periods $ U_F^2\sim \prod_j \rxj e^{-i\sum_jh^z_j \sigma^z_j} \prod_j \rxj e^{-i\sum_jh^z_j \sigma^z_j} \propto e^{- i\sum_jh^z_j (-\sigma^z_j)}e^{-i\sum_jh^z_j \sigma^z_j}$. Thus, strong disorders in the Ising-even interaction $J_{jk}\in[J-W/2, J+W/2]$ is needed. The resulting spectrum is expected to host extensive numbers of pairwise cat eigenstates, which can show large variation in their IPR values. As an illustration, in the second column of Table~\ref{tab:dtc_compare}, we choose $JT = JW = \pi/4$ similar to the parameters in Ref.~\cite{Mi2022}, with eigenstate IPRs and quasienergy for a representative disorder sample shown in the bottom row. Cat eigenstates of largest overlap with target GHZ state subspace (specified in the caption) are highlighted by larger dots.
    
    \item 
    {\bf Prethermalization with Landau's symmetry breaking (pLSB). }\\
    In the pLSB Floquet operator $U_F = e^{-iH_{\text{preth}}}e^{-i\pi\sum_j\sigma^x_j/2}$, upon factoring out a global spin flip, remaining terms in $H_{\text{preth}}$ satisfy two conditions. First, the interaction $J_{jk}$ is much weaker than driving frequency $J_{jk}T\ll1$ (high-frequency limit), thereby giving a prethermal time scale $\sim e^{1/J_{jk}T}$ before significant heating occurs. Second, $H_{\text{preth}}$ undergoes LSB and induces an ordered ground state. To stabilize the LSB, a one-dimensional system should be long-range interacting. The extended interaction range typically suppresses ground states of inhomogeneous spin patterns akin to magnetic frustration. In the bottom of the third column for Table~\ref{tab:dtc_compare}, we adopt the interaction $J_{jk} = J/|j-k|^{1.5}$ with exponent $\sim1.5$ similar to Ref.~\cite{Kyprianidis2021} for trapped ions. The spectrum of $H_{\text{preth}}$ has a small bandwidth compared with Floquet quasienergy window $2\pi/T$  due to high-frequency constraint $JT\ll1$. This spectrum is duplicated into two pieces by $e^{-i\pi\sum_j\sigma^x_j/2}$. The two cat eigenstates $\sim |\boldsymbol{s}\rangle \pm |\bar{\boldsymbol{s}}\rangle $ of highest IPRs are made of uniform ferromagnetic spin patterns $|\boldsymbol{s}\rangle = |000\dots\rangle, |\bar{\boldsymbol{s}}\rangle = |111\dots\rangle$.
    
    \item 
    {\bf Cat scars.}\\
    The cat scars are stabilized by strong Ising interaction $JT \sim 1$. Under this condition, eigenstates are separated to different subspaces according to effective total domain wall numbers. In particular, there are two rare non-ergodic subspaces, each involving only two Fock states. Sign structure in $J_{j,j+1} \equiv J_j$ specifies two pairs of cat scars in these two subspaces, with spin patterns listed below,
    \begin{align}\label{smeq:Js}
        J_{j} = (2s_j-1) (2s_{j+1}-1)J  \Rightarrow
        \begin{array}{l}
        \begin{cases}
        \frac{1}{\sqrt{2}}(|\bar{s}_1 s_2 \bar{s}_3 s_4 \dots \bar{s}_{N-1} s_N\rangle + e^{i\gamma'}  |s_1 \bar{s}_2 s_3 \bar{s}_4\dots s_{N-1} \bar{s}_N \rangle )
        &
        \epsilon_+ = E_1
        \\
        \frac{1}{\sqrt{2}} (|\bar{s}_1 s_2 \bar{s}_3 s_4 \dots \bar{s}_{N-1} s_N\rangle - e^{i\gamma'}  |s_1 \bar{s}_2 s_3 \bar{s}_4\dots s_{N-1} \bar{s}_N \rangle )
        &
        \epsilon_- = E_1 + \pi
        \end{cases}
        \\
        \begin{cases}
            \frac{1}{\sqrt{2}}(|\bar{s}_1 \bar{s}_2 \bar{s}_3 \bar{s}_4\dots \bar{s}_{N-1} \bar{s}_N \rangle + e^{i\gamma } |s_1s_2s_3s_4\dots s_{N-1}s_N\rangle ), 
            & \epsilon_+ = E_2, 
            \\
            \frac{1}{\sqrt{2}}(|\bar{s}_1 \bar{s}_2 \bar{s}_3 \bar{s}_4\dots \bar{s}_{N-1} \bar{s}_N \rangle - e^{i\gamma } |s_1s_2s_3s_4\dots s_{N-1}s_N\rangle )
            &
            \epsilon_- = E_2+\pi 
        \end{cases}
        \end{array}
    \end{align}
    where $\bar{s}_j  = (1-s_j)$, $E_1\approx E_{\text{Ising}}(\bar{s}_1s_2\dots \bar{s}_{N-1}s_N), E_2\approx E_{\text{Ising}}(\bar{s}_1 \bar{s}_2\dots \bar{s}_{N-1} \bar{s}_N)$, $\gamma, \gamma'$ are constants that can be absorbed into the Fock states as gauge choices. Meanwhile, the scaling of scar IPRs can be obtained order-by-order via Floquet perturbation theory~\cite{Huang2023}, with the dominant order given by
    \begin{align}\label{smeq:ipr_analytical}
        \text{IPR} = \frac{1}{2} \frac{1}{(1+\bar{V}^2 \lambda^2 N)^2} + O((\lambda^2N)^2).
    \end{align}
    Here, $\lambda_1=\lambda_2=\lambda$ is the perturbation strength, and $\bar{V}^2$ is the coupling constant which can be extracted from numerical data at a single parameter point of a certain $\lambda, N$, as will be elaborated later in Table~\ref{tab:iprs}. 
\end{enumerate}

Based on the features of three schemes to engineer cat eigenstates, we next consider their suitability in the practical task of preserving and controlling a targeted GHZ state. This task involves two crucial requirements. First, the scheme should allow for a deterministic selection of the spin patterns for cat eigenstates matching that of the targeted GHZ state. Second, there should be a quantitative control of the quality for relevant cat eigenstates, i.e. in terms of IPR values, so as to effectively protect the GHZ states, especially if the spin patterns for GHZ states are switched during evolution. From these requirements, we see the challenges in exploiting preexisting experimental platforms based on MBL or prethermalization here. MBL DTCs are affluent in the number of cat eigenstates, but the wide range of IPR distribution implies a large variation of the quality for different cat eigenstates. This issue will be discussed further in Figs.~\ref{fig:mbl_scar}  and \ref{fig:ea_ipr_sc}. Also, the prethermal scheme in one dimension chiefly produces uniform (ferromagnetic) spin patterns for cat eigenstates and lacks tunability. In the following, we focus on the cat-scarred DTCs. To lighten notations, we use $J_j$ to denote $J_{j,j+1}$ for both MBL and cat scar cases in the following. Also, we set $T=1$ to lighten notations below, i.e. $J_jT \rightarrow J_j$, although $T$ will be mentioned for bookkeeping driving periods.

\subsection{Properties of cat scar DTCs }

In this subsection, we show in more details the essential two properties for cat scars, namely, the capability to engineer different cat scar spin patterns as specified in Eq.~\eqref{smeq:Js}, and the characterization of scar IPRs given in Eq.~\eqref{smeq:ipr_analytical}.

\subsubsection{Pattern control}
\begin{figure}
[ht]
\includegraphics[width=18cm]{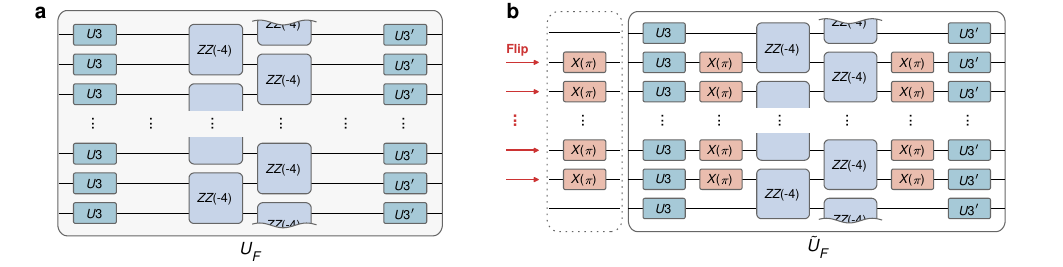}
\caption{\label{fig:uf_circuit} 
    Editing the scar patterns by locally changing the sign of Ising interaction, which can be achieved via single-qubit drivings.
}
\end{figure}

First, we discuss the circuit implementation of cat scar DTCs and demonstrate the method to flexibly switch spin patterns for cat scars. As illustrated in \mbox{Fig.~\ref{fig:uf_circuit}a}, we start from a uniform circuit,
\begin{align}\nonumber
    &U_F = (U3') ZZ(-4) (U3),
    \\ \label{eq:ufexpt}
    & ZZ(-4) =  \prod_{j=1}^N e^{-4i (\sigma^z_j/2) (\sigma^z_{j+1}/2) }, 
    \qquad
    U3' = \prod_{j=1}^N e^{i\varphi (\sigma^z_j/2) } 
    e^{-i\lambda_2 (\sigma^y_j/2) } , 
    \quad 
    U3 = \prod_{j=1}^N e^{ i\lambda_2 (\sigma^y_j/2)  }
    e^{-i\varphi' (\sigma^z_j/2) } 
    e^{-i(\pi-\lambda_1) (\sigma^x_j/2) }
\end{align}  
Written in the notation $U3(\alpha,\phi,\theta)$ of Eq.~\eqref{eqs:single_gate}, we have 
$\alpha = \lambda_2,  
\phi = \frac{\pi}{2} - \varphi, 
\theta = -\varphi$ 
in $U3'$, and 
$\alpha = 2\arccos\sqrt{\frac{1-\cos\lambda_2 \cos\lambda_1 - \sin\lambda_2 \sin\lambda_1 \sin\varphi'}{2}}$, 
$\phi = -\frac{\pi}{2} - \varphi' + 2\arg\left(1-ie^{i\varphi'} \cot\frac{\lambda_2}{2} \cot\frac{\lambda_1}{2} \right),
\theta = \varphi' - 2\arg\left(1 + ie^{i\varphi'} \tan\frac{\lambda_2}{2} \cot\frac{\lambda_1}{2} \right)
$
in $U3$
up to global phases.
Note the identities $e^{-i\lambda_2 \sigma^y_j/2} \sigma^z_j e^{i\lambda_2 \sigma^y_j/2} = \cos(\lambda_2) \sigma^z_j + \sin(\lambda_2) \sigma^x_j $, $e^{-i (\pi/4)\sigma^z_j} \sigma^x_j e^{i(\pi/4)\sigma^z_j} = \sigma^y_j $, and perform a gauge transformation $ e^{-i\varphi \sum_{j=1}^N (\sigma^z_j/2)} U_F e^{i \varphi \sum_{j=1}^N (\sigma^z_j/2)}$, we arrive at the physical model
\begin{align} \nonumber
    U_F &= U_2 U_1,
    \\ \nonumber
    U_2 &= e^{-i \sum_{j=1}^N J \left(\cos(\lambda_2) \sigma^z_j + \sin(\lambda_2)\sigma^x_j \right) \left(\cos(\lambda_2) \sigma^z_{j+1} + \sin(\lambda_2) \sigma^x_{j+1} \right) }
    \\ 
    U_1 &= U_P(\yuanphi, \lambda_1, \yuanhz) e^{i\pi\sum_{j=1}^N (\sigma^x_j/2) },
    \qquad
    U_P(\yuanphi, \lambda_1, \yuanhz) = \prod_{j=1}^N
    e^{-i \yuanphi (\sigma^z_j/2)} e^{i\lambda_1 (\sigma^y_j/2)} e^{-i \yuanhz (\sigma^z_j/2)}.
\end{align}
With a uniform interaction $J_{j} = J$ here, two pairs of cat eigenstates with ferromagnetic and antiferromagnetic patterns are engineered according to the general relation in Eq.~\eqref{smeq:Js} (for simplicity, we choose the gauge that extra phase factors are absorbed into the Fock states which is allowed when only one pair of cat eigenstates are involved at a time, and also illustrate the dominant terms for the perturbed cat scars corresponding to their unperturbed anchor point form),
\begin{align}\nonumber
    &\text{Type 1: homogeneous --- } 
    \text{No extra $X(\pi)$ attached}
    \\ \label{smeq:homoscar}
    &
    \Rightarrow 
    J_{j} = J
    \Rightarrow \quad \sbbs_1=\sbbs_2=\dots=\sbbs_N=1
    \quad 
    \Rightarrow
    \begin{cases}
    \frac{1}{\sqrt{2}} (|0101\dots01\rangle + |1010\dots10\rangle ),\\
    \frac{1}{\sqrt{2}} (|0101\dots01\rangle - |1010\dots10\rangle );\\
\frac{1}{\sqrt{2}} (   |0000\dots00\rangle + |1111\dots 11\rangle), \\
    \frac{1}{\sqrt{2}} (|0000\dots00\rangle - |1111\dots11\rangle  )
    \end{cases},
\end{align}
where the pair with antiferromagnetic pattern is exemplified in \mbox{Fig.~\ref{fig:uf_circuit}a} and used in our experiment. 

To edit the scar spin patterns, we could change the signs of Ising interaction by attaching a pair of $X(\pi)$ gates before and after the $ZZ(-4)$ gates only at certain qubits, as illustrated in \mbox{Fig.~\ref{fig:uf_circuit}b}. That results in a dressed $\tilde{U}_F$. Suppose a pair of $X(\pi)$ gates are attached to $Q_j$, note $e^{-i\pi \sigma^x_j/2} = -i \sigma^x_j$ and $ \sigma^x_j \sigma^z_j \sigma^x_j = -\sigma^z_j$, we can effectively change the sign of $J$ in $U_F$ into a position-dependent one $J_j$ in $\tilde{U}_F$,
\begin{align}
    U_F\sim e^{-iJ [\sigma^z_j\sigma^z_{j+1} ]} 
    \quad\rightarrow\quad
    \tilde{U}_F \sim e^{-i\pi (\sigma^x_j/2) } 
    e^{-iJ [ \sigma^z_{j}  \sigma^{z}_{j+1}] } 
    e^{-i\pi (\sigma^x_j/2)} = -e^{-iJ [ (-\sigma^z_j) \sigma^z_{j+1} ]} \equiv -e^{-iJ_j [\sigma^z_j\sigma^z_{j+1} ]}.
\end{align}
Since the signs of $J_j$ depends on both qubits $Q_j$ and $Q_{j+1}$, generically it is edited into
\begin{align}\label{smeq:Jgates}
    J_j = (2x_{j}-1) (2x_{j+1}-1)J,
\end{align}
where $x_j = 1$ denotes no extra $X(\pi)$ gates are attached to $Q_j$, and $x_j = 0$ means a pair of $X(\pi)$ is attached to $Q_j$ sandwiching the $ZZ(-4)$ gates as in \mbox{Fig.~\ref{fig:uf_circuit}b}. Comparing Eq.~\eqref{smeq:Jgates} with Eq.~\eqref{smeq:Js}, we find the identification
\begin{align}
    x_{j} = \sbbs_j,
\end{align}
namely, if we want to flip the spin for cat scars with respect to Eq.~\eqref{smeq:homoscar} at certain qubits, just sandwich the $ZZ(-4)$ gates at corresponding qubits with pairs of $X(\pi)$ gates. The fact that scar spin patterns can be edited by single-qubit drivings $X(\pi)$ alone means that fast and accurate programming of scars into patterns compatible with GHZ states, which can even be switched during evolution, becomes possible. Also note that because both ferromagnetic and antiferromagnetic scar patterns are achieved simultaneously in Eq.~\eqref{smeq:homoscar}, at most $N/2$ pairs of $X(\pi)$ are needed to achieve arbitrary patterns. For intuitiveness, we exemplify the results by giving explicitly the gate configurations for the two types of scars in Fig.~4 of the main text, 
\begin{align}
    \nonumber
    & \text{Type 2: flip 1 at $Q_{19}$ --- }
    \text{A pair of $X(\pi)$ gates at $Q_{19}$}
    \\
    &
     \Rightarrow  
    \begin{cases}
        J_{18}=J_{19}=-J
        \\
        J_{j\neq 18, 19} = J
    \end{cases} 
    \Rightarrow 
    \sbbs_{19} = 0, \sbbs_{j\neq19} = 1
    \Rightarrow
    \begin{cases}
        \frac{1}{\sqrt{2}} (  |0101\dots01 {\color{red}\bf 1_{19}} 1 01\dots 01\rangle + |1010\dots10{\color{red}\bf 0_{19}}010\dots 10 \rangle),\\
        \frac{1}{\sqrt{2}} ( |0101\dots01 {\color{red}\bf 1_{19}} 1 01\dots 01\rangle - |1010\dots10{\color{red}\bf 0_{19}}010\dots 10 \rangle  );
\\
        \frac{1}{\sqrt{2}} (|0000\dots00{\color{red}\bf 1_{19}}000\dots 00\rangle + |1111\dots11{\color{red}\bf 0_{19}} 111\dots 11\rangle   ), \\
        \frac{1}{\sqrt{2}} (  |0000\dots00{\color{red}\bf 1_{19}}000\dots 00 \rangle - |1111\dots11{\color{red}\bf 0_{19}} 111\dots 11 \rangle).
    \end{cases}
    \\
    \nonumber
    & \text{Type 3: maximal deviation --- }
    \text{18 pairs of $X(\pi)$ gates at qubits $Q_{4m+2}, Q_{4m+3}, m=0,1,\dots8$}
    \\
    &
    \Rightarrow 
    \begin{cases}
        J_{4m+1} = J_{4m+3} = -J
        \\
        J_{4m+2} = J_{4m+4} = J
    \end{cases} 
    \Rightarrow\quad 
    \SBBS = 10011001\dots 1001
    \Rightarrow
    \begin{cases}
    \frac{1}{\sqrt{2}} (|0{\color{red}\bf01}1 0{\color{red}\bf01}1 \dots 0{\color{red}\bf01}1\rangle + |1{\color{red}\bf10}0 1{\color{red}\bf10}0 \dots 1{\color{red}\bf10}0\rangle), \\
        \frac{1}{\sqrt{2}} (|0{\color{red}\bf01}1 0{\color{red}\bf01}1 \dots 0{\color{red}\bf01}1\rangle - |1{\color{red}\bf10}0 1{\color{red}\bf10}0 \dots 1{\color{red}\bf10}0\rangle);
\\
        \frac{1}{\sqrt{2}} (   |0{\color{red}\bf11}0 0{\color{red}\bf11}0 \dots 0{\color{red}\bf11}0\rangle + |1{\color{red}\bf00}1 1{\color{red}\bf00}1 \dots 1{\color{red}\bf00}1\rangle), \\
        \frac{1}{\sqrt{2}} (   |0{\color{red}\bf11}0 0{\color{red}\bf11}0 \dots 0{\color{red}\bf11}0\rangle - |1{\color{red}\bf00}1 1{\color{red}\bf00}1 \dots 1{\color{red}\bf00}1\rangle).
    \end{cases}
\end{align}
where the highlighted qubits involve pairs of $X(\pi)$ gates.

\subsubsection{Quality control}

Preserving a GHZ state of $N$-qubits imposes stringent and unusual requirements on the DTCs accommodating it. On the one hand, any defective region can bring down the number of qubits being effectively entangled in a GHZ state after several periods of evolution. Meanwhile, a GHZ state made of two Fock state components chiefly overlaps with only two cat eigenstates in the relevant subspace. Thus, compared with properties averaged over all eigenstates, the protective effects on GHZ states depend more sensitively on the quality of a few relevant cat eigenstates.  In the following, we numerically and analytically benchmark the scaling of cat scars with the change of system size and perturbation strength. Also, we benchmark the results with MBL DTCs whose properties have been more extensively studied in previous works.

\begin{figure}
    [ht]
    \includegraphics[width=18cm]{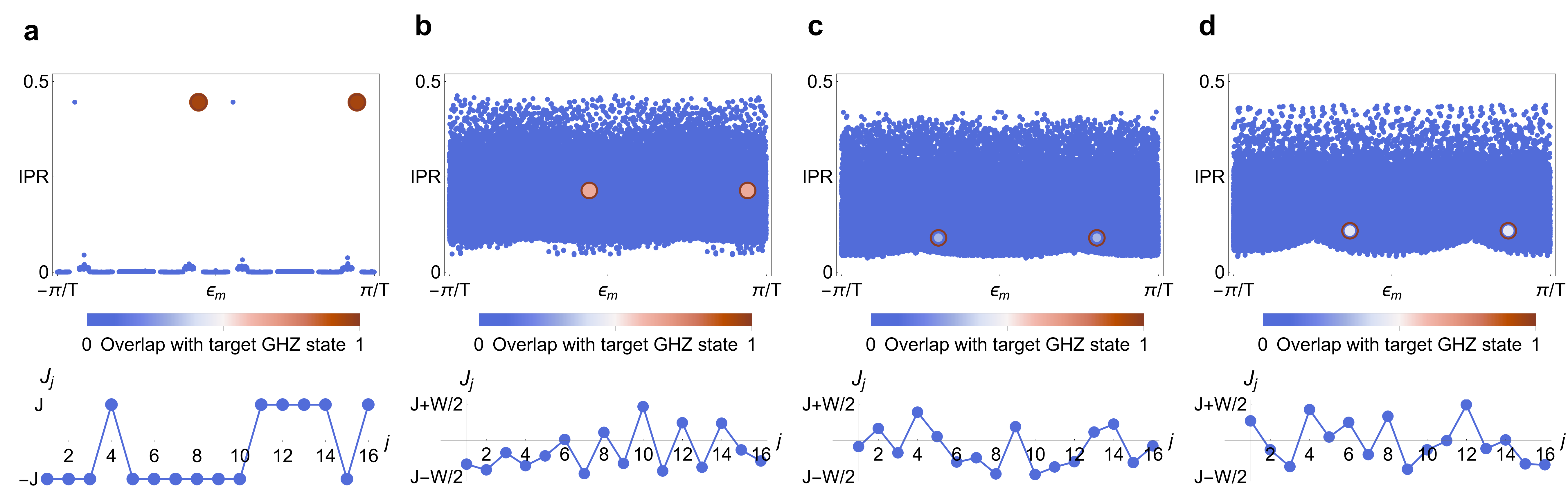}
    \caption{\label{fig:mbl_scar} Comparison of cat scar and more MBL samples for generating cat eigenstates to accommodate a targeted GHZ state. An arbitrary relevant subspace is spanned by $|\SBBS\rangle = |101001010101000001\rangle$ and $ |\bar{\SBBS}\rangle = |0101101010111110\rangle $, the same as that in the caption of Table~\ref{tab:dtc_compare}. Two eigenstates overlapping most strongly with the target GHZ state subspace are highlighted by red circles.   }
\end{figure}

Let us introduce the procedure of scaling analysis more concretely in Fig.~\ref{fig:mbl_scar}. For both cat scar and MBL cases, we fix an arbitrary pair of spin patterns $|\boldsymbol{s}\rangle, |\bar{\boldsymbol{s}}\rangle $ spanning the targeted GHZ state subspace at each system size $N$, and highlight the IPR for the eigenstate $|\epsilon_m\rangle $ of largest overlap $|\langle \boldsymbol{s}| \epsilon_m\rangle|^2 +|\langle \bar{\boldsymbol{s}}| \epsilon_m\rangle|^2 $ with the subspace. For cat scars, we take a single landscape of interaction $J_j$ specified by Eq.~\eqref{smeq:Js}, as illustrated in Fig.~\ref{fig:mbl_scar}a for the same result in Table~\ref{tab:dtc_compare}. For MBL cases, we fix the same GHZ state subspace as cat scars, and take different disordered sample $J_j$ as exemplified in \mbox{Fig.~\ref{fig:mbl_scar}b -- d}. In each sample, we find the eigenstate of largest overlap with the target GHZ state subspace, which are highlighted by red circles. Here for the MBL cases, we can readily see that the IPR of cat eigenstates relevant to a target GHZ state depend sensitively on different disorder realizations. Also, although in each sample there are high IPR eigenstates, there lacks a definitive scheme to ensure that those eigenstates are relevant to a given GHZ state.

\begin{figure}
    [ht]
    \includegraphics[width=18cm]{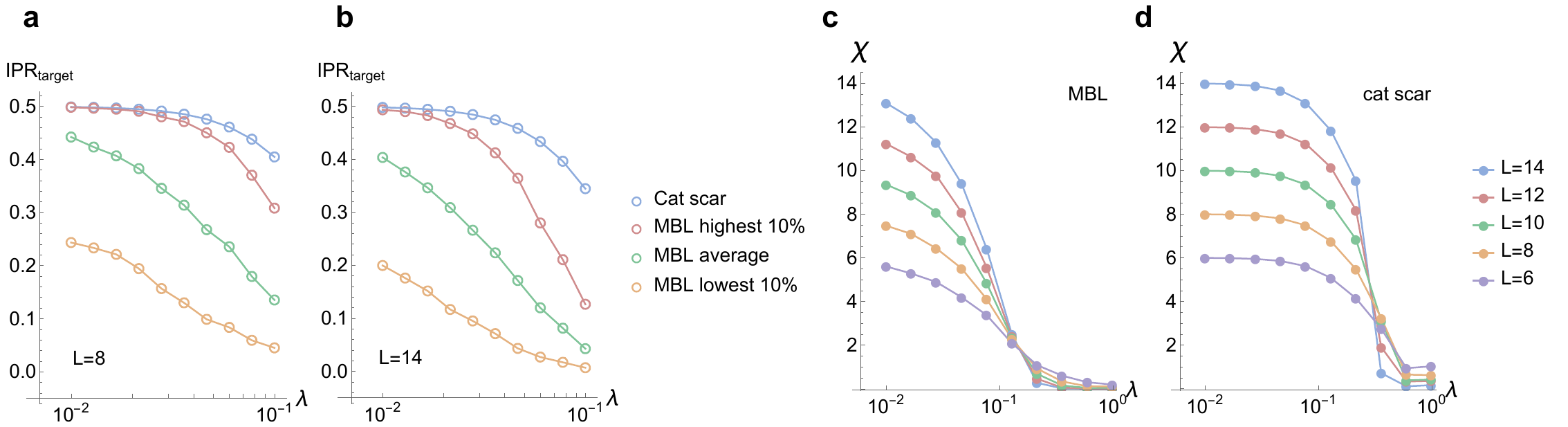}
    \caption{\label{fig:ea_ipr_sc}
    Comparison of scaling behaviors for cat scars and MBL cases. {\bf a} and {\bf b}, IPR of the targeted cat eigenstate in cat scar and MBL cases, where we fix $|\SBBS\rangle = |10011010\rangle$ for $N=8$ and $|\boldsymbol{s}\rangle = |01101110001001\rangle $ for $N=14$. For cat scars, we take a single landscape of interaction specified by Eq.~\eqref{smeq:Js}, and identify the IPR for the eigenstate with largest overlap with $|\SBBS\rangle, |\bar{\boldsymbol{s}}\rangle$. For MBL, we take $10^3$ samples independently for each $(\lambda,N)$. In each sample, we pick out the IPR for the cat eigenstate of largest overlap with $|\SBBS\rangle, |\bar{\SBBS}\rangle$, and form a set of $\text{IPR}_{\text{target}}$ contributed by different samples. From the set of $\text{IPR}_{\text{target}}$ at each $(\lambda, N)$, we obtain the average of highest $10\%$ IPRs, lowest $10\%$ IPRs, and overall averaged IPR.
    {\bf c} and {\bf d}, Edwards-Anderson parameter for estimating stable DTC regimes. In both cases, we average over $10^3$ samples at each $(\lambda, N)$ data point. For MBL, each sample involves a random profile of $J_j$ and we average over all eigenstates. For cat scars, each sample is specified by different target GHZ state subspace $|\SBBS\rangle, |\bar{\SBBS}\rangle$. We take compatible $J_j$ specified in Eq.~\eqref{smeq:Js} and the cat scar relevant to the target GHZ subspace.
    }
\end{figure}

In \mbox{Fig.~\ref{fig:ea_ipr_sc}a -- b} we show the result of scaling for two system sizes $N=8$ and $N=14$. Here we fix the target GHZ state subspace $|\SBBS\rangle, |\bar{\SBBS}\rangle $ in the caption, and compare the IPRs of relevant cat eigenstates indicated by the largest overlap with GHZ state subspace. The cat scar takes a definite $J_j$ landscape given by Eq.~\eqref{smeq:Js}. To compare with disordered cases more comprehensively, MBL cases take $10^3$ samples at each $(\lambda, N)$ of $J_j\in[J-W/2, J+W/2]$ with $J=W=\pi/4$ as in Table~\ref{tab:dtc_compare}. From each sample we locate the eigenstate of largest overlap with target GHZ state subspace and form a set of $\text{IPR}_{\text{target}}$ at each $(\lambda,N)$ contributed by all $10^3$ samples. In Fig.~\ref{fig:ea_ipr_sc}a and b, we see that the IPRs of relevant cat eigenstates for MBL cases distribute widely among different samples, signaled by the large separation between highest $10\%$ IPRs and lowest $10\%$ ones. On average the IPRs of cat scars are notably higher than the relevant cat eigenstates in MBL cases, a trend that tends to be enhanced with the increase of system size. Therefore, the data suggests that the cat scar scheme can consistently produce relevant cat eigenstates belonging to those of highest quality in a deterministic way.

Next, we estimate the parameter regimes where cat scars are stable against perturbation. Here we exploit the Edwards-Anderson parameter
\begin{align}
    \chi(\epsilon_m) = \frac{1}{N-1} \sum_{j \neq k} \langle \epsilon_m| \sigma^z_j \sigma^z_k |\epsilon_m \rangle^2,
\end{align}
which has been used to benchmark stability of MBL DTCs~\cite{Ippoliti2021}. A finite value of $\chi$ indicates a spatial long-range correlation for the associated eigenstates $|\epsilon_m\rangle$, while for delocalized eigenstates its values approach 0. Note that there are $N(N-1)$ terms in the summation $j\neq k$. For an ideal cat eigenstate, each term contributes 1, giving $\chi = N$ proportional to system size. Thus, $\chi$ is an extensive quantity, in contrast to IPR being an intensive one approaching a fixed value $1/2$ for ideal cat eigenstates. Unlike IPRs for measuring the cat eigenstate quality, the usefulness of $\chi$ is to give an estimation of transition points between DTC and thermal regimes. In the time-crystalline ordered regime larger system sizes give larger values of $\chi \xrightarrow{\lambda\rightarrow0} N$. Contrarily, in the ergodic limit each term in the summation approaches a value inversely proportional to the Hilbert space size, which decreases exponentially with $N$, so larger $N$ gives smaller $\chi$. That means the scaling curves will have a crossing point for transitions between the localized DTC regime and the ergodic thermalizing regime with the change of perturbation strength $\lambda$. The scaling of $\chi(\epsilon_m)$ is shown in Fig.~\ref{fig:ea_ipr_sc}c and d for MBL and cat scars respectively. In both cases, we take $10^3$ samples at each $(\lambda, N)$ parameter point. For MBL cases, we follow Ref.~\cite{Ippoliti2021} to select random $J_j\in[\pi/8, 3\pi/8]$ and average over all eigenstates. For the cat scar case, a target GHZ state subspace $|\boldsymbol{s}\rangle, |\bar{\boldsymbol{s}}\rangle $ is randomly selected in each sample. Then, the compatible $J_j$ given in Eq.~\eqref{smeq:Js} gives the relevant cat scars with associated $\chi$. In both cases we observe the expected crossing point, namely, $\lambda_c\approx 0.129$ for MBL and $ \lambda_c \approx 0.36 $ for cat scars. Thus, our choice of perturbation strength $\lambda=0.05$ resides deeply within the DTC regime.

Numerical evidence above has shown qualitatively the cat scar stability and the parameter regimes that can host such robust scars. Experimentally, our system sizes have far transcended the limit for classical computers to perform unbiased simulations (i.e. exact diagonalization) and extract eigenstate properties. As an alternative way to provide quantitative estimations of cat scar quality, we apply IPR predictions by analytical perturbation theory~\cite{Huang2023} as given in Eq.~\eqref{smeq:ipr_analytical}.

\begin{table}
    [ht]
    \caption{\label{tab:iprs} Comparison of analytical (Eq.~\eqref{smeq:ipr_analytical}) and numerical results of cat scar IPRs for perturbation strength $\lambda_1=\lambda_2 \equiv \lambda$ and sizes $N$. }
    \begin{tabular}{clllllllllll}
        \hline\hline
        $\lambda$ & 0.01 & 0.01292 &  0.01668 & 0.02154 & 0.02783 & 0.03594 & 0.04642 & {\bf 0.05} & 0.05995 & 0.07743 & 0.1 
        \\ \hline
        numerical $(N=8)$ & 0.498820 & 0.4980353 & 0.496731 & 0.49457& 0.49101& 0.4852& 0.4758& {\bf 0.4721} & 0.4609 & 0.438& 0.405 \\
        analytical $(N=8)$ & (fit $\bar{V}^2$) & 0.4980345 & 0.496728 & 0.49456 & 0.49097 & 0.4851 & 0.4755 & {\bf 0.4717} &0.4601 & 0.436 & 0.400
        \\
        est. err. $ (\lambda^2N)^2$ & -- & $0.000002$ & $ 0.000005 $ & $ 0.00001$ & $0.00004$ & $0.0001$ & $ 0.0003$ & {\bf $0.0004$} & $0.0008$ & $ 0.002$ & $ 0.006$
        \\ \hline
        numerical $(N=14)$ & 0.497937 & 0.496566 & 0.494293 & 0.490536 & 0.484367 & 0.47434 & 0.45832 & {\bf 0.45213} & 0.4334 & 0.3963 & 0.344 \\
        analytical $(N=14)$ & 0.497938 & 0.496568 & 0.494295 & 0.490537 & 0.484364 & 0.47432 & 0.45826 & {\bf 0.45205} & 0.4332 & 0.3958 & 0.343
        \\
        est. err. $(\lambda^2N)^2$ & $0.000002$ & $0.000006$ & $0.00002$ & $0.00004$ & $0.0001$ & $0.0003$ & $0.0009$ & {\bf $0.001$} & $0.002$ & $0.007$ & $0.02$
        \\ \hline\hline 
    \end{tabular}
\end{table}

We first benchmark the accuracy by comparing the analytical results with the numerical data in Fig.~\ref{fig:mbl_scar}a and b for cat scars, as shown in Table~\ref{tab:iprs}. We extract the coupling strength $\bar{V}^2 = [(2\text{IPR})^{-1/2} - 1]\lambda^{-2}N^{-1} \approx 1.47762 $ using a single data point of $\lambda=0.01, N=8$, as indicated in the third row and second column of Table~\ref{tab:iprs}. Then, we can proceed to obtain IPRs for all other $(\lambda, N)$. In Table~\ref{tab:iprs}, we see the quantitative agreement between analytical results given by $(1/2) (1+\bar{V}^2\lambda^2N)^{-2}$ in Eq.~\eqref{smeq:ipr_analytical}, and the numerical data corresponding to the cat scar IPRs in Fig.~\ref{fig:ea_ipr_sc}a and b, with differences residing within the estimated errors $(\lambda^2N)^2$. This error corresponds to the amplitudes of next leading order term in the perturbation series when the most generic perturbations are present. Thus, when considering a specific model, $(\lambda^2N)^2$ usually overestimates the error and can serve as an upper bound for the errors of analytical predictions.

\begin{table}
    [ht]
    \caption{\label{tab:iprL36} Numerical and analytical results of the IPRs for cat scars with spin patterns $ |0101\dots01\rangle$, $|1010\dots10\rangle$, and $\lambda_1=\lambda_2=0.05$. Here parameters are the same as the experimental cases.}
    \begin{tabular}{cllllllll}
        \hline\hline 
        $N$ & 8 & 10 & 12 & 14 & 16 & 18 & 20 & {\bf 36 (expt.)}
        \\ \hline 
        numerical &0.481725 & 0.47726 & 0.47284 & 0.4685 & 0.4641 & 0.4598 & 0.4556 & 
         ---
        \\ \hline 
        analytical & (fit $\bar{V}^2$) & 0.47731 & 0.47296 & 0.4687 & 0.4644 & 0.4603 & 0.4561 & {\bf 0.4251}
        \\ \hline 
        est. err. $(\lambda^2N)^2$ & --- & 0.0006 & 0.0009 & 0.001 & 0.002 & 0.002 & 0.003 & {\bf 0.008}
        \\ \hline \hline
    \end{tabular}
\end{table}

Next, we estimate the IPR for experimentally prepared cat scars in a 36-qubit DTC. Here, to access more system sizes, we consider the translation-invariant setup with scars given by Eq.~\eqref{smeq:homoscar}. This is the major case investigated in the main text. In Table~\ref{tab:iprL36}, we take the experimentally used parameters and vary the system size $N$ where we compare the numerical and analytical results for IPRs of cat scars in the target GHZ state subspace. Again, agreements are found, with errors within the range of next-leading-order strength $(\lambda^2N)^2$. Based on this, we obtain the IPR of cat scars for the $N=36$ case in experiment via analytical method as 
\begin{align}
    \text{IPR}_{\text{scar}}(N=36) = 0.4251 \pm 0.008.
\end{align}
This value will be used to analyze experimental data for the dynamical macroscopic quantum coherence in Sec.~\ref{sec:data_dance}.

\subsection{Macroscopic quantum coherence in static and DTC systems\label{sec:macro_coh}}

Our experiments are based on a quantitative characterization of macroscopic coherence for GHZ states. Traditionally, several measurement protocols have been developed to measure the static macroscopic coherence, including parity and multiple-quantum-coherence (MQC) measurements. They both share the similar feature of quantifying the far-off-diagonal elements in the density matrix related to the subspace spanned by the two Fock state components in GHZ states. We would start from the static case, review the essential concepts, and generalize these protocols to the dynamical regime for characterizing unconventional DTC features exhibited by the GHZ states.

We first review how the macroscopic coherence for a static GHZ state is defined and measured. Suppose a target ideal GHZ state of $N$ particles is
\begin{align}\label{smeq:GHZideal}
    |\spps\rangle_N = \frac{1}{\sqrt{2}} \left(
    |\SBBS\rangle + e^{-i\Phi} |\bar{\SBBS}\rangle
    \right),
\end{align}
with Fock states $|\SBBS\rangle = |\sbbs_1\sbbs_2\dots \sbbs_N\rangle$ and $|\bar{\SBBS}\rangle = |(1-\sbbs_1)(1-\sbbs_2)\dots (1-\sbbs_N)\rangle$, $\sbbs_j=1,0$ as before. If a pure state $|\Psi\rangle $ is generated, the fidelity of such a state compared with the ideal $|\spps\rangle_N$ can be straightforwardly defined as the overlap ${\cal F} = |\langle \Psi|\spps\rangle_N|^2 = \,_N\langle \spps| \rho_{\text{expt}} |\spps\rangle_N \in[0,1]$, where $ \rho_{\text{expt}} = |\Psi\rangle \langle \Psi| $, and ${\cal F}=1$ means perfect overlap with highest fidelity. 
In real experiments, a mixed state represented by a generic density matrix $\rho_{\text{expt}}$ can be expanded in the Fock basis, where the subspace relevant to GHZ state patterns are spanned simply a two-by-two matrix
\begin{align}\label{smeq:GHZ}
    \rho_{\text{expt}} = 
    \begin{pmatrix}
        |\SBBS\rangle & |\bar{\SBBS}\rangle
    \end{pmatrix}
    \begin{pmatrix}
        P_{\SBBS} & a_{2}
        \\
        a_2^* & P_{\bar{\SBBS}}
    \end{pmatrix}
    \begin{pmatrix}
        \langle \SBBS | \\ \langle \bar{\SBBS} |
    \end{pmatrix}
    .
\end{align}
Then, the fidelity of a mixed state $\rho_{\text{expt}}$ compared with target GHZ state $|\spps\rangle_N$ can be similarly defined as 
\begin{align}\label{smeq:fidelitya}
    {\cal F} = \,_N\langle \spps | \rho_{\text{expt}} | \spps \rangle_N = \frac{1}{2} \left(
    P_{\SBBS} + P_{\bar{\SBBS}}
    \right) + \text{Re}(a_2 e^{-i\Phi}).
\end{align}
Here, real numbers $P_{\SBBS}$ and $P_{\bar{\SBBS}}$ in diagonal parts of $\rho_{\text{expt}}$ represent the probability of observing particle distribution of patterns $|\SBBS\rangle $ and $ |\bar{\SBBS}\rangle$ respectively. Crucially, the off-diagonal element $\text{Re}(a_2e^{-i\Phi})$ stands for macroscopic coherence of a quantum state, and will be our focus in the following. 

First, from the above discussion, it can be understood why ${\cal F}>0.5$ is usually quoted as a benchmark of macroscopic coherence for experimentally generated GHZ states. This is because a density matrix has unit trace $\text{tr}(\rho_{\text{expt}}) = 1$, which means within GHZ subspace $ \frac{1}{2} \left(
P_{\boldsymbol{s}} + P_{\bar{\boldsymbol{s}}}
\right) \leqslant 0.5$ in the first term of Eq.~\eqref{smeq:fidelitya}. For instance, an unentangled Fock state $|1\rangle^{\otimes N}$ gives the density matrix $\rho_{\text{expt}} = |1\rangle^{\otimes N} \,^{\otimes N}\langle 1| $, and therefore $
P_{\SBBS}=1, a_2= P_{\bar{\SBBS}}=0$, giving ${\cal F}=0.5$. Thus, the criterion ${\cal F}>0.5$  guarantees $|a_2|\neq 0$, which then verifies macroscopic coherence for a quantum state $\rho_{\text{expt}}$.

Strictly speaking, ${\cal F}>0.5$ is a sufficient but unnecessary condition for the existence of macroscopic coherence $|a_2|\neq 0$ itself, because the diagonal components  $\frac{1}{2} \left(
P_{\SBBS} + P_{\bar{\SBBS}}\right)$ of Eq.~\eqref{smeq:fidelitya} can fairly be smaller than $0.5$, implying that a quantum state has partially leaked out from the GHZ pattern subspace, but the remaining components can still exhibit nonzero macroscopic coherence $|a_2|\neq0$.
In conventional experiments, without further aids, usually the coherence $a_2$ decays much faster than the probability distributions $P_{\SBBS}, P_{\bar{\SBBS}}$. Namely,  usually $a_2$ quickly decays to zero before probability $\frac{1}{2} \left(
P_{\SBBS} + P_{\bar{\SBBS}}\right)$ notably deviates from $0.5$ during dynamical processes, and therefore using ${\cal F}>0.5$ is a good approximation to verify $a_2\neq0$ there. However, macroscopic coherence in DTCs can be abnormally robust, as $a_2\neq0$ is maintained even when the depolarizing noise (which necessarily causes leakage $ \frac{1}{2}\left( P_{\SBBS} + P_{\bar{\SBBS}}\right)<0.5$) already plays significant roles. Thus, we opt to observe the off-diagonal elements $a_2$ of density matrices directly in our experiments, unlike ${\cal F}$ contributed by both diagonal and off-diagonal ones.

In the following, we review two widely exploited methods of measuring $|a_2|$ directly, namely, the parity and the multiple-quantum-coherence (MQC) methods. They constitute the conceptual basis for our design of Schr\"{o}dinger cat interferometry to quantify dynamical macroscopic coherence for evolved GHZ states.

Both methods are built upon the intuition that for a GHZ state of the form in Eq.~\eqref{smeq:GHZideal}, if one applies a phase rotation $\phi$ on each constituent spin ($\boldsymbol{s}=0101\dots$),
\begin{align}\label{smeq:phase_rotate}
    e^{i\sum_{j=1}^N (-1)^j \phi \sigma^z_j/2} | \spps \rangle_N = e^{iN \phi/2} |\spn\rangle_N,
\end{align}
the relative phase $\Phi$ between two components in Eq.~\eqref{smeq:GHZideal} is rotated by an enhanced amount proportional to $N$. Such a phase sensitivity is often exploited to perform ultra-sensitive sensing of external magnetic fields using GHZ states, where a small change in field strength $\phi$ is amplified by a factor $N$. This is to be contrasted against other short-range entangled states, i.e. a local product state, where phase rotations on different sites need not purely add up but may cancel each other as large numbers of different Fock states are involved. Thus, the factor $N$, or equivalently speaking the $\sim1/N$ periodic response to phase rotation $\phi$ is a unique character of GHZ states made of two maximally different Fock states.

Following the operation in Eq.~\eqref{smeq:phase_rotate}, there are two options to reveal the period-$\pi/N$ phase rotation response. One option is the {\bf parity} measurement, where the parity operator along, i.e. $x$-axis, reads
\begin{align}
    {\cal P}_x = \prod_{j=1}^N \rxj, 
\end{align}
It has eigenstates
\begin{align}
    {\cal P}_x|\pm, \SBBS\rangle_N = \pm |\pm, \SBBS\rangle_N,
    \qquad
    |\pm, \SBBS\rangle_N = \frac{1}{\sqrt{2}}(| \SBBS \rangle \pm |\bar{\SBBS}\rangle).
\end{align}
Correspondingly, an ideal GHZ state can be expressed as
\begin{align}
    |\spn\rangle_N = e^{-i(\Phi + N \phi)/2} \left(
    \cos\frac{\Phi + N \phi}{2} |+, \SBBS\rangle_N + i\sin\frac{\Phi + N \phi}{2} |-,\SBBS \rangle_N 
    \right)
\end{align}
Thus, the probability of obtaining $+1$ for the parity measurement is
\begin{align}
    P_{|+,\SBBS\rangle} = \cos^2 \frac{\Phi + N \phi}{2} = \frac{1}{2}(1+\cos(\Phi + N \phi)).
\end{align}
The macroscopic coherence $|a_2|$ can then be read out from the period-$2\pi/N$ oscillation amplitude of $P_{|+,\SBBS\rangle}$ with the change of $\phi$. In practice, one can combine the phase rotation in Eq.~\eqref{smeq:phase_rotate} and a global spin-flipping exchanging spin $x$ and $z$ basis, namely, $U_{\text{parity}} = e^{i\frac{\pi}{2} \sum_{j=1}^N \left(\cos(\phi) \rxj + \sin(\phi) \ryj \right)/2} $, $U_{\text{parity}}^\dagger \rho_{\text{expt}} U_{\text{parity}}$. Then, the $P_{+1}$ probability for parity corresponds to the probability of observing the ground state $|0\rangle^{\otimes N}$ in the spin-$z$ basis after the operation $U_{\text{parity}}$. 

Another option is the {\bf MQC} method more suitable for systems equipped with digital gates. MQC operation involves three consecutive steps, which we illustrate using the ideal GHZ state in Eq.~\eqref{smeq:GHZideal}. First, one similarly performs a phase rotation as in Eq.~\eqref{smeq:phase_rotate}, ending up with the GHZ state $|\spn \rangle_N$. Second, different from the parity protocol involving simultaneous spin flips, here one consecutively acts on neighboring qubits with CNOT gates $ |11\rangle \leftrightarrow |10\rangle, |01\rangle \leftrightarrow |01\rangle, |00\rangle \leftrightarrow |00\rangle  $, which is the inverse of the scheme to generate a GHZ from a product state. Then, the GHZ state is disentangled to
\begin{align}
    |\spn\rangle_N \rightarrow \frac{1}{\sqrt{2}} 
    \left(|1\rangle + e^{-i(\Phi + N \phi)}|0\rangle \right) 
    \otimes 
    |0\rangle^{\otimes (N-1)},
\end{align}
namely, the whole phase factor is deposited to the first qubit. 
Finally, one performs a spin-rotation on the first qubit alone $e^{i\frac{\pi}{4} \sigma^x_1}$, and measure the ground state probability
\begin{align}
    P_{|000\dots0\rangle} = \cos^2\frac{\Phi + N \phi}{2} = \frac{1}{2}(1+\cos(\Phi + N \phi)),
\end{align}
which is of the same character as $P_+$ in parity measurement. For practical purposes, a spin-flipping $\pi$-pulse $\prod_{j=1}^N \rxj$ is usually applied after the generation of GHZ states, before performing the above three steps for MQC measurements. The pulse serves to cancel the dephasing noise accumulated during the GHZ state generation with the corresponding phase noise in MQC measurement processes.

Compared with the static macroscopic coherence measurement, the additional task we need to perform in DTC systems is to reveal the dynamics of coherent phase $\Phi$ in GHZ state $|\spps\rangle_N$. In an ideal noise-free setting, one can simply observe the oscillation via, i.e. the traditional MQC measurements
    $P_{|000\dots0\rangle} \sim \cos\left[(\Phi + N \phi) (-1)^{t/T}\right].$
However, in practical experiments, the DTC evolution $U_F^{t/T}$ is accompanied by dephasing noise that would endow an additional time-dependent factor $
    P_{|000\dots0\rangle} \sim \cos\left[(\Phi  + N \phi) (-1)^{t/T} + \Phi_{\text{noise}}(t)\right].$
The dephasing noise contributes a random factor in each cycle, and therefore a definition of the DTC order in terms of the value of the phase factor at a single $\phi$ is invalid. Importantly, what really serves as the definitive evidence of macroscopic coherence for GHZ state is the $\sim 1/N $ periodicity of the phase response to single-qubit rotation, namely, the $N \phi$ term of the phase factor, rather than the absolute value of $\Phi$ that can be redefined via a gauge transformation. 

Thus, we define the Schr\"{o}dinger cat interferometry protocol in the following way, and illustrate its effects on an ideal GHZ state $|\spps\rangle_N$.
\begin{enumerate}
    \item Generate a GHZ state $|\spps \rangle_N$. Apply phase rotation $\phi$ on each qubit to levitate the coherent phase for GHZ state into $|\spn \rangle_N$.
    \item Evolve with DTC unitary for certain periods, ending up with the state $|(\Phi + N \phi)(-1)^{t/T} + \Phi_{\text{noise}}(t), \SBBS \rangle_N$, 
    \item Apply a spin $\pi$-pulse to reduce the effects of dephasing noise, giving $|-(\Phi + N \phi)(-1)^{t/T} - \Phi_{\text{noise}}(t), \SBBS \rangle_N $
    \item 
    Crucially, perform again single-qubit phase rotation with opposite values $-\phi$, so the GHZ state becomes
    \begin{align}
        \left|-N \phi (1 + (-1)^{t/T}) -\left(\Phi (-1)^{t/T} + \Phi_{\text{noise}}(t) \right) , \SBBS \right\rangle_N .
    \end{align}
    \item 
    Apply the reversed circuit for generating GHZ state similar to the last step in MQC measurements, so the ground state probability corresponds to the phase information
    \begin{align}\label{smeq:kphi_ideal}
        \mathcal{K}'(\phi ,t) \equiv P_{|000\dots 0\rangle} = 
        \cos^2 \left[N \phi(1+(-1)^{t/T}) + \delta \Phi(t)\right]
    \end{align}
    where $\delta\Phi(t) = \Phi(-1)^{t/T} + \Phi_{\text{noise}}(t)$, which does not carry the factor $N$.
\end{enumerate}
Now, we see that the spatiotemporal structure of cat eigenstates is simultaneously revealed by the first term in the coherent phase factor $-N \phi(1+(-1)^{t/T})$, without referring to the absolute value of the whole phase factor. Here, the $\pi/N$-periodicity in $\phi$ at even period ends benchmarks the $N$-qubit macroscopic coherence. Meanwhile, the disappearance and revival of oscillation in $\phi$ alternating between odd and even periods respectively confirm the period-$2T$ oscillation, and therefore the quasienergy spectral pairing gap $\pi/T$ for cat scars. To extract the macroscopic coherent part $\sim N \phi$ from the whole phase factor, we perform a Fourier transformation
\begin{align}
    \fuliyeK = \left| \frac{1}{M} \sum_{k=0}^{M-1} e^{i (qN) \phi(k) } \mathcal{K}'(\phi(k),t) \right|,
    \qquad
    \phi(k) = -\frac{\pi}{N} + \frac{2\pi}{NM} k, \quad k=0,1,\dots, (M-1),
\end{align}
where $M$ is the number of phase angles $\phi$ measured in the experiment. Then, the $\pi/N$-periodicity in $\phi$ is reflected by the Fourier component $\fuliyeK = \mathcal{K}^\prime_f(-2N,t)$. For an ideal GHZ state evolving in an unperturbed DTC where $\mathcal{K}^\prime(\phi,t)$ is given by Eq.~\eqref{smeq:kphi_ideal},
\begin{align}
    \fuliyeK = \frac{1}{4} \frac{1+(-1)^{t/T}}{2} = 
    \begin{cases}
        0.25, & \text{ even period $t/T$}
        \\
        0, & \text{ odd period $t/T$}
    \end{cases}.
\end{align}

\subsection{Analysis of main text and additional experimental data }

\subsubsection{Phase oscillation of GHZ states \label{sec:data_dance}}

To connect the previous noise-free results with experimental data, we exploit a phenomenological model to simulate the effects of noise. 
Suppose the error rate on each qubit and over each Floquet driving cycle is $e_p$.  The macroscopic quantum coherence $\mathcal{K}'(\phi,t)$ involves checking simultaneously all qubits, i.e. the ground state probability $ P_{|000\dots0\rangle}$. Thus, $\fuliyeK$ in an $N$-qubit system would experience an extra noise-induced decay
\begin{align}
    (1-e_p)^{Nt} ,
    \qquad
    e_p = \begin{cases}
        0.007, & \text{DTC case with two-qubit gate errors};
        \\
        0.003, & \text{Rabi case without two-qubit gate errors}
    \end{cases},
\end{align}
where the values of $e_p$ are estimated by comparing with experimental data, which are close to the gate error rates.

For cat scar DTCs, the cat scar wave function under perturbation can be written as $ \frac{\alpha_0}{\sqrt2}(|\boldsymbol{s}\rangle \pm |\bar{\boldsymbol{s}}\rangle) + ...$, where the delocalized fractions have a negligible contribution to IPR, while the dominant part effective for protecting a GHZ state is approximately related to the scar IPR$\approx \alpha_0^4/2$. Therefore, the cat scar $\text{IPR}$ is converted to the relevant fraction of the wave function amplitude as $\alpha_0^2 = \sqrt{2}\cdot \sqrt{\text{IPR}} \approx 0.922 $, where we use the analytical value $\text{IPR}_{\text{scar}} = 0.4251$ for 36 qubits from Table~\ref{tab:iprL36}. Thus,
\begin{align}\label{smeq:KDTC}
    \frac{\fuliyeK}{\fuliyeKchu} =  (1-e_p)^{Nt} \times (92\%) .
\end{align}
Thus, the evolution of $\fuliyeK$ is dominated by a linear-exponential decay caused by noise effect.

For a Rabi oscillator, its evolutions are given by single-qubit rotations,
\begin{align}\nonumber
    U_F &= U_1 = (-i)^N \prod_{j=1}^N \left( e^{-i \yuanphi \rzj/2 } e^{i \lambda_1 \ryj/2} e^{-i \yuanhz \rzj/2} \rxj \right)
     \qquad
     \Rightarrow
     \quad
     U_F^2 \equiv (-1)^N \prod_{j=1}^N u_j^2,\qquad
     u_j = e^{-i(2\alpha) \hat{\boldsymbol{n}} \cdot \boldsymbol{\sigma}_j/2},
\end{align}
where the detuning from echoes ($u_j^2=1$) for every two periods at each qubit  reads
\begin{align}\nonumber
    u_j^2 &= \left( e^{-i \yuanphi \rzj/2} e^{+i \lambda_1 \ryj/2} e^{-i \yuanhz \rzj/2}  \right)
    \left( e^{+i \yuanphi \rzj/2} e^{-i \lambda_1 \ryj/2} e^{+i \yuanhz \rzj/2}  \right)
    =
    e^{i\lambda_1(\ryj \cos\yuanphi - \rxj \sin\yuanphi)/2}
    e^{-i\lambda_1 (\ryj\cos(\yuanhz) - \rxj \sin(\yuanhz) )/2}
    \\ \label{smeq:uj}
    &=
    \cos^2\frac{\lambda_1}{2} + \sin^2\frac{\lambda_1}{2} \cos(\yuanphi- \yuanhz)
    -
    i\sin\frac{\lambda_1}{2}\cos\frac{\lambda_1}{2} 
    \left(
    \rxj (\sin\yuanphi - \sin \yuanhz) + \ryj(-\cos\yuanphi + \cos \yuanhz)
    \right) 
    -i\rzj \sin^2\frac{\lambda_1}{2} \sin(\yuanphi - \yuanhz).
\end{align}
That gives the detuning angles
\begin{align}\label{smeq:detuneangle}
    \alpha = \arccos\left(1- \sin^2\frac{\lambda_1}{2} (1 - \cos(\yuanphi - \yuanhz)) \right),
    \qquad
    \hat{n}_z = \pm \sin\left[
    \arctan \left(
    \tan\frac{\lambda_1}{2} \cos\frac{\yuanphi - \yuanhz}{2}
    \right)
    \right],
\end{align}
where $\pm$ sign corresponds to $\yuanphi - \yuanhz >0$ or $<0$. Then, at even periods $t/T$, we have
\begin{align}
    u_j^{t/T} = e^{-i(\alpha t/T) \hat{\boldsymbol{n}} \cdot \boldsymbol{\sigma}_j/2 } = 
    \cos\frac{\alpha t}{2T} - i\sin\frac{\alpha t}{2T} \hat{n}_z \rzj - i\sin\frac{\alpha t}{2T} (\hat{n}_x\rxj + \hat{n}_y \ryj)
\end{align}
Acting on a GHZ state, the perturbed Rabi oscillator performs a rotation (at even period ends $t/T$)
\begin{align}\nonumber
    &
    \prod_{j=1}^N u_j^{t/T} \frac{|\sbbs_1 \sbbs_2\dots \sbbs_N\rangle + e^{i\Phi} |\bar{\sbbs}_1 \bar{\sbbs}_2 \dots \bar{\sbbs}_N \rangle}{\sqrt2}
    \\
    &=
    \frac{1}{\sqrt2} \left[
    \left(
    (\cos\frac{\alpha t}{2T} - i \sbbs_j \sin\frac{\alpha t}{2T} \hat{n}_z) |\sbbs_j\rangle + (|\bar{\sbbs}_j \rangle \text{ component})
    \right)^{\otimes N}  
    +e^{i\Phi}
    \left(
    (\cos\frac{\alpha t}{2T} - i \sbbs_j \sin\frac{\alpha t}{2T} \hat{n}_z) |\bar{\sbbs_j} \rangle + (|\sbbs_j \rangle \text{ component})
    \right)^{\otimes N}  
    \right]
\end{align}
Thus, the probability for the evolved state to stay in the original subspace $|\sbbs_j\rangle^{\otimes N}, |\bar{\sbbs}_j\rangle^{\otimes N}$ is
\begin{align}\label{smeq:rabiprob1}
    P_{|\SBBS\rangle + \exp(i\Phi) | \bar{\SBBS} \rangle} = \left( \cos^2\frac{\alpha t}{2T} + \sin^2\frac{\alpha t}{2T} \hat{n}_z^2 \right)^N.
\end{align}	
Intuitively, we can look at two limits. The strongest perturbation effect is achieved at $\yuanphi - \yuanhz = \pi \mod 2\pi$, where
\begin{align}\nonumber
    \alpha = \lambda_1,  \hat{n}^2_z = 0 
    \quad 
    \Rightarrow 
    P_{|\SBBS\rangle + \exp(i\Phi) | \bar{\SBBS} \rangle} &= \cos^{2N} \frac{\lambda_1 t}{2T} 
    \xrightarrow{N\gg1, \lambda_1 t/2T\ll 1 }  \left( 1- \frac{1}{2!}\frac{\lambda_1^2t^2}{(2T)^2}
    \right)^{2N} 
    \\
    &\approx e^{-N(\lambda_{1}/2)^2 (t/T)^2}
\end{align}
Contrarily, in the limit $\yuanphi - \yuanhz = 0 \mod 2 \pi$,  $
\alpha = 0 \Rightarrow P_{|\SBBS\rangle + \exp(i\Phi) | \bar{\SBBS} \rangle} = 1$, a single-spin echo $u_j^2=1$ is approached, as can be directly verified by Eq.~\eqref{smeq:uj}. For other values of $\yuanphi - \yuanhz$, we can use the unified form for probability of evolved state to stay in the original GHZ subspace
\begin{align}\label{smeq:rabiprob2}
    P_{|\SBBS\rangle + \exp(i\Phi) | \bar{\SBBS} \rangle} \approx e^{-N\lambda_{\text{eff}}^2 (t/T)^2},
    \qquad
    \lambda_{\text{eff}}^2 = \left(\frac{\alpha}{2}\right)^2(1-\hat{n}^2_z),
\end{align}
where $\alpha, \hat{n}_z$ are given by Eq.~\eqref{smeq:detuneangle}. Several exemplary effective decay rate $\lambda_{\text{eff}}$ and the comparison of expressions for probability in rigorous Eq.~\eqref{smeq:rabiprob2} and the approximated Eq.~\eqref{smeq:rabiprob2} are shown.
\begin{figure}
    [ht]
    \includegraphics[width=18cm]{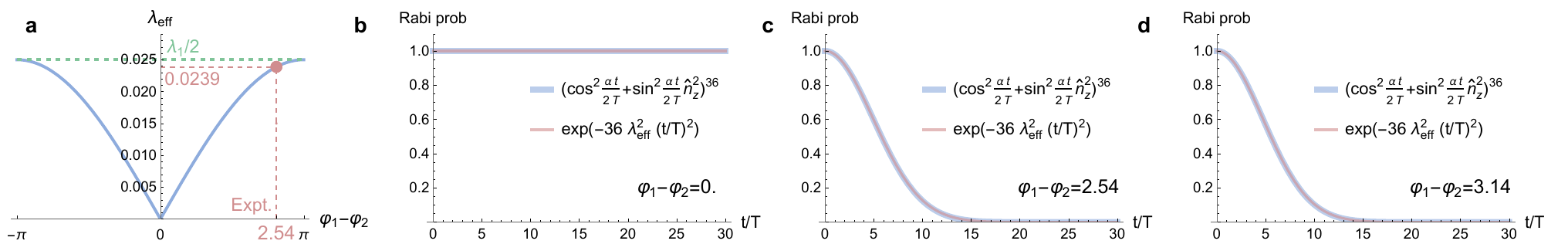}
    \caption{\label{fig:rabi} {\bf a}, Illustration of the effective decay rate $\lambda_{\text{eff}}$ for macroscopic quantum coherence evolved by Rabi oscillators. The fine-tuned single-spin echo occurs at $\yuanphi - \yuanhz = 0$, while close to $\yuanphi - \yuanhz = \pi $ it approaches the maximal value of $\lambda_1/2$. Experimentally we choose a parameter $\approx 2.54$ far detuned from the single-spin echo.  
        {\bf b} -- {\bf d}, Several examples of probability for a GHZ state to remain in the original Fock subspace as evolved by Rabi oscillators. Experimental situation corresponds to {\bf c}. The agreement between rigorous Eq.~\eqref{smeq:rabiprob1} and the approximated \eqref{smeq:rabiprob2} verifies the super-exponential decay. }
\end{figure}

\begin{figure}
    [ht]
    \includegraphics[width=18cm]{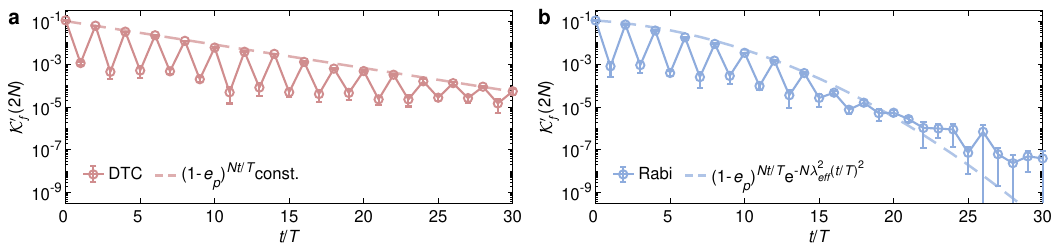}
    \caption{\label{fig:kphi} 
        A reproduction of Fig.~3 in the main text. 
        {\bf a}, the DTC only experience a linear-exponential decay $\sim e^{-t}$ due to external depolarizing noise, which is represented by a straight line in the log-scale plot. 
        {\bf b}, In contrast, the Rabi oscillator involve a super-exponential decay $e^{-t^2}$ due to the leakage of quantum states from the target GHZ state subspace via internal coherent quantum dynamics. Thus, the Rabi case shows a sharp distinction of being represented by an accelerated downward-curved decay. Here we model the noise effect by the depolarization channel, where DTC experiences a stronger effect $e_{p}=0.007$ due to dominant two-qubit gate errors compared with the Rabi case $e_{p}=0.003$ with weaker single-qubit errors and environment noise only.
        }
\end{figure}

Based on this result, we obtain the analytical formula for Rabi oscillators
\begin{align}\label{smeq:KRabi}
    \frac{\fuliyeK}{\fuliyeKchu} = (1-e_p)^{Nt/T} \times e^{-N\lambda_{\text{eff}}^2 (t/T)^2}
\end{align}
Crucially, the evolution of macroscopic quantum coherence for Rabi oscillators is dominated by a super-exponential decay $\sim e^{-t^2}$ caused by coherent quantum dynamics taking particles out of the target GHZ state subspace. Eq.~\eqref{smeq:KRabi} is to be contrasted against the DTC situation (Eq.~\eqref{smeq:KDTC}) with a linear-exponential decay $\sim e^{-t}$ caused simply by external noise. The qualitatively different behaviors lead to sharp distinctions when we plot the evolution of $\fuliyeK$ in a log-scale figure. As shown in Fig.~\ref{fig:kphi} (or Fig.~3 in the main text), the DTC case corresponds to a straight line, while the Rabi case shows an accelerated downward-curved decay due to both external perturbation by noise and internal quantum state leakage via coherent quantum dynamics. 

\subsubsection{Butterfly velocity for lightcone propagation}

Light cones in a cat scar DTC arise from local thermalization processes under the approximate conservation of total domain wall numbers. Thus, we can derive the butterfly velocity for light cone propagation speed by considering the effects of spin flips by perturbations. Here, large parameters involve Ising interaction $J$, the spin $\pi$ pulse, and longitudinal fields $\yuanphi, \yuanhz$. To extract the detuning information that is characterized by small parameters $\lambda_1, \lambda_2$, we consider first the evolution at double-period ends $U_F^2$ to cancel the perfect spin-flip $\pi$-pulse $\exp\left(-i\pi \rxj/2\right) = -i\rxj$. Using $ \rxj (\rxj, \ryj, \rzj) \rxj = (\rxj, -\ryj, -\rzj) $, and perform a gauge transformation $U_F^2 \rightarrow  e^{i\lambda_2\sum_{j=1}^N \ryj/2} U_F^2 e^{-i\lambda_2 \sum_{j=1}^N \ryj/2}$ (which amounts to shifting time origin for periodic drivings), we have
\begin{align}\nonumber
    U_F^2 
    &= e^{-iJ\sum_{j=1}^N \rzj \rzjp}
    \times [ U_P(0,\lambda_2,0) U_P(\yuanphi,\lambda_1, \yuanhz)
    e^{-iJ\sum_{j=1}^N (-\rzj \cos\lambda_2 + \rxj \sin\lambda_2 ) (- \rzjp \cos\lambda_2 + \rxjp \sin\lambda_2 ) }
    U_p^\dagger(\yuanphi, \lambda_1, \yuanhz) U_P^\dagger (0,\lambda_2, 0)]
    \\ 
    &\quad \times 
    \left[ U_P(0,\lambda_2, 0) U_P(\yuanphi, \lambda_1, \yuanhz)
    U_P(-\yuanphi, -\lambda_1, -\yuanhz) U_P^\dagger (0,\lambda_2,0)  \right],
\end{align}
where $U_P(\yuanphi,\lambda_1, \yuanhz) = \prod_{j=1}^N e^{-i\yuanphi \rzj/2} e^{i\lambda_1 \ryj/2} e^{-i\yuanhz \rzj/2}$.
Now, we combine the perturbation to two-qubit rotations into
\begin{align}\nonumber
    &
    U_P(0,\lambda_2,0)  U_P(\yuanphi,\lambda_1,\yuanhz) \left(\rzj\cos\lambda_2 - \rxj \sin\lambda_2 \right) U_P^\dagger(\yuanphi,\lambda_1,\yuanhz) U_P^\dagger (0,\lambda_2,0)
    \equiv
    \alpha_2 \rzj + \beta_2 \rxj + \gamma_2 \ryj,
    \\ \nonumber
    \alpha_2 
    &=
    \cos(\lambda_1) \cos^2(\lambda_2)  - \sin(\lambda_1) \sin(\lambda_2)\cos(\lambda_2) [\cos(\yuanhz) + \cos(\yuanphi) ] 
    +
    \sin^2(\lambda_2) [\sin(\yuanhz)\sin(\yuanphi) - \cos(\lambda_1) \cos(\yuanhz) \cos(\yuanphi) ],
    \\	\nonumber
    \beta_2
    &=
    \sin(\lambda_2) \cos(\lambda_2) \left[\sin(\yuanhz)\sin(\yuanphi) - \cos(\lambda_1)(1+\cos(\yuanhz)\cos(\yuanphi)) \right]
    -
    \sin(\lambda_1) \left[\cos^2 (\lambda_2) \cos(\yuanphi) - \sin^2 (\lambda_2) \cos(\yuanhz) \right],
    \\ \label{smeq:b2g2}
    \gamma_2
    &= 
    -\left[\sin(\lambda_1)\cos(\lambda_2) + \cos(\lambda_1)\sin(\lambda_2) \cos(\yuanhz) \right] \sin(\yuanphi)  
    - \sin(\lambda_2) \sin(\yuanhz) \cos(\yuanphi).
\end{align}
Also, for single-spin perturbations
\begin{align}\nonumber
    &
    U_P(0,\lambda_2, 0) U_P(\yuanphi, \lambda_1, \yuanhz)
    U_P(-\yuanphi, -\lambda_1, -\yuanhz) U_P^\dagger (0,\lambda_2,0)
    \equiv 
    \prod_{j=1}^N \left( \alpha_0 + i\rzj \alpha_1 + i\rxj \beta_1 + i\ryj \gamma_1\right) ,
    \\ \nonumber
    \alpha_0 &= 
    \cos^2\frac{\lambda_1}{2} + \sin^2\frac{\lambda_1}{2} \cos(\yuanhz-\yuanphi),
    \qquad
    \alpha_1 =
    \sin^2\frac{\lambda_1}{2}\cos\lambda_2 \sin(\yuanhz-\yuanphi)
    +
    \sin\frac{\lambda_1}{2} \cos\frac{\lambda_1}{2} \sin\lambda_2 (\sin(\yuanhz) - \sin(\yuanphi))
    \\ \nonumber
    \beta_1
    &=
    - \sin^2\frac{\lambda_1}{2} \sin\lambda_2 \sin(\yuanhz-\yuanphi) 
    +
    \sin\frac{\lambda_1}{2} \cos\frac{\lambda_1}{2} \cos\lambda_2 (\sin\yuanhz - \sin\yuanphi) 
    \\ \label{smeq:b1g1}
    \gamma_1
    &= 
    - 
    \sin\frac{\lambda_1}{2} \cos\frac{\lambda_1}{2} (\cos(\yuanhz) - \cos(\yuanphi))  
\end{align}
Then, the perturbation effects are grouped into
\begin{align}\label{smeq:uf2}
    &U_F^2 = \left[ e^{-iJ\sum_{j=1}^N \rzj \rzjp} e^{-iJ \sum_{j=1}^N (\alpha_2 \rzj + \beta_2 \rxj + \gamma_2 \ryj) (\alpha_2 \rzjp + \beta_2 \rxjp + \gamma_2 \ryjp)} 
    \right]
    \left[
    \prod_{j=1}^N (\alpha_0 + i\rzj\alpha_1 + i\rxj\beta_1 + i\ryj\gamma_1 )
    \right]
    ,
\end{align}
where $\beta_2, \gamma_2$ characterizing two-spin flips and $\beta_1,\gamma_1$ for single-spin flip are all of perturbative strength $\sim \lambda_1, \lambda_2$. 

Up to this point, the derivations are rigorous. To practically extract the butterfly velocity, we make two approximations next. First, we assume the effects of spin flips by perturbations to accumulate independently in different driving cycles and also in the single- and two-qubit drivings separately. This will lead to overestimation of light cone spreading speed at late time because it neglects revivals during coherent quantum dynamics that may delay the thermalization, but would serve as good approximation at early time. Second, based on the eigenstructure of total domain wall separation, we assume that local spin flips during each driving cycle should also conserve the total domain wall numbers.

Based on the first approximation, the one-spin contribution can be estimated by checking the off-diagonal amplitudes $\sim \beta_1, \gamma_1$. Replacing $\rxj \rightarrow 1$, $\ryj \rightarrow i$ in the second part of Eq.~\eqref{smeq:uf2},
\begin{align}\label{smeq:vb1}
    v_B^{(1)} = |\beta_1 + i\gamma_1| .
\end{align}

For the two-spin terms, we further exploit the approximate conservation of total domain wall numbers. That means we can first replace $P_w\sum_{j=1}^N \rzj \rzjp P_w \rightarrow (N-2w)$, where $w$ is the total domain wall number of the initial state, and $P_w$ is a projection to the subspace with $w$ domain walls. Further, domain wall conservation means that a single spin flip can only occur when its two neighboring spins are along opposite directions, namely,
\begin{align}\nonumber
    \uparrow {\color{red}\uparrow} \uparrow 
    &\quad \leftrightarrow \quad
    \uparrow {\color{red}\downarrow} \uparrow 
    \qquad
    \text{domain wall $0\leftrightarrow 2$ \XSolidBrush}
    \\ \nonumber
    \downarrow {\color{red}\uparrow} \downarrow 
    &\quad \leftrightarrow \quad
    \downarrow {\color{red}\downarrow} \downarrow 
    \qquad
    \text{domain wall $2\leftrightarrow 0$ \XSolidBrush}
    \\ \nonumber
    \uparrow {\color{red}\uparrow} \downarrow 
    &\quad \leftrightarrow \quad
    \uparrow {\color{red}\downarrow} \downarrow 
    \qquad
    \text{domain wall $1\leftrightarrow 1$ \CheckmarkBold}
    \\  
    \downarrow {\color{red}\uparrow}  \uparrow
    &\quad \leftrightarrow \quad
    \downarrow {\color{red}\downarrow}  \uparrow
    \qquad
    \text{domain wall $1\leftrightarrow 1$ \CheckmarkBold}
\end{align} 
Such a constraint implies that $ P_w\rxj (\rzjm + \rzjp) P_w = 0 = \ryj (\rzjm + \rzjp)$, as $\sigma^z_{j+1}$ and $\sigma^z_{j-1}$ take opposite values.
Then, the two-spin perturbations are reduced to
\begin{align}\nonumber
    & \quad 
    e^{-iJ\sum_{j=1}^N \rzj \rzjp} e^{-iJ \sum_{j=1}^N (\alpha_2 \rzj + \beta_2 \rxj + \gamma_2 \ryj) (\alpha_2 \rzjp + \beta_2 \rxjp + \gamma_2 \ryjp)}
    \rightarrow
    \text{const.}\times 
    \prod_{j=1}^N e^{-iJ(\beta_2 \rxj + \gamma_2 \ryj) (\beta_2 \rxjp + \gamma_2 \ryjp)}
    \\ 
    &= 
    \text{const.} \times \prod_{j=1}^N
    \left[ 
    \cos\left(J(\beta^2_2 + \gamma^2_2) \right) - i \sin\left(J(\beta_2^2 + \gamma_2^2) \right) 
    \times \frac{(\beta_2 \rxj + \gamma_2 \ryj) (\beta_2 \rxjp + \gamma_2 \ryjp)}{\beta_2^2+\gamma_2^2}
    \right]
\end{align}
Similarly, we consider the off-diagonal amplitude on each site and replace $\rxj\rightarrow 1, \ryj \rightarrow i$, thereby obtaining the contribution from two-spin perturbation
\begin{align}\label{smeq:vb2}
    v_B^{(2)} = \sin(J(\beta_2^2+\gamma_2^2)) 
\end{align}
Altogether, the total butterfly velocity for lightcone propagation from Eqs.~\eqref{smeq:b1g1}, \eqref{smeq:b2g2}, \eqref{smeq:vb1} and \eqref{smeq:vb2} reads
\begin{align}\nonumber
    &v_B = v_B^{(1)} + v_B^{(2)},
    \\ \nonumber			
    &
    v_B^{(1)} = \frac{1}{2} \left|
    -(1-\cos(\lambda_1)) \sin(\lambda_2) \sin(\yuanhz-\yuanphi)   
    + 
    \sin(\lambda_1) [\cos(\lambda_2) (\sin(\yuanhz) - \sin(\yuanphi))
    -i(\cos(\yuanhz) - \cos(\yuanphi)] \right|
    \\ \nonumber
    &
    v_B^{(2)} = \sin(J(\beta_2^2+\gamma_2^2)) ,			
    \\ \nonumber
    & 
    \beta_2 = \sin(\lambda_2) \cos(\lambda_2) \left[\sin(\yuanhz)\sin(\yuanphi) - \cos(\lambda_1)(1+\cos(\yuanhz)\cos(\yuanphi)) \right]
    - 
    \sin(\lambda_1) \left[\cos^2 (\lambda_2) \cos(\yuanphi) - \sin^2 (\lambda_2) \cos(\yuanhz) \right],
    \\ \label{smeq:vb}
    &
    \gamma_2 = -(\sin(\lambda_1)\cos(\lambda_2) + \cos(\lambda_1)\sin(\lambda_2) \cos(\yuanhz)) \sin(\yuanphi) 
    - 
    \sin(\lambda_2) \sin(\yuanhz) \cos(\yuanphi).
\end{align}

\begin{figure}
    [ht]
    \includegraphics[width=18cm]{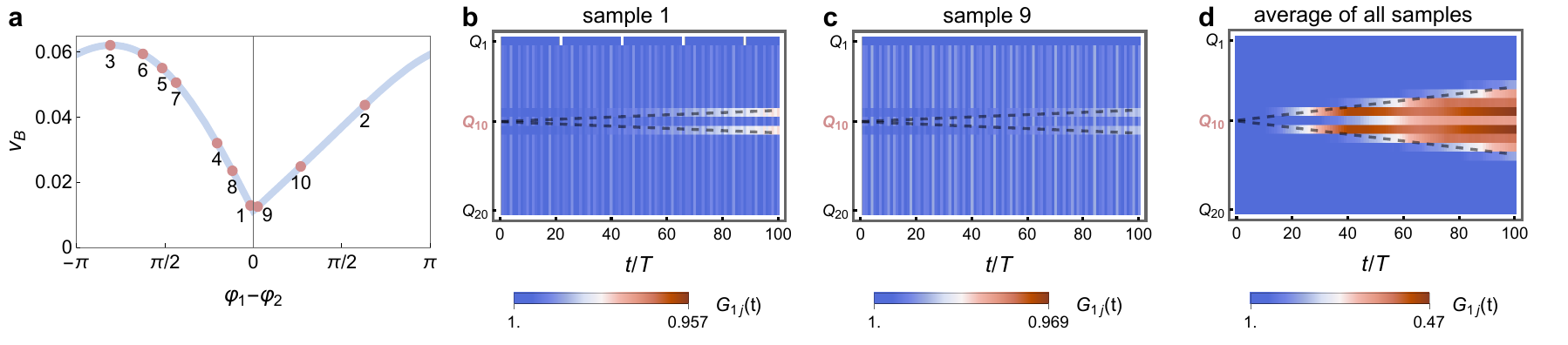}
    \caption{\label{fig:vbecho} 
    {\bf a}, The butterfly velocity given by Eq.~\eqref{smeq:vb} (blue curve). Red dots corresponds to the 10 samples in Table~\ref{tab:phis}.
    {\bf b}, {\bf c}, Numerical simulation of $G_{1j}(t)$ for the two samples close to single-spin echoes, where we see relatively small light cone and slow thermalization rate.
    {\bf d}, Numerical simulation of $G_{1j}(t)$ averaged over all 10 samples.
    }
\end{figure}

Note that when $\lambda_2=0$, the single-spin contribution is reduced to $v_B^{(1)} = \frac{1}{2}\sin\lambda_1 | e^{-i\yuanhz} - e^{-i\yuanphi} | $, which has a full-scale single-spin echo condition $\yuanhz - \yuanphi = 0\mod 2\pi$ without light cone propagation. This corresponds to what Fig.~\ref{fig:rabi}a indicates as the effective decay rate $\lambda_{\text{eff}}$ vanishes at the same single-spin echo point. To avoid such a fine-tuned echo and benchmark many-body effects, we experimentally take 10 random samples 
The deviation from single-spin echo	is summarized in Table~\ref{tab:phis}. The associated butterfly velocity for the 10 cases are plotted in Fig.~\ref{fig:vbecho}a, where numerical simulation of $G_{1j}(t)$ (to be introduced later) for the two cases (samples 1 and 9) close to echoes are shown in Fig.~\ref{fig:vbecho}b, c. A very small lightcone is generated here compared with Fig.~\ref{fig:vbecho}d for the averaged $G_{1j}(t)$ over all samples.
\begin{table}
    [ht]
    \caption{\label{tab:phis} Values of $\varphi_1$ for different samples, so as to check whether single-spin echoes ($|\yuanphi - \yuanhz|=0 \mod 2\pi$) occur. Here we generate 10 random samples of spatially uniform $\yuanphi$ and fix $\yuanhz=0.97$. Samples 1 and 9 are close to single-spin echoes, as we can see later in Fig.~\ref{fig:Gdata} that they produce small light cones. To avoid such non-interacting echoes, in most experiments we use $\yuanphi=-\pi/2$, close to the third sample, which stays far away from single-spin echoes in order to benchmark the interaction effects in stabilizing cat scars against generic perturbation.}
    \begin{tabular}{c c c c  c c c c c c c}
        \hline\hline 
        sample \# &  1 ($\sim$echo) & 2 & {\bf 3 ($\sim$expt)} & 4 & 5 & 6 & 7 & 8 & 9 ($\sim$echo) & 10\\
        \hline
        $\yuanphi$ & 1.04943 & 2.95514& {\bf -1.57008} & 0.32882 & -0.64870& -0.99062& -0.40049& 0.60043  & 0.92217 & 1.81469
        \\
        \hline 
        $ |\yuanphi-\yuanhz | $ & 0.07863 & 1.98434 & {\bf 2.54087} & 0.64198 & 1.61950 & 1.96141& 1.37129&0.37037 & 0.04863 & 0.84389 \\
        \hline\hline
    \end{tabular}
\end{table}

\begin{figure}
    [ht]
    \includegraphics[width=18cm]{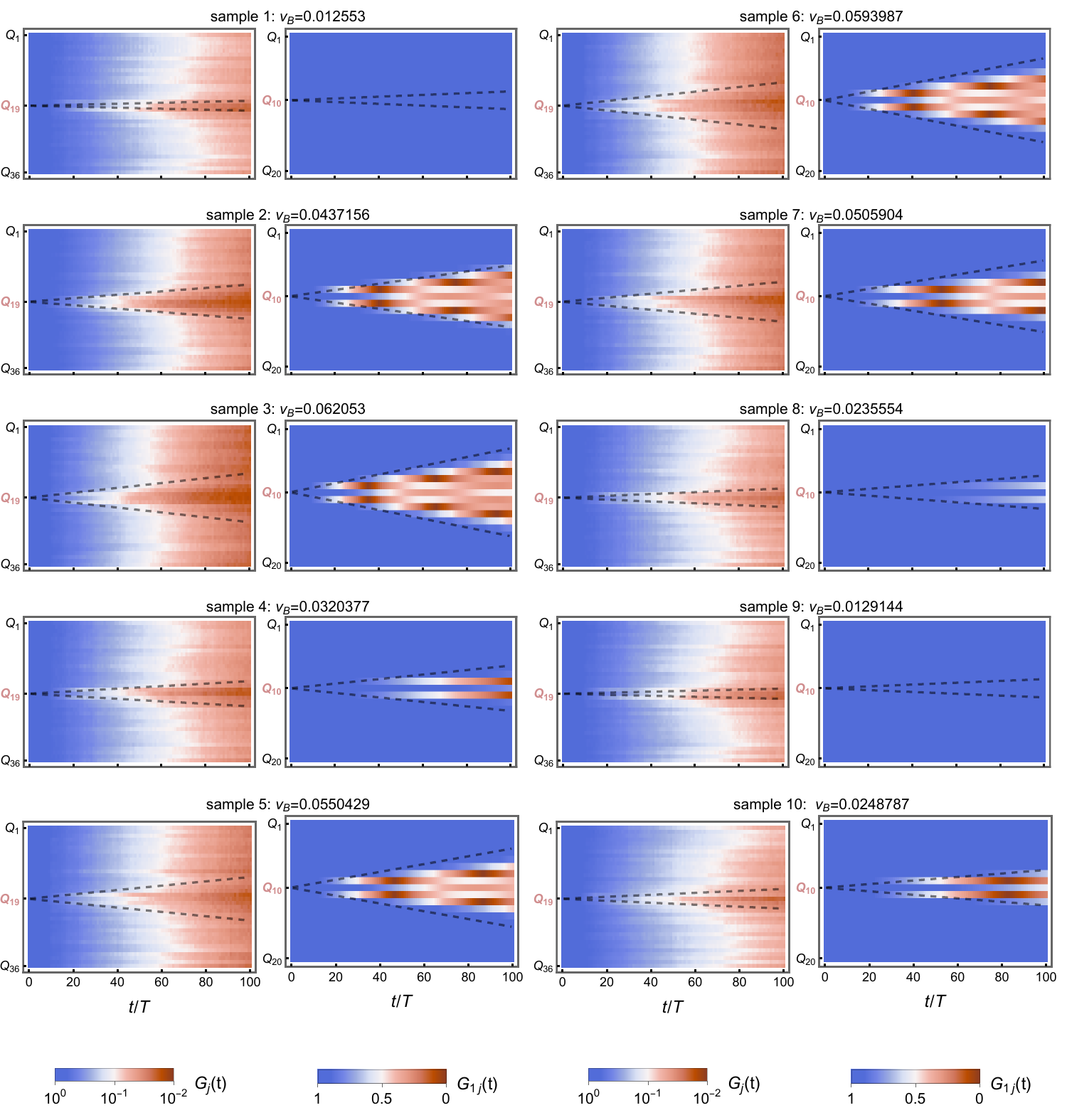}
    \caption{\label{fig:Gdata} 
    (The first and the third columns) Experimentally measured $G_j(t)$ (Eq.~\eqref{smeq:Gjt}) for the 10 samples of $\varphi_1$ in Table~\ref{tab:phis}. Here the initial state is the GHZ state with one qubit (denoted $Q_{19}$) flipped with respect to an antiferromagnetic pattern, where a thermalizing lightcone stems from the two  qubits $Q_{18}, Q_{20}$ centering around flipped $Q_{19}$. Similar to the main text Fig.~4a, b, and d, we sweep over 6 random physical qubits for the flipped $Q_{19}$. 
    (The second and the fourth columns) 
    Numerical simulation for $G_{1j}(t)$ where in the noise-free case we choose $Q_1$ as the observer qubit. The initial state is a GHZ state with one qubit $Q_{10}$ flipped on top of an antiferromagnetic pattern, and similarly lightcones spread out from $Q_{9}, Q_{11}$ surrounding it. Note that samples 1 and 9 close to single-qubit echoes are further shown in Fig.~\ref{fig:vbecho}b and c.
    In all cases, we compare the thermalizing lightcone with analytical butterfly velocity $v_B$ (dashed lines for $Q_{19}\pm v_Bt$ in experimental and $Q_{10}\pm v_Bt$ in numerical data) estimated by Eq.~\eqref{smeq:vb}. 
    }
\end{figure}

With the analytical estimation of butterfly velocity in Eq.~\eqref{smeq:vb}, we next compare the results given by experiments and numerical simulations in Fig.~\ref{fig:Gdata}. We would chiefly focus on the initial GHZ state with the pattern where one spin (at $Q_{10}$ for numerics or $Q_{19}$ for experiments) is flipped on top of the original antiferromagnetic pattern. The flipped spin generates a mini-ferromagnetic domain of three sites surrounding it, and thermalization would occur starting from the two neighboring spins. Such a spatial structure of thermalization can be revealed by the connected correlation function
\begin{align}
    G_{jk}(t) = \left|
    \langle \rzj\rzk \rangle - \langle \rzj	\rangle \langle \rzk \rangle 
    \right|.
\end{align}
For a noise-free numerical simulation without qubit quality differences, we can choose an ``observer" spin, i.e. $Q_1$, and examine its correlation with other qubits. 
In the second and fourth columns of Fig.~\ref{fig:Gdata}, we plot the value of $G_{1j}(t)$. A qualitative agreement for the shape of lightcone between the analytical prediction of $Q_{10}\pm v_Bt$ is observed, with $v_B$ given by Eq.~\eqref{smeq:vb}. In parallel, we compare the analytically derived results with experimental data for each sample in the first and third columns of Fig.~\ref{fig:Gdata}. Here, to cancel the error differences in different physical qubits, we opt to check the averaged correlation for each site with all the remaining ones
\begin{align}\label{smeq:Gjt}
    G_j(t) = \frac{1}{N-1}\sum_{k=1\dots N; k\neq j} G_{jk}(t).
\end{align}
Starting from an initial GHZ state where qubit $Q_{19}$ is flipped on top of an overall antiferromagnetic pattern, a lightcone is similarly generated starting from $Q_{18}, Q_{20}$, with the propagation speed qualitatively in agreement with analytical predictions $Q_{19}\pm v_Bt$, with $v_B$ in Eq.~\eqref{smeq:vb}, because thermalization is a local dynamics.

\begin{figure}[t]
    \includegraphics[width=18cm]{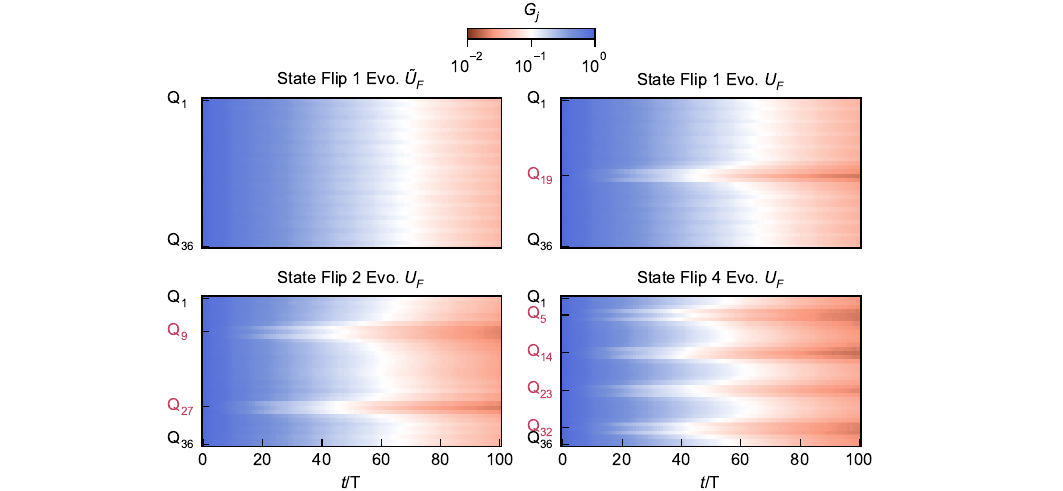}
    \caption{{\bf $G_{j}(t)$ for different initial GHZ states evolving in cat scar DTCs.} Here, for each panel, we take an initial GHZ state, whose qubit pattern involves flipping one (for both of the upper two panels) or more (as denoted by red qubit labels in the lower two panels) qubits on top of an antiferromagnetic pattern. In the upper two panels, the initial state involves $Q_{19}$ flipped on top of an antiferromagnetic GHZ state, which is evolved by a compatible $\tilde{U}_F$ or incompatible (with antiferromagnetic cat scars) $U_F$ respectively. The flipped $Q_{19}$ is swept over 12 equal-spacing physical qubits to cancel qubit error differences. In the lower-left panel, the flipped pair $Q_{9}$-$Q_{27}$ are swept over 9 equal-spacing physical qubits. For the $Q_5$-$Q_{14}$-$Q_{23}$-$Q_{32}$ quadruplet on the lower-right panel, we sweep over all physical qubits with 9 possibilities. } 
    \label{fig:cones}
\end{figure}

Finally, we present additional experimental data for the light cone propagation as revealed by $G_{jk}(t)$ starting from other initial states or evolved by different $\tilde{U}_F$ in Fig.~\ref{fig:cones}. Here, to focus on the thermalization ignited by coherent internal quantum dynamics, we take more sweeps over different physical qubits (see caption for details) to cancel the effects of qubit error differences. On the upper left panel, we first show $G_j(t)$ for the initial state with one qubit $Q_{19}$ flipped on top of an antiferromagnetic pattern, but the initial state is evolved by a compatible $\tilde{U}_F$ (in parallel with Fig.~4d in the main text), where we see suppression of lightcone by cat scar DTCs. In contrast, a lightcone spreads out on the upper right panel for the same initial state evolved by the original $U_F$ hosting incompatible antiferromagnetic cat scars. Following this line of investigation, we continue flipping more qubits for the initial states (i.e. 2 for the lower left and 4 for the lower right panels) and evolving each of them under the same $U_F$ with antiferromagnetic cat scars. We indeed see that more sources of thermalization are ignited, each producing a similar light cone until these light cones tend to merge at late time. These results help verify the expectation that the initial GHZ state with qubit pattern $00110011\dots$, where all individual qubits are sources of thermalization (related to Fig.~4e of the main text), undergoes the most rapid thermalization in a cat scar DTC with an incompatible antiferromagnetic cat scars.

\bibliography{catdtc.bib}